\documentclass[twocolumn]{aastex63}
\turnoffedit

\newcommand{\hb}{H$\beta$}

\newcommand{\hg}{H$\gamma$}

\newcommand{\oiii}{[\ion{O}{3}]}
\newcommand{\oii}{[\ion{O}{2}]}
\newcommand{\feii}{\ion{Fe}{2}}
\newcommand{\hei}{\ion{He}{1}}
\newcommand{\heii}{\ion{He}{2}}

\newcommand{\mgii}{\ion{Mg}{2}}
\newcommand{\civ}{\ion{C}{4}}

\usepackage[bottom]{footmisc}
\usepackage[caption=false]{subfig}
\usepackage[utf8]{inputenc}
\usepackage[T1]{fontenc}
\usepackage{appendix}
\usepackage{graphicx,graphbox}

\hypersetup{
   colorlinks,
   linkcolor={blue!88!black!80},
   citecolor={blue!88!black!80},
   urlcolor={blue!88!black!80}}

\shorttitle{LAMP2016: Velocity-Resolved \hb~Lags}
\shortauthors{U et al.}

\begin{document}

\title{The Lick AGN Monitoring Project 2016: Velocity-Resolved
  H$\beta$~Lags in Luminous Seyfert Galaxies}

\correspondingauthor{Vivian U}
\email{vivianu@uci.edu}

\author[0000-0002-1912-0024]{Vivian U}
\affiliation{Department of Physics and Astronomy, 4129 Frederick Reines Hall, University of California, Irvine, CA 92697, USA}
\affiliation{Department of Physics and Astronomy, University of California, Riverside, CA 92521, USA}

\author[0000-0002-3026-0562]{Aaron J. Barth}
\affiliation{Department of Physics and Astronomy, 4129 Frederick Reines Hall, University of California, Irvine, CA 92697, USA}

\author{H. Alexander Vogler}
\affiliation{Department of Physics and Astronomy, 4129 Frederick Reines Hall, University of California, Irvine, CA 92697, USA}
\affiliation{Department of Physics and Astronomy, University of California, 1 Shields Avenue, Davis, CA 95616, USA}

\author[0000-0001-8416-7059]{Hengxiao Guo}
\affiliation{Department of Physics and Astronomy, 4129 Frederick Reines Hall, University of California, Irvine, CA 92697, USA}

\author[0000-0002-8460-0390]{Tommaso Treu}
\affiliation{Department of Physics and Astronomy, University of California, Los Angeles, CA 90095-1547, USA}

\author[0000-0003-2064-0518]{Vardha N. Bennert}
\affiliation{Physics Department, California Polytechnic State University, San Luis Obispo CA 93407, USA}

\author[0000-0003-4693-6157]{Gabriela Canalizo}
\affiliation{Department of Physics and Astronomy, University of California, Riverside, CA 92521, USA}

\author[0000-0003-3460-0103]{Alexei V. Filippenko}
\affiliation{Department of Astronomy, University of California, 501 Campbell Hall, Berkeley, CA 94720-3411, USA}
\affiliation{Miller Institute for Basic Research in Science, University of California, Berkeley, CA 94720, USA}

\author[0000-0002-3739-0423]{Elinor Gates}
\affiliation{Lick Observatory, P.O. Box 85, Mt. Hamilton, CA 95140, USA}

\author{Frederick Hamann}
\affiliation{Department of Physics and Astronomy, University of California, Riverside, CA 92521, USA}

\author[0000-0003-0634-8449]{Michael D. Joner}
\affiliation{Department of Physics and Astronomy, N283 ESC, Brigham Young University, Provo, UT 84602, USA}

\author{Matthew A. Malkan}
\affiliation{Department of Physics and Astronomy, University of California, Los Angeles, CA 90095-1547, USA}

\author{Anna Pancoast}
\affiliation{Harvard-Smithsonian Center for Astrophysics, 60 Garden Street, Cambridge, MA 02138, USA}

\author{Peter R. Williams}
\affiliation{Department of Physics and Astronomy, University of California, Los Angeles, CA 90095-1547, USA}

\author[0000-0002-8055-5465]{Jong-Hak Woo}
\affil{Astronomy Program, Department of Physics and Astronomy, Seoul National University, 1 Gwanak-ro, Gwanak-gu, Seoul 08826, Korea}
\affil{SNU Astronomy Research Center, Seoul National University, 1 Gwanak-ro, Gwanak-gu, Seoul 08826, Republic of Korea}

\author{Bela Abolfathi}
\affiliation{Department of Physics and Astronomy, 4129 Frederick Reines Hall, University of California, Irvine, CA 92697, USA}

\author[0000-0002-8860-1032]{L. E.~Abramson}
\affiliation{Carnegie Observatories, 813 Santa Barbara Street, Pasadena, CA 91101, USA}

\author{Stephen F. Armen}
\affiliation{Department of Astronomy, San Diego State University, San Diego, CA 92182-1221, USA}
  
\author{Hyun-Jin Bae}
\affil{Astronomy Program, Department of Physics and Astronomy, Seoul National University, 1 Gwanak-ro, Gwanak-gu, Seoul 08826, Korea}

\author{Thomas Bohn}
\affiliation{Department of Physics and Astronomy, University of California, Riverside, CA 92521, USA}

\author[0000-0001-6301-570X]{Benjamin D. Boizelle}
\affiliation{Department of Physics and Astronomy, N283 ESC, Brigham Young University, Provo, UT 84602, USA}
\affiliation{Department of Physics and Astronomy, 4129 Frederick Reines Hall, University of California, Irvine, CA 92697, USA}

\author[0000-0002-4924-444X]{Azalee Bostroem}
\affiliation{Department of Physics and Astronomy, University of California, 1 Shields Avenue, Davis, CA 95616, USA}
\affiliation{DiRAC Institute, Department of Astronomy, University of Washington, 3910 15th Avenue, NE, Seattle, WA 98195, USA}

\author{Andrew Brandel}
\affiliation{Department of Physics and Astronomy, 4129 Frederick Reines Hall, University of California, Irvine, CA 92697, USA}

\author[0000-0001-5955-2502]{Thomas G. Brink}
\affiliation{Department of Astronomy, University of California, 501 Campbell Hall, Berkeley, CA 94720-3411, USA}


\author{Sanyum Channa}
\affiliation{Department of Physics, University of California, Berkeley, CA 94720, USA}
\affiliation{Department of Physics, Stanford University, Stanford, CA 94305, USA}

\author{M. C. Cooper}
\affiliation{Department of Physics and Astronomy, 4129 Frederick Reines Hall, University of California, Irvine, CA 92697, USA}

\author[0000-0002-2248-6107]{Maren Cosens}
\affiliation{Physics Department, California Polytechnic State University, San Luis Obispo CA 93407, USA}
\affiliation{Physics Department, University of California, San Diego, 9500 Gilman Drive, La Jolla, CA 92093 USA}
\affiliation{Center for Astrophysics and Space Sciences, University of California, San Diego, 9500 Gilman Drive, La Jolla, CA 92093 USA}

\author{Edward Donohue}
\affiliation{Physics Department, California Polytechnic State University, San Luis Obispo CA 93407, USA}
\affiliation{Booz Allen, 1615 Murray Canyon Road, Suite 8000, San Diego, CA 92108, USA}

\author[0000-0002-8425-0351]{Sean P. Fillingham}
\affiliation{Department of Physics and Astronomy, 4129 Frederick Reines Hall, University of California, Irvine, CA 92697, USA}

\author[0000-0002-9280-1184]{Diego Gonz\'{a}lez-Buitrago}
\affiliation{Department of Physics and Astronomy, 4129 Frederick Reines Hall, University of California, Irvine, CA 92697, USA}
\affiliation{Universidad Nacional Aut\'onoma de M\'exico, Instituto de Astronom\'ia, AP 106,  Ensenada 22860, BC, M\'exico}

\author[0000-0002-7232-101X]{Goni Halevi}
\affiliation{Department of Astronomy, University of California, 501 Campbell Hall, Berkeley, CA 94720-3411, USA}
\affiliation{Department of Astrophysical Sciences, Princeton University, 4 Ivy Lane, Princeton, NJ 08544, USA}

\author{Andrew Halle}
\affiliation{Department of Physics, University of California, Berkeley, CA 94720, USA}

\author[0000-0003-0034-5909]{Carol E. Hood}
\affiliation{Department of Physics, California State University, San Bernardino, 5500 University Parkway, San Bernardino, CA 92407, USA}

\author[0000-0003-1728-0304]{Keith Horne}
\affiliation{SUPA Physics and Astronomy, University of St~Andrews, North Haugh, St~Andrews, KY16 9SS, Scotland, UK}

\author{J. Chuck Horst}
\affiliation{Department of Astronomy, San Diego State University, San Diego, CA 92182-1221, USA}

\author{Maxime de Kouchkovsky}
\affiliation{Department of Astronomy, University of California, 501 Campbell Hall, Berkeley, CA 94720-3411, USA}

\author{Benjamin Kuhn}
\affiliation{Space Telescope Science Institute, 3700 San Martin Drive, Baltimore, MD 21218, USA}
\affiliation{Department of Astronomy, San Diego State University, San Diego, CA 92182-1221, USA}

\author[0000-0001-8367-7591]{Sahana Kumar}
\affil{Department of Astronomy, University of California, 501 Campbell Hall, Berkeley, CA 94720-3411, USA}
\affil{Department of Physics, Florida State University, 77 Chieftan Way, Tallahassee, FL 32306, USA}

\author[0000-0001-7839-1986]{Douglas C. Leonard}
\affiliation{Department of Astronomy, San Diego State University, San Diego, CA 92182-1221, USA}

\author{Donald Loveland}
\affiliation{Physics Department, California Polytechnic State University, San Luis Obispo CA 93407, USA}
\affiliation{Lawrence Livermore National Laboratory, 7000 East Avenue, Livermore, CA 94550}

\author{Christina Manzano-King}
\affiliation{Department of Physics and Astronomy, University of California, Riverside, CA 92521, USA}

\author{Ian McHardy}
\affiliation{University of Southampton, Highfield, Southampton, SO17 1BJ, UK} 

\author[0000-0003-1263-808X]{Ra\'ul Michel}
\affiliation{Instituto de Astronom\'ia, Universidad Nacional Aut\'onoma de M\'exico, AP 877, Ensenada, Baja California, C.P. 22830 M\'exico}

\author{Melanie Kae B. Olaes}
\affiliation{Department of Astronomy, San Diego State University, San Diego, CA 92182-1221, USA}
  
\author[0000-0001-9877-1732]{Daeseong Park}
\affiliation{Department of Astronomy and Atmospheric Sciences, Kyungpook National University, Daegu, 41566, Republic of Korea}
\affiliation{Korea Astronomy and Space Science Institute, Daejeon, 34055, Republic of Korea}

\author{Songyoun Park}
\affil{Astronomy Program, Department of Physics and Astronomy, Seoul National University, 1 Gwanak-ro, Gwanak-gu, Seoul 08826, Korea}

\author{Liuyi Pei}
\affiliation{Department of Physics and Astronomy, 4129 Frederick Reines Hall, University of California, Irvine, CA 92697, USA}

\author{Timothy W. Ross}
\affiliation{Department of Astronomy, University of California, 501 Campbell Hall, Berkeley, CA 94720-3411, USA}

\author[0000-0003-4852-8958]{Jordan N. Runco}
\affiliation{Department of Physics and Astronomy, University of California, Los Angeles, CA 90095-1547, USA}

\author[0000-0002-8429-4100]{Jenna Samuel}
\affiliation{Department of Astronomy, The University of Texas at Austin, 2515 Speedway, Stop C1400, Austin, TX 78712, USA}
\affiliation{Department of Physics and Astronomy, University of California, 1 Shields Avenue, Davis, CA 95616, USA}

\author[0000-0003-3136-9532]{Javier S\'{a}nchez}
\affiliation{Department of Physics and Astronomy, 4129 Frederick Reines Hall, University of California, Irvine, CA 92697, USA}
\affiliation{Fermi National Accelerator Laboratory, Kirk Rd. \& Pine St, Batavia, IL 60510, USA}
\affiliation{Kavli Institute for Cosmological Physics, 5640 South Ellis Avenue, Chicago, IL 60637, USA}

\author{Bryan Scott}
\affiliation{Department of Physics and Astronomy, University of California, Riverside, CA 92521, USA}

\author[0000-0003-3432-2094]{Remington O. Sexton}
\affiliation{Department of Physics and Astronomy, University of California, Riverside, CA 92521, USA}
\affiliation{U.S. Naval Observatory, 3450 Massachusetts Ave NW, Washington, DC 20392-5420, USA}
\affiliation{Department of Physics and Astronomy, George Mason University, 4400 University Dr, Fairfax, VA 22030-4444, USA}

\author{Jaejin Shin}
\affil{Astronomy Program, Department of Physics and Astronomy, Seoul National University, 1 Gwanak-ro, Gwanak-gu, Seoul 08826, Korea}

\author{Isaac Shivvers}
\affiliation{Department of Astronomy, University of California, 501 Campbell Hall, Berkeley, CA 94720-3411, USA}

\author[0000-0002-4202-4188]{Chance L. Spencer}
\affiliation{Physics Department, California Polytechnic State University, San Luis Obispo CA 93407, USA}
\affiliation{Department of Physics, California State University Fresno, Fresno, CA 93740-8031, USA}

\author[0000-0002-3169-3167]{Benjamin E. Stahl}
\affiliation{Department of Astronomy, University of California, 501 Campbell Hall, Berkeley, CA 94720-3411, USA}
\affiliation{Department of Physics, University of California, Berkeley, CA 94720, USA}

\author{Samantha Stegman}
\affiliation{Department of Astronomy, University of California, 501 Campbell Hall, Berkeley, CA 94720-3411, USA}
\affiliation{Department of Chemistry, University of Wisconsin, Madison, WI 53706, USA}

\author[0000-0001-9685-7049]{Isak Stomberg}
\affiliation{Physics Department, California Polytechnic State University, San Luis Obispo CA 93407, USA}
\affiliation{Deutsches Elektronen-Synchrotron DESY, 22607 Hamburg, Germany}

\author{Stefano Valenti}
\affiliation{Department of Physics and Astronomy, University of California, 1 Shields Avenue, Davis, CA 95616, USA}

\author[0000-0002-1961-6361]{L. Villafa\~{n}a}
\affiliation{Department of Physics and Astronomy, University of California, Los Angeles, CA 90095-1547, USA}

\author[0000-0002-1881-5908]{Jonelle L. Walsh}
\affiliation{George P. and Cynthia W. Mitchell Institute for Fundamental Physics and Astronomy, Department of Physics \& Astronomy, Texas A\&M University, 4242 TAMU, College Station, TX 77843, USA}

\author{Heechan Yuk}
\affiliation{Department of Astronomy, University of California, 501 Campbell Hall, Berkeley, CA 94720-3411, USA}
\affiliation{Department of Physics and Astronomy, University of Oklahoma, 440 W. Brooks St., Norman, OK 73019, USA} 

\author{WeiKang Zheng}
\affiliation{Department of Astronomy, University of California, 501 Campbell Hall, Berkeley, CA 94720-3411, USA}



\begin{abstract}

We carried out spectroscopic monitoring of 21 
low-redshift Seyfert 1 galaxies using the Kast double spectrograph on the 3\,m
Shane telescope at Lick Observatory from April 2016 to May
2017. Targeting active galactic nuclei (AGN) with luminosities 
of $\lambda L_{\lambda}$(5100\,\AA) $\approx 10^{44}$ erg s$^{-1}$ and predicted \hb~lags of
$\sim 20$--30 days or black hole masses of $10^7$--$10^{8.5}$
M$_\odot$, our campaign probes luminosity-dependent trends in broad-line region (BLR) structure and dynamics as well as to improve calibrations for
single-epoch estimates of quasar black hole masses. Here
we present the first results from the campaign, including
\hb~emission-line light curves, integrated \hb~lag times (8--30 days) measured
against $V$-band continuum light curves, velocity-resolved
reverberation lags, line widths of the broad \hb~components, and virial black
hole mass estimates ($10^{7.1}$--$10^{8.1}$ M$_\odot$). Our results add significantly to the number of existing
velocity-resolved lag measurements and reveal a diversity of BLR gas
kinematics at moderately high AGN luminosities. AGN continuum
luminosity appears not to be correlated with the type of kinematics that its BLR gas
may exhibit. Follow-up direct modeling of this dataset will
elucidate the detailed kinematics and provide robust dynamical black
hole masses for several objects in this sample.
\end{abstract}



\section{Introduction} 
\label{sec:intro}
It has been known for about two decades that black hole (BH) mass
exhibits a tight correlation with the stellar velocity dispersion of
the galactic bulge within which it resides, and this relation holds
over several orders of magnitude in black hole
mass~\cite[e.g.,][]{Ferrarese00,Gebhardt00}. This  
relation suggests that supermassive black holes and their host galaxies are tightly linked throughout their lifetimes. A
fundamental understanding of the growth of supermassive black holes
and their interaction with their immediate vicinity will 
provide key constraints on cosmological models of galaxy
evolution~\cite[][]{Springel05,Croton06}. A crucial component in
these numerical recipes is therefore the accurate determination of
black hole masses as a function of cosmic history. 

Robust black hole mass measurements have been limited mostly to
nearby galaxies owing to the exquisite angular resolution needed to
spatially resolve a black hole's sphere of influence.
The technique of temporally resolving the structure of the
broad-line region (BLR) around an actively accreting black hole, 
called reverberation mapping~\cite[RM;][]{Blandford82,Peterson93},
provides a viable alternative --- or in the case of the distant
universe, the only currently available --- tool for directly measuring black hole masses. 

Black hole scaling relations, such as that between the BLR size and the active galactic nuclei (AGN)
rest-frame luminosity at 5100\,\AA~\cite[the radius--luminosity
relationship;][]{Wandel99,Kaspi00,Kaspi05,Bentz06,Bentz09b,Bentz13,Shen13},
enable ``single-epoch'' mass-determination methods for estimating
black hole masses in broad-lined AGN out to high
redshifts. The calibrations for such mass estimates, however, rest
upon the presumptions that distant quasars are similar to the
low-redshift reverberation-mapped AGN, while in fact they exhibit
higher luminosities and larger Eddington ratios~\cite[][]{Richards11,Shen13}. Whether 
virial assumptions for BLR gas dynamics based on local Seyferts 
can be extended to distant quasars must be tested against a
broad range of black hole masses and luminosities. RM, as a tool that probes
BLR gas structure, provides the means for such a test.

The RM method is successful for black hole
mass determination because the emission lines in the BLR gas are found
to respond to the stochastic continuum variations in an AGN via the
transfer function~\cite[][]{Peterson93}
\begin{equation}
  L(v_z,t) = \int^\infty_{-\infty} \Psi(v_z,\tau)~C(t-\tau)~d\tau \quad,
\end{equation}
where $L(v_z,t)$ is the emission-line luminosity at line-of-sight
velocity $v_z$ at observed time $t$, $C(t)$ is the continuum light
curve, and $\Psi(v_z,\tau)$ is the transfer function that maps
continuum variability to the emission-line response at $v_z$ after
some time delay $\tau$. 
This simple model assumes a linear response and no background light; fitting AGN light curves requires the linearized echo model that subtracts off the reference levels, e.g. $L(v_z,t) - L_0(v_z)$~\cite[see discussion in][]{Horne21}. 
Monitoring both the continuum and the
emission-line light curves provides data that can allow the determination of $\Psi(v_z,\tau)$, also
known as the velocity-delay map, whose shape depends on the structure and
kinematics of the BLR~\cite[][]{Horne04}. 
In practice, precise velocity-resolved information in the form of the
transfer function demands intense monitoring with frequent 
sampling, high signal-to-noise ratio (S/N) data, and long duration. As
a result, the number of AGN with such data available to-date has been limited. 

With the goal of refining black hole scaling relations through
expanding the local reverberation-mapped AGN database by which they
are anchored, we embarked on a campaign aimed to monitor AGN at 
higher luminosity than samples targeted in the 2008 and 2011 campaigns of the Lick AGN 
Monitoring Project~\cite[LAMP, e.g.,][]{Bentz09c,Barth15}.
Higher-luminosity AGN tend to have longer lags. Coupled with the fact
that they have weaker variability amplitudes owing to the ample fuel supply in their accretion disks~\cite[][]{vandenBerk04,Wilhite08,MacLeod10}, lag recovery is more
challenging for these objects. The advent of large, multi-object
programs such as SDSS-RM~\cite[][]{Shen15} and OzDES~\cite[][]{Yuan15}
provides a new landscape where large numbers of reverberation
lags for \hb, \mgii, and \civ~in quasars are determined over a broad redshift
range~\cite[][]{Shen16,Grier17,Grier19}.~\cite{Grier17} measured
\hb~time lags for 44 AGN with luminosities $\log$[$\lambda
L_\lambda$(5100\,\AA)/$L_\sun] \approx 43$--45.5 at redshifts $z = 0.12$--1. 
Early results from the SDSS-RM campaign determined \civ~lags in 52
quasars~\cite[][]{Grier19}, whereas those from OzDES found two \civ-based 
black hole masses to be among the highest redshift ($z = 1.9$--2.6) and highest mass black 
holes [$M_\mathrm{BH} = (3.3$--4.4) $\times 10^9$ M$_\sun$] measured thus far with RM studies~\cite[][]{Hoormann19}.
Complementing these large multifiber spectroscopic studies, our
campaign targeted AGN at low redshifts and aimed to yield high-fidelity
data for velocity-resolved lag measurements and dynamical modeling.

During the past several years, different groups have determined resolved 
velocity-delay reverberation signatures for dozens of unique objects 
among local Seyferts, changing-look AGN, and those with
\hb~asymmetry~\cite[e.g.,][]{Bentz10a,Grier13a,Pancoast14b,DeRosa18,Du18b,Williams18,Williams20,Lu19,Zhang19,Sergeev20,Lu21,Bentz21}. 
Our primary intent is to use a large, relatively broad sample to investigate 
statistical luminosity-dependent trends in BLR structure and gas dynamics using
velocity-resolved reverberation, which could directly impact the
accuracy of virial mass estimates. Here we present our first results that
focus on new \hb~velocity-resolved measurements along with integrated 
\hb~lags and virial black hole masses. Our dataset will allow for
forward-modeling work using, for instance, the CARAMEL code~\cite[][Villafa\~{n}a et al., in prep.]{Pancoast11,Pancoast14a} to directly determine black hole masses.

This paper is organized as follows. The
sample selection is described in Section~\ref{sec:sample}. In Section~\ref{sec:speccampaign}, we present our
observational program at Lick Observatory and details of the data
reduction and processing work. Section~\ref{sec:photcampaign} reports
on our photometric campaign to provide the 
continuum light curves for our AGN sources. In Section~\ref{sec:lc}, we
illustrate our emission-line light curves and subsequent integrated
and velocity-resolved \hb~lag detections, including an
assessment of the lag significance and the observed variety
of BLR kinematics. Section~\ref{sec:bh} depicts our line-width 
measurements and derived virial black hole masses, within the larger
context of how our results compare with the existing AGN
radius--luminosity relation.

Throughout the paper, we have adopted
$H_0 = 67.8$\,km\,s$^{-1}$\,Mpc$^{-1}$, $\Omega_{\rm m}$ = 0.308, and 
 $\Omega_{\rm vac}$ = 0.692
~\cite[][]{Planck16}.

\section{Sample Selection}
\label{sec:sample}
The past Lick AGN Monitoring Project (LAMP) campaigns targeted AGN with \hb~lags of $\sim 3$--15 days~\citep{Bentz09c,Bentz10b,Barth15}.
The chief objective of this campaign is to investigate BLR
kinematics of moderate-luminosity AGN, 
so we targeted primarily Seyfert 1s with 
extinction-corrected
$\log$[$\lambda L_\lambda$(5100\,\AA)/$L_\sun$] $\approx 43.5$--43.9.
Based on the radius--luminosity 
relation~\cite[e.g.,][]{Bentz13}, the targeted AGN continuum luminosity range corresponds to an \hb~lag range of 20$-$30 days.
We note, however, that recent results from the SEAMBH~\citep{Du16a,Du18a,Du19} and SDSS-RM~\citep{Grier17} programs have shown that some AGN have lags shorter than would be expected from earlier versions of the radius--luminosity relation~\citep{Bentz13}.

We applied this 20--30-day criterion in \hb~lag or equivalently $\lambda L_{\lambda}$(5100\,\AA) to
various catalogs of Type 1 AGN~\cite[e.g.,][]{BG92,Marziani03,Peterson04,Vestergaard06,Bachev08,Winter10,Shen11,Zu11,Joshi12,Bennert15,Sun15}. 
Our selection was further narrowed with a redshift limit at $z < 0.08$ to
ensure that the \hb~and \oiii~lines fall blueward of the cutoff wavelength
of the dichroic used in the Kast spectrograph (5500\,\AA). Sources were selected
with declination $\delta \geq$ $-$5\degr~and magnitude $<$ 17 in the optical 
$V$ or $r$ band. We also
assessed the short-term variability in previous light curves from the Catalina
Real-time Transient Survey~\cite[][]{Drake09} for all our candidate
sources and selected those with at least 0.1 magnitude of variations. Targets having high-quality
velocity-resolved lag measurement from previous RM
campaigns were excluded \cite[e.g.,][]{Denney10,Grier12,Du14,Wang14a}, as were those no
longer featuring broad hydrogen recombination lines in archival Lick optical spectra.

In order to ensure that lags can be measured accurately, we placed a
further constraint requiring that the monitoring duration of an AGN, defined as
the longest continuous period when the AGN can be observed at
airmass $<$ 2 during our campaign, be at least 3 times
larger than the expected \hb~lag based on simulations. Given that our initial campaign duration was designed to be 9 months long (which extended to 1\,year later on), this constraint resulted in a 
sample of 29 objects with duration-to-lag ratio
$>$ 3. Two sources with slightly shorter \hb~lag estimates
(Mrk 315: 13 days; Mrk 704: 16 days) were included to better fill 
the right-ascension range of the sample.

After an initial probationary period of $\sim1$ month, we discarded 8 sources showing low continuum variability amplitude, retaining a sample of 21 objects for ongoing monitoring. This final sample of Seyfert 1s and their properties are listed in
Table~\ref{tbl:sample}.  

\begin{deluxetable*}{llcccc}[htb]
\tablecaption{Sample Properties \label{tbl:sample}}
\tablecolumns{6}
\tablewidth{0pt}
\tablehead{
\colhead{Object} &
\colhead{Other Name(s)} &
\colhead{Right Ascension} &
\colhead{Declination} &
\colhead{Redshift} &
\colhead{Reference} \\
\colhead{} & 
\colhead{} & 
\colhead{(J2000)} & 
\colhead{(J2000)} & 
\colhead{} & 
\colhead{} 
}
\startdata
Zw 535-012 & & 00:36:20.983 & $+$45:39:54.08 & 0.04764 & 1 \\
I Zw 1 & UGC 00545, Mrk 1502, PG 0050+124 & 00:53:34.940 & $+$12:41:36.20 & 0.05890 & 2 \\
Mrk 1048 & NGC 985, VV 285 & 02:34:37.769 & $-$08:47:15.44 & 0.04314 &  2 \\
Ark 120 & UGC 03271, Mrk 1095 & 05:16:11.421 & $-$00:08:59.38 & 0.03271 & 3 \\
Mrk 376 & KUG 0710+457, IRAS 07105+4547 & 07:14:15.070 & $+$45:41:55.78 & 0.05598 &  3 \\
Mrk 9 & & 07:36:56.979 & $+$58:46:13.43 & 0.03987 &  1,3 \\
Mrk 704 & CGCG 091-065, MCG +03-24-043 & 09:18:26.005 & $+$16:18:19.22 & 0.02923 &  4 \\
MCG +04$-$22$-$042 & & 09:23:43.003 & $+$22:54:32.64 & 0.03235 &  5 \\
Mrk 110 & PG 0921+525 &  09:25:12.870 & $+$52:17:10.52 & 0.03529 &  6,7 \\
RBS 1303 & CGS R14.01 & 13:41:12.904 & $-$14:38:40.58 & 0.04179 &  4 \\
Mrk 684 & & 14:31:04.783 & $+$28:17:14.11 & 0.04608 &  1 \\
Mrk 841 & J15040+1026 & 15:04:01.201 & $+$10:26:16.15 & 0.03642 &  5 \\
Mrk 1392 & 1505+0342 & 15:05:56.553 & $+$03:42:26.32 & 0.03614 &  8 \\
SBS 1518+593 & & 15:19:21.650 & $+$59:08:23.70 & 0.07810 &  9 \\
3C 382 & CGCG 173-014 & 18:35:03.390 & $+$32:41:46.80 & 0.05787 & 2 \\
NPM1G+27.0587 & 2MASX J18530389+2750275 & 18:53:03.874 & $+$27:50:27.72 & 0.06200 & 10 \\
RXJ 2044.0+2833 & & 20:44:04.500 & $+$28:33:12.10 & 0.05000 & 10 \\
PG 2209+184 & II Zw 171 & 22:11:53.889 & $+$18:41:49.86 & 0.07000 & 2 \\
PG 2214+139 & Mrk 304 & 22:17:12.262 & $+$14:14:20.89 & 0.06576 &  2 \\
RBS 1917 &  2MASX J22563642+0525167 & 22:56:36.500 & $+$05:25:17.20 & 0.06600 &  2 \\
Mrk 315 & & 23:04:02.622 & $+$22:37:27.53 & 0.03887 &  1
\enddata
\tablecomments{Objects in this and subsequent tables and figures are listed in RA order. Redshifts are from NED.
References for $\lambda L_{\lambda}$(5100\,\AA) used to estimate the \hb~lag in our sample selection: (1) Previous Lick spectra; (2)~\cite{Marziani03}; (3)~\cite{Joshi12}; (4)~\cite{Barth15}; (5)~\cite{Winter10}; (6)~\cite{Peterson04}; (7)~\cite{Zu11}; (8)~\cite{Bennert15}; (9)~\cite{Sun15}; (10)~\cite{Bachev08}}
\end{deluxetable*}

\section{The Spectroscopic Campaign at Lick} 
\label{sec:speccampaign}

\subsection{Overview of the Observational Program}
We conducted a spectroscopic monitoring program of 1\,yr duration at Lick
Observatory on Mount Hamilton, California. 
Prior to the start of the campaign, we ran simulations to determine the minimum threshold for successful lag recovery among monitoring periods of varying lengths and cadences.
Our estimated \hb~lag range
required a sampling cadence of two nights per week over the course of a year to
sufficiently resolve time-dependent emission-line variations and to
extend the temporal baseline for detecting robust reverberation signal
across the entire extent of the BLR, accounting for anticipated weather
losses and seasonal gaps.  

This project was allocated 100 nights at the Lick 3\,m Shane telescope
across three observing semesters between 2016 April 28 and 2017 May 6 (UT). The observing runs were distributed mostly with a frequency of
$\sim 2$ nights per week as requested, with the exception of bright time
and occasional scheduling constraints owing to other time-sensitive
observing programs. Partial nights were occasionally exchanged
with or obtained from other programs to 
facilitate the sampling cadence of certain targets during this
monitoring period.   

Spectroscopic data were taken using the Kast Double
Spectrograph~\cite[][]{Miller94}. The red CCD detector was upgraded 
halfway into our campaign (September 17--20,
2016). The upgrade from the Reticon $400 \times 1200$ pixel CCD to the
Hamamatsu $4096 \times 1024$ pixel CCD significantly
improved the quantum efficiency, up to a factor of 2 at the red
end, and removed severe fringing effects. Images taken with the new red CCD suffered from a high incidence rate of cosmic-ray hits and alpha-particle hits due to radioactive material in the new dewar window. These defects were removed at the reduction stage where frames were combined before the spectrum was extracted.
 
We employed the following setup for our AGN observations: the
D55 dichroic with the 600/4310 grism (nominal coverage of
3300--5520\,\AA~at 1.02\,\AA/pixel) on the blue side and the 600/7500
grating (nominal coverage of 4000--11,000\,\AA~at 2.35\,\AA/pixel
pre-upgrade, and 3800--10,000\,\AA~at 1.31\,\AA/pixel post-uprade) on the
red side, a compromise between broad spectral coverage and 
moderate spectral resolution. The plate scale was 0\farcs43/pixel
for both the blue and the new red detectors, and 0\farcs78/pixel for the old red detector. A slit width of 4\arcsec~was employed to minimize slit losses due to seeing variations between different nights~\citep{Filippenko82}. A fixed position angle, optimized for each individual target, was chosen so that the spectroscopic aperture sampled the same portion of the host galaxy across the duration of the monitoring campaign. 

Exposure times were typically up to 30\,min per object each night,
split into 2 or 3 separate exposures (pre- and post-upgrade of the red
CCD, respectively) to facilitate cosmic-ray cleaning. Standard
calibrations including bias exposures, arc frames (using the Ar, He,
Hg, Cd, and Ne lamps), and dome flats were taken during the afternoon,
and well-calibrated flux standard stars (G191B2B, Feige 34, BD+284211,
HZ44, BD+262606, and/or BD+174708) were observed during the
nights. Before the red CCD upgrade, red-side observations suffered
from severe fringing at wavelengths longward of 7000\,\AA. In order to
remove the fringe pattern, dome flats at the position of every object
were taken before or after each standard star and AGN observation on
the red side. This time-consuming procedure was no longer needed after
the upgrade that eliminated the fringing effect. The observing
parameters for our sample are listed in 
Table \ref{tbl:observing}.  

The wet 2016--2017 winter season at Lick hampered our observing
effort, particularly during December through February.
Overall, our spectroscopic campaign achieved an observational success
rate of $\lesssim 70$\%, where 30 nights were lost entirely to weather or
dome issues. For another 12 nights, only 6 or fewer AGN were
observed when typically 12--15 objects would have been observed on a night with good conditions. We were confident that at least fourteen nights of observations were carried out under photometric
conditions. Of the 70 nights when data were obtained, half were done
with the old red CCD and half with the upgraded detector. The number
of total epochs observed for each object ranges from 22 to 50, with a
median of 38 (see Table \ref{tbl:observing}).  

\begin{deluxetable*}{lcccccc}[htb]
\tablecaption{Observing Parameters \label{tbl:observing}}
\tablecolumns{7}
\tablewidth{0pt}
\tablehead{
\colhead{Object} &
\colhead{Slit PA} &
\colhead{$t_{\rm exp}$} &
\colhead{Monitoring Period} &
\colhead{$N_{\rm obs}$} &
\colhead{S/N} &
\colhead{$N_{\rm phot}$} \\
\colhead{} & 
\colhead{(deg)} & 
\colhead{(s)} & 
\colhead{(UT)} & 
\colhead{} &
\colhead{} &
\colhead{} 
}
\startdata
Zw 535$-$012 & 100 & 1800 & 2016/06/27$-$2017/03/04 & 41 & 57 & 10 \\
I Zw 1 & 45 & 1600 & 2016/07/03$-$2017/02/01 & 34 & 103 & 6 \\
Mrk 1048 & 0 & 1800 & 2016/08/08$-$2017/02/16 & 27 & 88 & 5 \\
Ark 120 & 0 & 1800 & 2016/08/15$-$2017/03/26 & 34 & 124 & 5 \\
Mrk 376 & 70 & 1800 & 2016/05/01$-$2017/05/01 & 31 & 84 & 6 \\
Mrk 9 & 90 & 1800 & 2016/05/01$-$2017/05/01 & 33  & 78 & 5 \\
Mrk 704 & 130 & 1800 & 2016/10/20$-$2017/05/01 & 23 & 106 & 3 \\
MCG +04$-$22$-$042 & 145 & 1800 & 2016/05/01$-$2017/05/01 & 34 & 54 & 7 \\
Mrk 110 & 44 &  1200 & 2016/05/01$-$2017/05/01 & 41 & 74 & 6 \\
RBS 1303 & 20 & 1200 & 2016/05/01$-$2017/05/01 & 22 & 67 & 5 \\
Mrk 684 & 60 & 1800 & 2016/05/01$-$2017/05/01 & 43 & 98 & 12 \\
Mrk 841 & 50 & 800 & 2016/05/01$-$2017/05/01 & 45 & 77 & 11 \\
Mrk 1392 & 50 & 1800 & 2016/05/01$-$2017/05/01 & 39 & 55 & 10 \\
SBS 1518+593 & 90 & 1800 & 2016/05/01$-$2017/05/01 & 48 & 44 & 8 \\
3C 382 & 70 & 1800 & 2016/05/01$-$2016/12/03 & 50 & 81 & 12 \\
NPM1G+27.0587 & 60 & 1800 & 2016/05/01$-$2016/12/03 & 38 & 55 & 7 \\
RXJ 2044.0+2833 & 60 & 1800 & 2016/05/01$-$2016/12/31 & 46 & 58 & 9 \\
PG 2209+184 & 60 & 1800 & 2016/05/01$-$2016/12/31 & 40 & 32 & 9 \\
PG 2214+139 & 60 & 1800 & 2016/05/18$-$2016/12/31 & 43 & 81 & 10 \\
RBS 1917 & 30 & 1800 & 2016/06/01$-$2016/12/31 & 32 & 39 & 9 \\
Mrk 315 & 60 & 1800 & 2016/06/07$-$2016/12/31 & 35 & 59 & 9
\enddata
\tablecomments{$t_{\rm exp}$ represents the typical total on-source
  exposure time for each AGN every night it was observed, usually split
  among 2--4 frames. $N_{\rm obs}$ represents the total number of
  spectroscopic observations for each source. S/N represents the median
  S/N per pixel in the continuum at (5100--5200)\,($1+z$)\,\AA.  $N_{\rm phot}$ represents the number of
  photometric nights for each source.}
\end{deluxetable*}

\subsection{Spectroscopic Data Reduction}
The Kast spectroscopic data were reduced using a combination of
standard routines in IRAF 
and IDL, following the procedure outlined
by~\cite{Barth15}.  Here we provide a brief description of the
reduction process of the blue-side data and pre-/post-upgrade red-side data. 

In general, the data were processed via bias
subtraction, flat fielding, cosmic-ray cleaning, one-dimensional (1D) extraction using an unweighted boxcar extraction region,
wavelength calibration, and flux calibration. Error spectra were extracted and propagated through all subsequent calibration steps. Multiple spectra taken on the same night were combined. The width adopted for the spectral extraction was
10\arcsec, though it was widened to $\sim15$--20\arcsec~for some
sources (Mrk 315, Mrk 9, MCG +04$-$22$-$042, and NPM1G+27.0587) and
during particular nights where the seeing was exceptionally poor. 

Most of the AGN were flux calibrated using the standard-star exposure that was closest in airmass. 
However, the flux-calibrated spectra for a few epochs exhibited some abnormal slopes and features close to the dichroic cutoff. We were unable to conclusively determine the cause of these anomalies, but they were most likely related to the dichroic. Since the \oiii~$\lambda$5007 line, situated close to the dichroic cutoff, was assumed to be constant throughout the observing campaign and was used to normalize the nightly spectra, this peculiarity identified in several spectra added extra scatter to the \hb~light curves. In these cases, we attempted to correct the problem by calibrating the AGN with standard stars that were taken close in time for six nights. As a result, this flux-calibration issue was partially rectified but remained responsible for some residual scatter in the emission-line light curves.
\edit1{For eight sources (Ark 120, RBS 1303, Mrk 9, RXJ 2044.0$+$2833, Zw 535$-$012, SBS 1518$+$593, Mrk 684, and Mrk 110), we further divided the \oiii~light curve by the median \oiii~flux, and subsequently normalized the \hb~light curve with the residual \oiii~light curve. This normalization helped to mitigate the residual scatter remaining in the \hb~light curves. We suspect that this flux-calibration anomaly due to the dichroic was occurring at a lower level throughout much of the campaign, which caused scatter larger than what might be typically achievable in other RM datasets.}

For red-side data taken prior to 2016 Sep. 20, each observation
was flattened using a dome-flat exposure taken at the same telescope
position. For red-side data
taken between 2016 Sep. 20 and 2017 Jan. 31, the Kast red
dewar was slightly tilted and thus the two-dimensional spectra had to be
adjusted with a rotation of 0.9\degr~before spectral
extraction. 

\subsection{Photometric and Spectral Scaling} 
To ensure that all the spectra for each AGN are on consistent
flux and wavelength scales across the monitoring period, a modified
version of the procedure described by 
\citet[hereafter VGW92]{VGW92} was applied to the blue-side data for internal
calibration~\cite[see][for more
details]{Barth15,Fausnaugh17_mapspec}. Having the spectra
exhibit consistent spectral resolution throughout the temporal series is important particularly
for extracting velocity-resolved measurements. First, we computed the shifts in wavelength among the time-series spectra from cross correlation. We applied these shifts to align the spectra and calculated an initial mean spectrum that was then designated as the reference. The \oiii~$\lambda$5007 fluxes, presumed to
be intrinsically constant throughout the course of the observing campaign, were
measured from spectra taken during photometric nights and were
averaged to estimate the true absolute flux of \oiii. The number of photometric nights designated per object ranged from 3 to 12, with a median of 8 nights. 
The resulting reference spectrum was then scaled to match this measurement of the \oiii~flux, with a median uncertainty of 8\%.

Each night's spectrum was then aligned and scaled to the reference by
matching the \oiii~$\lambda$5007 emission-line profile following our modified 
VGW92 scaling method. In order to minimize variations outside
of intrinsic AGN variability, we aimed to achieve a uniform spectral
resolution across the full time series of spectra. Where VGW92 ignores
resolution corrections for epochs observed at lower resolution than
the reference spectrum, we adopted instead a modified approach similar to that 
described by~\cite{Fausnaugh17_mapspec}: we first applied a Gaussian
broadening kernel to the reference spectrum to ensure that all the rescaled spectra match a single
resolution, typically the worst among the spectra. Excluding the occasional
extreme outliers, this broadening kernel had a typical $\sigma$ of $\sim 1.5$--3\,\AA~with a median $\sigma$ of 2\,\AA~for each AGN.
Each of the spectra was then aligned by wavelength, scaled in flux, and broadened
in spectral resolution to match the \oiii~line profile of the broadened reference spectrum.

We performed a comparison of the modified VGW92 spectral scaling method to that using the Python code
\texttt{mapspec}~\cite[][]{Fausnaugh17_mapspec} with the same
wavelength windows and broadened reference spectrum.  Differences
between the two approaches include the following: (i) VGW92 calls for using a
Gaussian kernel for smoothing, while \texttt{mapspec} offers an additional option of
using Gauss-Hermite polynomials; and (ii) our modified VGW92 approach fits for free parameters 
via $\chi^2$~minimization, while \texttt{mapspec} uses a
Bayesian framework to optimize rescaling parameters and estimate
model uncertainties. The runtime for the modified VGW92 method was
typically $\sim 100$ times shorter than for \texttt{mapspec}. 
We ran several tests comparing the level of scatter in the light curves of the \oiii~line that resulted from both methods and concluded that modified VGW92 performed better for most of the AGN in our sample. We subsequently adopted the spectra scaled with the modified VGW92 approach for the ensuing analysis.

\begin{figure*}[bht!]
  \centering
  \includegraphics[angle=90,width=\textwidth]{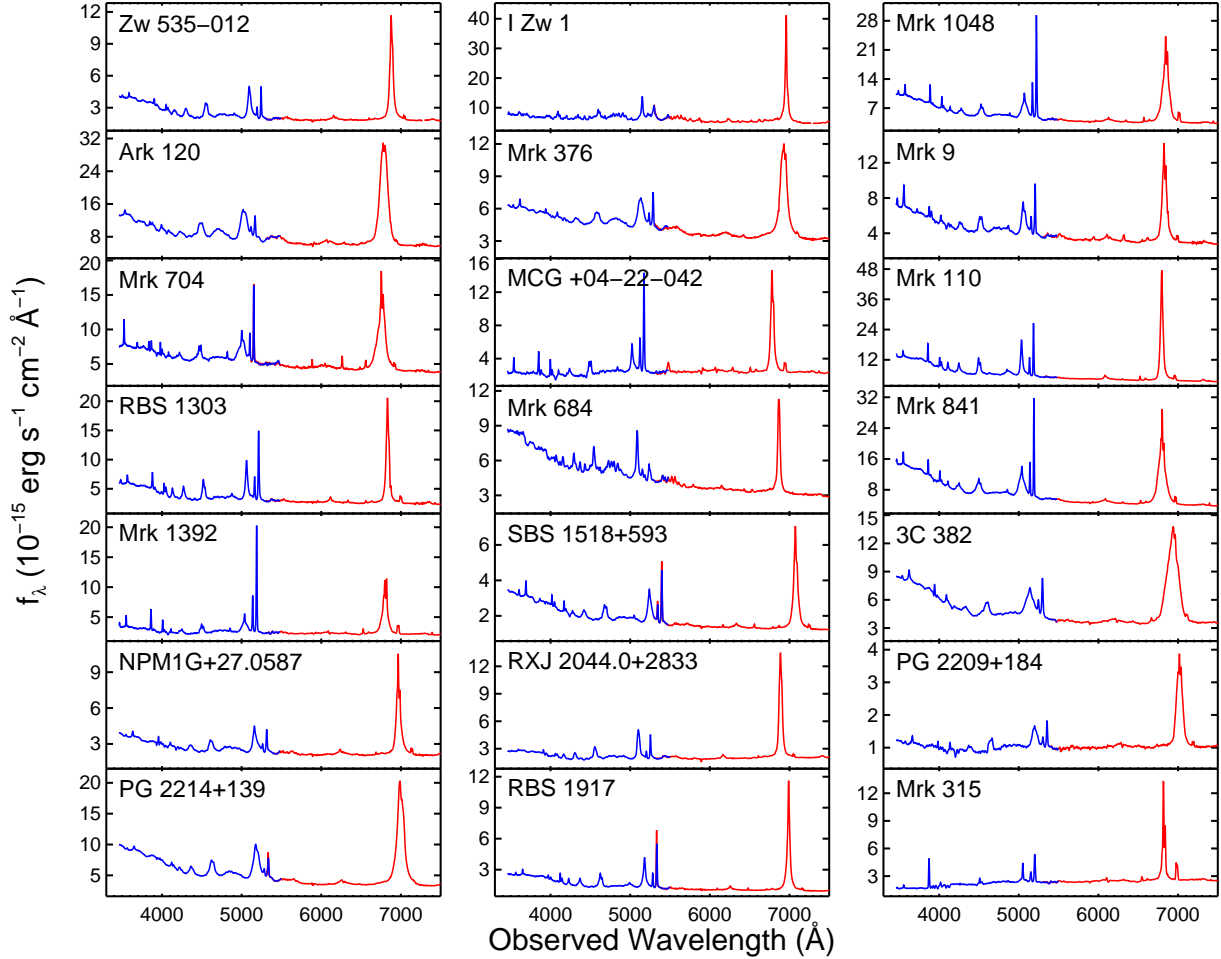}
\caption{Mean spectra for the LAMP2016 sample including both blue- and
  red-side data. Here, as in Table~\ref{tbl:sample}, the AGN are shown in order of increasing right ascensions.\label{fig:1dspec}}
\end{figure*}

The scaling of the red-side data took a different approach. To
normalize the red-side flux scale, red-side spectra were stitched to the
corresponding blue-side spectra via aligning their overlapping regions ($\sim
5300$--5500\,\AA). All spectra were first aligned to the reference spectrum in wavelength. Because the ends of the spectra 
tended to be noisy, we used a weighted average to determine the
overall multiplicative scaling factor. This
technique worked well in most cases where the spectra were smooth or
well behaved at the ends.  
The average scaled spectra for the sample are presented in Figure \ref{fig:1dspec}. 

\subsection{Noise-Corrected RMS Spectra}
Relative variability across the spectrum can be visualized using the root-mean-square
(rms) spectrum. The~\cite{Peterson04} 
procedure of taking the standard deviation of flux values at each
wavelength element over all the epochs may inadvertently bias the rms
spectrum given the inclusion of various noise
factors in addition to genuine AGN variability \cite[][]{Barth15}. To remove the contribution due to
photon-counting noise from the rms spectrum,~\cite{Pei17} suggested an
approach to produce the ``excess rms'' (e-rms hereafter) spectrum
defined per wavelength element $\lambda$ and epoch $i$ as
\begin{equation}
\textrm{e-rms}_\lambda = \sqrt{\frac{1}{N-1} \sum^N_{i=1}[(F_{\lambda,i} -
  \langle F_\lambda \rangle )^2 - \delta^2_{\lambda,i}]} \quad, 
\label{eqn:erms}
\end{equation}
where $N$ is the number of epochs in the time series, $\langle
F_\lambda \rangle$ is the
flux averaged over the time series, and $F_{\lambda,i}$ and
$\delta_{\lambda,i}$ refer to the flux and rms uncertainty in the flux at each $\lambda$ and
$i$, respectively. This was applied to the set of scaled spectra for each AGN.

\begin{figure*}[htb!]
  \centering
  \includegraphics[angle=90,width=\textwidth]{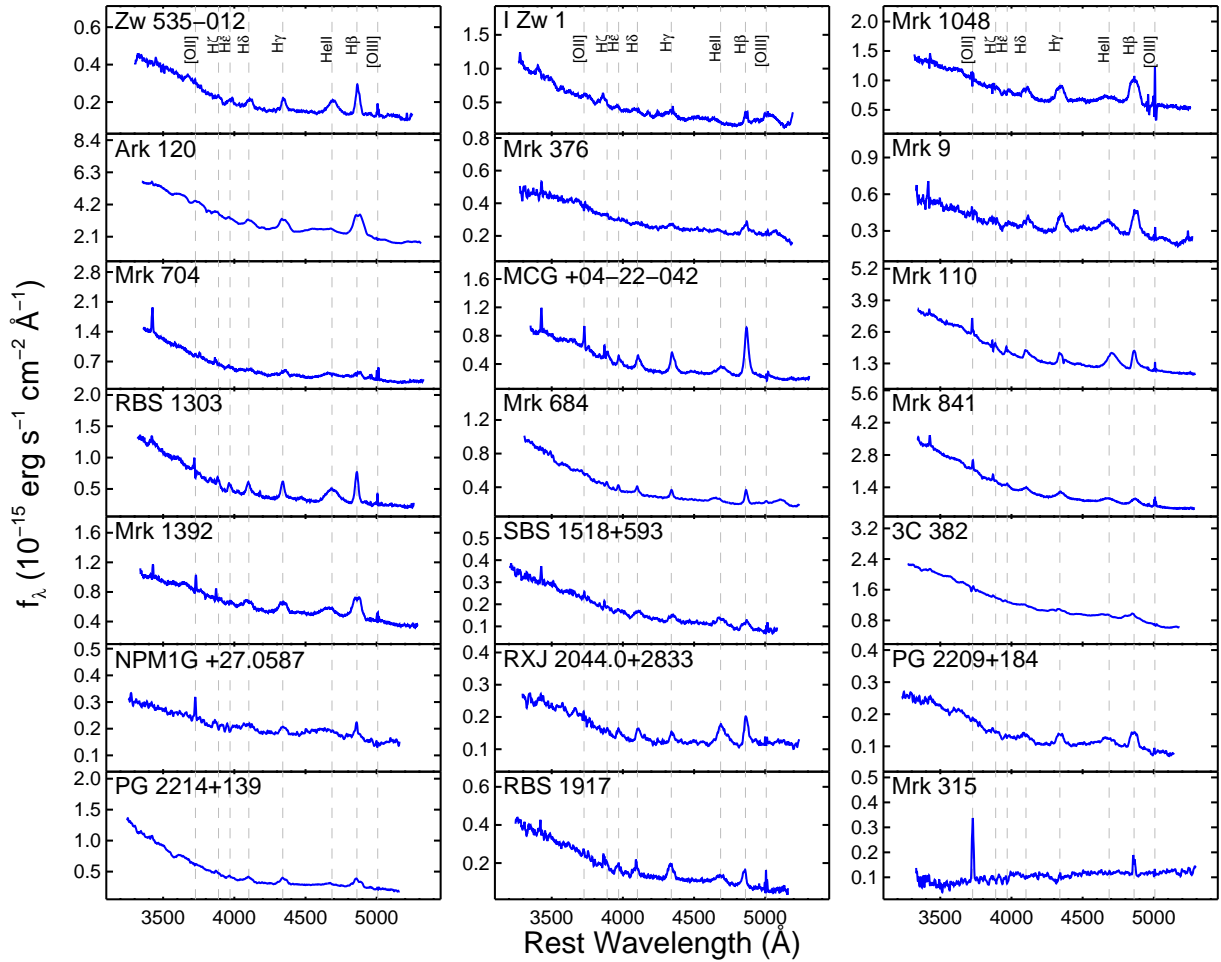}
\caption{The blue-side e-rms spectra for the VGW92-scaled 
  time-series data featuring strong blue continuum and broad-line emission in the Balmer series, overlaid with dashed lines indicating the wavelengths of the prominent emission features (labeled in the top panels). 
  The absence of residual narrow \oiii~features demonstrates the effectiveness of our internal flux calibration at the red end, but other residual narrow-line features (such as the \oii~$\lambda$3727 feature in Mrk 315) is most likely due to
  a combination of underlying continuum variability and imperfect spectrophotometric calibration errors at the blue end.
\label{fig:1derms}}
\end{figure*}

In the case of an AGN with strongly varying broad-line profiles, broad
features are expected to dominate the e-rms spectra in the relevant
spectral regions. The narrow \oiii~lines, on the other hand, generally
vanish in the rms given that they are assumed to be constant within the
time-series data. Presented in Figure~\ref{fig:1derms}, the e-rms spectra generally feature a strong blue continuum (except for Mrk 315), broad-line emission in the Balmer series and (usually) in \heii. The absence of residual narrow \oiii~features confirms that our internal flux calibration worked well near this wavelength, but other residual narrow-line features are likely artifacts of imperfect flux calibration errors that increase at bluer wavelengths. This is particularly the case for Mrk 315, which has a red e-rms continuum slope and prominent \oii~$\lambda$3727.

\subsection{Spectral Decomposition}
\label{sec:specdec} 

The traditional approach of measuring broad emission-line fluxes using a simple linear continuum subtraction is subject to several inadequacies. A linear continuum model is unable to separate blended emission-line features, such as \ion{He}{2} or \ion{Fe}{2} lines that often overlap the broad H$\beta$ profile. Furthermore, a linear fit is a poor approximation to the actual continuum, which includes both AGN and host-galaxy components. Residual errors from continuum subtraction can be particularly problematic for velocity-resolved RM in the faint high-velocity wings of broad emission lines, where oversubtraction or undersubtraction of the continuum could lead to biased inferences on BLR structure and kinematics. 
In recent years, spectral decomposition approaches have been applied to the data from some RM campaigns \citep[e.g.,][]{Barth13,Barth15,Hu2015}. By fitting a multicomponent model to each night's spectrum, the contributions of individual line and continuum components can be \edit1{better} isolated. 
\edit1{We find that in most of our objects, the \hb~line profiles isolated using the two methods do not differ dramatically except in the red wings, but spectral decomposition is useful for separating blended line components and removes the starlight component for objects with high starlight fraction much more effectively.}

\begin{figure*}[htbp]
  \centering
  \includegraphics[width=.95\textwidth]{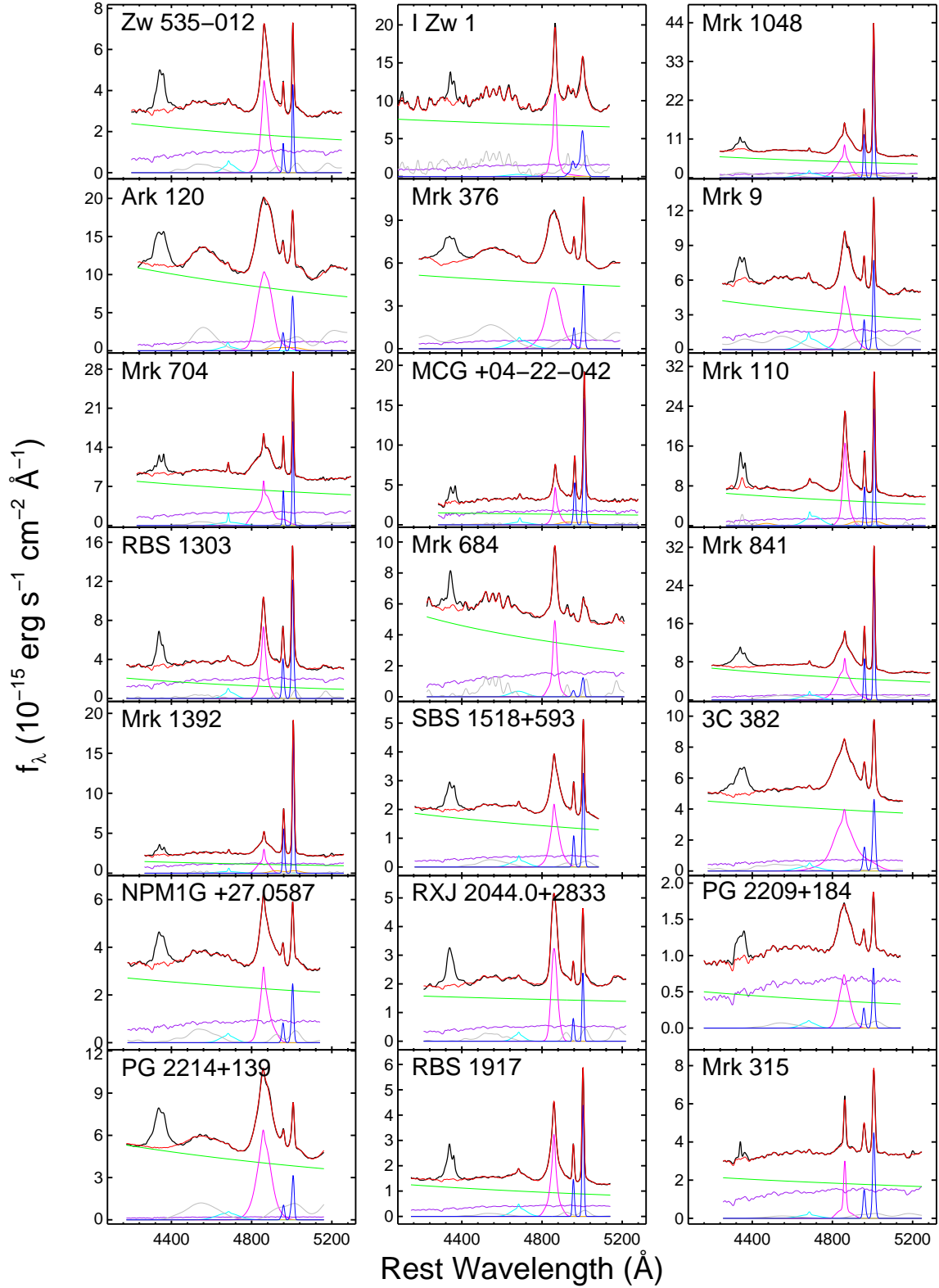}
\caption{Spectral decomposition results. The observed mean
  spectrum (black) for each galaxy is plotted alongside the
  decomposed model components: starlight (purple), AGN power-law
  continuum (green), \hb~(magenta), \heii~(cyan), \hei~(orange),
  \feii~(grey), and \oiii~(blue). The sum of the fits is represented by the red line. The \hei~component is negligible in
  most cases. The blend of \hg~and \oiii~$\lambda$4363 is excluded
  from the fit. The~\cite{Kovacevic10} \feii~template was applied to 
  all sources except for I~Zw~1, which was fitted with the~\cite{BG92}
  template. \label{fig:specdecomp}} 
\end{figure*}

Here we have adopted and applied the spectral fitting method described by \cite{Barth15} for the H$\beta$ spectral region. We refer the reader to \citet{Barth15} for a detailed description, and provide a brief overview of the method here. The model components include a power-law AGN continuum, starlight from a single-burst old stellar population at solar
metallicity \citep{BC03} convolved with a Gaussian velocity broadening, and emission lines including \oiii\ (narrow), \hb\ (broad and narrow), \heii\ (broad and narrow),
\hei\ (broad), and an \feii\ emission template convolved with a Gaussian velocity broadening. Emission-line profiles were modeled using a fourth-order Gauss-Hermite function, except for the \hei\ and \heii\ lines for which a Gaussian was used. Each line's velocity centroid was allowed to vary independently, except for the \oiii\ $\lambda\lambda$4959, 5007 lines which were required to have the same velocity profile and a 1:3 flux ratio. For the \feii\ blends, we tested different template spectra as described by~\cite{BG92},~\cite{Veron04}, and~\cite{Kovacevic10}. In the near future, we will employ a promising, new \feii\ template spectrum based on Mrk 493 (Park et al., in prep.). 
Similar to~\citet{Barth15}, the multicomponent \citet{Kovacevic10} template provided the
best fit to the \feii~lines in all of our AGN and was used for the final fits,  except for the case of I~Zw~1. For  I~Zw~1 we found that  the~\cite{BG92} template, which is based on I~Zw~1 itself, provided the best fit with minimized residuals. As part of the fitting process, a~\cite{Cardelli89} reddening model is applied to the model spectrum, allowing $E(B-V)$ to be a free parameter. Fitting was carried out over a rest-wavelength range of approximately 4200--5200\,\AA, with the exact range tailored to the data for each AGN depending on its redshift.  The H$\gamma$ + \oiii\ $\lambda4363$ blend was masked out from the fit, because decomposing this blend would add several additional free parameters to the model.  The full model, including the multicomponent \citet{Kovacevic10} iron template, includes 33 free parameters.  Model fits were optimized using a Levenberg-Marquart algorithm as implemented by \citet{Markwardt2009}. For each AGN, the model was first fitted to the mean spectrum, and the fit parameters from the mean spectrum fit were used as the starting parameter estimates for the fit to each individual night's spectrum.
The overall mean spectrum as fitted with the different model components
for each galaxy is illustrated in Figure \ref{fig:specdecomp}.

\section{The Photometric Campaign}
\label{sec:photcampaign}
AGN continuum light curves were measured from imaging data. We chose the $V$ band since it provides a fairly clean continuum measurement with relatively little contamination from broad emission lines for low-redshift sources.
To maximize the chances of detecting the delayed response of the emission-line variations to those of the continuum, we began the photometric monitoring campaign two months before the start of the spectroscopic program. From
February 2016 to May 2017, our team used a network
of eight telescopes across the world and obtained high-fidelity $V$-band
images with up to nightly cadence within each 
object's monitoring season. 

\subsection{Telescopes and Cameras}

Our photometric campaign included observations with the following
telescopes: the 0.76\,m Katzman Automatic Imaging Telescope
(KAIT)~\cite[][]{Filippenko01} and the Anna Nickel telescope at Lick
Observatory on Mount Hamilton, California; the Las Cumbres Observatories Global Telescope (LCOGT)
network~\cite[][]{Brown13,Boroson14}; the 2\,m Liverpool Telescope at the
Observatorio del Roque de Los Muchachos on the Canary island of La
Palma, Spain~\cite[][]{Steele04}; the 1\,m Illinois Telescope at Mount Laguna Observatory~\cite[MLO;][]{Smith69} in the Laguna Mountains, California; the San Pedro
M\'{a}rtir Observatory (SPM) 1.5\,m Johnson telescope~\citep{Butler12,Watson12} at the Observatory Astron\'{o}mico
Nacional located in Baja California, M\'{e}xico; the Fred Lawrence
Whipple Observatory 1.2\,m telescope on Mount Hopkins, Arizona; and the 0.9\,m West Mountain
Observatory (WMO) telescope at Utah Lake in Utah. KAIT,
LCOGT, SPM-1.5\,m, and Liverpool are fully robotic telescopes. 
Table \ref{tbl:scope} summarizes the telescope and detector properties. 

\begin{deluxetable*}{lccccccc}[htb]
\tablecaption{Telescope and CCD Properties for Imaging Observations\label{tbl:scope}}
\tablecolumns{8}
\tablewidth{0pt}
\tablehead{
\colhead{Telescope} &
\colhead{$D_{\rm mirror}$} &
\colhead{Detector} & 
\colhead{Field of View} &
\colhead{Pixel Scale} &
\colhead{Gain} &
\colhead{Read Noise} &
\colhead{Binning} \\
\colhead{} &
\colhead{(m)} & 
\colhead{} &
\colhead{} &
\colhead{(\arcsec/pix)} & 
\colhead{($e^-$/ADU)} & 
\colhead{($e^-$)} &
\colhead{} 
}
\startdata
FLWO-1.2\,m & 1.2 & Fairchild CCD 486 & 23\farcm1 $\times$ 23\farcm1 & 0.336 & 4.45 & 7.18 & 2 $\times$ 2 \\ 
KAIT & 0.76 & Apogee AP7 & 6\farcm8 $\times$ 6\farcm8 & 0.80 & 4.5 & 12.0 & 1  $\times$ 1 \\ 
LCOGT-2\,m & 2.0 & Merope & 4\farcm7 $\times$ 4\farcm7 & 0.467 & 1.0 & 9.0 & 2 $\times$ 2 \\ 
LCOGT-1\,m & 1.0 & Fairchild Imaging & 26\farcm5 $\times$ 26\farcm5 &  0.389 & 1.0 & 13.5 & 1 $\times$ 1 \\ 
Liverpool & 2.0 & e2V CCD 231 & 10\arcmin $\times$ 10\arcmin & 0.304 & 1.62 & 8.0 &  2 $\times$ 2 \\ 
MLO-1\,m & 1.0 & Fairchild 446 & 13\farcm3 $\times$ 13\farcm3 & 0.718 & 2.13 & 3.88 & 1 $\times$ 1 \\ 
Nickel & 1.0 & Loral & 6\farcm3 $\times$ 6\farcm3 & 0.368 & 1.7 & 8.3 & 2 $\times$ 2 \\ 
SPM-1.5\,m & 1.5 & Fairchild 3041 & 5\farcm4 $\times$ 5\farcm4 &  0.32 & 4.2 & 14.0 & 2 $\times$ 2 \\ 
WMO-0.9\,m & 0.9 & Finger Lakes PL-09000 & 25\farcm2
$\times$ 25\farcm2 & 0.61 & 1.37 & 12.0 & 1 $\times$ 1 \\ 
\enddata
\end{deluxetable*}

\subsection{Photometry Measurements and Continuum Light Curves}

All images were processed with reduction procedures applied as part of the standard pipeline for each facility, including overscan subtraction and flat-fielding. For telescopes that did not include world coordinate system (WCS) solutions as part of their standard processing, we used \texttt{astrometry.net}~\cite[][]{Lang10} to add WCS information to the FITS headers.

Measurement of the AGN light curves was carried out using the automated aperture photometry pipeline described by \citet{Pei14}, which can be used to measure AGN light curves using data from any number of telescopes with diverse camera properties. This procedure, written in IDL and based on the photometry routines in the IDL Astronomy User's Library \citep{Landsman1993}, is designed to automate the process of identifying the AGN and a set of comparison stars in each image by their coordinates, measuring their instrumental magnitudes, and using the comparison-star measurements to obtain a consistent magnitude scale across the full time series for data from multiple telescopes. 

The aperture-photometry routine returns photometric uncertainties (in magnitudes) based on the photon-counting statistics, background uncertainty, and CCD readout noise. However, other error sources contribute to the actual error budget, such as imperfect flat-fielding and point-spread function (PSF) variations across the field of view. To account for these additional error sources, we measured the excess variance $\sigma_x^2$ in the light curves of the comparison stars for each AGN, for each telescope's data. Assuming that the excess variance in the comparison-star light curves is a good estimate of the additional error over and above the statistical uncertainties, we inflated the fractional flux uncertainties on the AGN photometry by addition in quadrature using $\sigma_\mathrm{tot}^2 = \sigma_\mathrm{phot}^2 + \langle\sigma_x^2\rangle$, where $\sigma_\mathrm{phot}$ is the original photometric uncertainty on a given data point, and $\langle\sigma_x^2\rangle$ is the average (over all comparison stars in the field) of the excess variance in the comparison-star light curves. This error adjustment was carried out separately for each telescope's data.

AGN light curves measured from different telescopes tend to have small offsets in magnitude relative to one another as a result of differences in filter transmission curves, even after normalizing them to the same set of comparison stars. To correct for these offsets, we applied an additive shift (in magnitudes) to bring each telescope's data into best average agreement with the data from the telescope having the most data points for each AGN.
Finally, the $V$-band magnitude scale for each AGN field was calibrated using observations of standard stars taken on photometric nights at WMO and cross-checked against the AAVSO Photometric All-Sky Survey~\cite[][]{Henden16}. 

Magnitudes were converted to $f_\lambda$ for measurement of broad emission-line lags.
The typical uncertainties of the photometric data points $\langle \delta_f
\rangle$ are at the 0.4\% level, not exceeding 0.7\% for any AGN.
The final $V$-band light curves are presented in Table
\ref{tbl:photlc} and shown alongside the emission-line light
curves in \S \ref{sec:lc}. 

\begin{deluxetable}{ccccc}[htb]
\tablecaption{$V$-band Light Curve Data \label{tbl:photlc}}
\tablecolumns{5}
\tablewidth{0pt}
\tablehead{
\colhead{Object} &
\colhead{HJD} &
\colhead{$f_\lambda$} &
\colhead{$\delta_f$} &
\colhead{Telescope} 
}
\startdata
              IZw1 & 7555.8493 &   7.124 &   0.020 &           LCOGT-fl \\ 
              IZw1 & 7557.8446 &   7.197 &   0.022 &           LCOGT-fl \\ 
              IZw1 & 7558.8839 &   7.210 &   0.023 &           LCOGT-fl \\ 
              IZw1 & 7559.8607 &   7.207 &   0.022 &           LCOGT-fl \\ 
              IZw1 & 7561.9280 &   7.195 &   0.021 &           LCOGT-fl \\ 
              IZw1 & 7563.9231 &   7.312 &   0.021 &           LCOGT-fl \\ 
              IZw1 & 7564.9379 &   7.237 &   0.018 &     LCOGT-McDonald \\ 
              IZw1 & 7566.9208 &   7.226 &   0.021 &           LCOGT-fl \\ 
              IZw1 & 7567.9343 &   7.194 &   0.022 &           LCOGT-fl \\ 
              IZw1 & 7570.9122 &   7.242 &   0.017 &     LCOGT-McDonald  
\enddata
\tablecomments{Dates are listed as HJD$-$2450000. Units for continuum
  flux density $f_\lambda$ are 10$^{-15}$ erg s$^{-1}$ cm$^{-2}$ \AA$^{-1}$. This table is
  published in its entirety in machine-readable format on the online
  version of the journal. Sample entries are shown here for guidance
  regarding the table's form and content.}
\end{deluxetable}

\section{Emission-Line Light Curves and Lag Determination}
\label{sec:lc}

We used results from spectral decomposition to derive \hb~emission-line light curves. The flexibility of the multicomponent modeling afforded us a number of ways to extract the \hb~flux. In the spectral region near the \hb~line, sources with strong \feii~emission were subjected to substantial degeneracy of spectral decomposition that could result in, from night to night, instability in model fitting in terms of how much flux gets assigned to the different model components. This instability results in additional noise in the \hb~light curve. To minimize the noise due to degeneracy from spectral fits, we used 
a version of the decomposed model spectra where only the AGN and stellar continuum components have been removed. This version consisted of the residuals of all present emission lines after continuum subtraction. We then selected a corresponding extraction window free of other emission lines for the calculation of \hb~flux (see Table \ref{tbl:windows}). A different way to extract \hb~flux is to use a version of the decomposed model spectra that only contained \hb~broad and narrow components, with other emission-line components as well as the AGN and stellar continua removed. The resulting \hb~light curves computed from the two methods were very similar for objects that have relatively clean \hb~spectral regions, but those for the emission-line residual versions were less noisy for objects with more complex line profiles within the \hb~spectral region. 
The final \hb~light-curve measurements for the full sample are presented in Table~\ref{tbl:linelc}. 
The light curves for the $V$ band and \hb~are shown in \edit1{Figures~\ref{fig:ccf}--\ref{fig:ccf7}}. The rest of the emission-line features will be presented in a forthcoming paper.

\begin{deluxetable}{lcccc}[htb]
\tablecaption{Emission-Line Extraction Windows in \AA~(observed frame), $F_{\rm var}$, and \oiii~Excess Scatter \label{tbl:windows}}
\tablecolumns{5}
\tablewidth{0pt}
\tablehead{
\colhead{Object} &
\colhead{\hb} &
\colhead{$F_{\rm var}$} &
\colhead{\oiii} & 
\colhead{$\sigma_\mathrm{[O~\textsc{iii}]}$}
}
\startdata
Zw 535$-$012 & 5018 $-$ 5181 & 0.050 & 5220 $-$ 5270 & 0.016 \\ 
I Zw 1 & 5099 $-$ 5199 & 0.044 & 5263 $-$ 5337 & 0.054 \\ 
Mrk 1048 & 5002 $-$ 5153  & 0.071 & 5200 $-$ 5252 & 0.004 \\ 
Ark 120 & 4931 $-$ 5102  & 0.195 & 5143 $-$ 5200 & 0.043 \\ 
Mrk 376 & 5069 $-$ 5201  & 0.064 & 5259 $-$ 5322 & 0.016 \\ 
Mrk 9 & 5007 $-$ 5127  & 0.069 & 5179 $-$ 5241 & 0.016 \\ 
Mrk 704 & 4945 $-$ 5059  & 0.046 & 5126 $-$ 5192 & 0.015 \\
MCG $+$04$-$22$-$042 & 4986 $-$ 5079  & 0.268 & 5149 $-$ 5203 & 0.012 \\ 
Mrk 110 & 4985 $-$ 5120  & 0.083 & 5161 $-$ 5208 & 0.015 \\ 
RBS 1303 & 5001 $-$ 5136  & 0.081 & 5185 $-$ 5251 & 0.019 \\ 
Mrk 684 & 5032 $-$ 5131  & 0.058 & 5209 $-$ 5262 &  0.048 \\ 
Mrk 841 & 4949 $-$ 5120  & 0.073 & 5167 $-$ 5218 & 0.006 \\ 
Mrk 1392 & 4958 $-$ 5108  & 0.173 & 5165 $-$ 5212 & 0.007 \\ 
SBS 1518$+$593 & 5191 $-$ 5310  & 0.051 & 5369 $-$ 5428 & 0.024 \\ 
3C 382 & 5030 $-$ 5226  & 0.142 & 5268 $-$ 5342 & 0.067 \\ 
NPM1G$+$27.0587 & 5092 $-$ 5246  & 0.051 & 5289 $-$ 5352 & 0.024 \\ 
RXJ 2044.0$+$2833 & 5040 $-$ 5187  & 0.055 & 5240 $-$ 5276 & 0.019 \\ 
PG 2209$+$184 & 5147 $-$ 5280  & 0.122 & 5329 $-$ 5387 & 0.052 \\ 
PG 2214$+$139 & 5110 $-$ 5260  & 0.028 & 5307 $-$ 5355 & 0.024 \\ 
RBS 1917 & 5111 $-$ 5261  & 0.028 & 5231 $-$ 5444 & 0.016 \\ 
Mrk 315 & 5002 $-$ 5111  & 0.109 & 5174 $-$ 5236 & 0.016  
\enddata
\end{deluxetable}

\begin{deluxetable}{ccccc}[htb]
\tablecaption{\hb~Light Curve Data \label{tbl:linelc}}
\tablecolumns{5}
\tablewidth{0pt}
\tablehead{
\colhead{Object} &
\colhead{HJD} &
\colhead{$f$} &
\colhead{$\delta_f$ (stat)} &
\colhead{$\delta_f$ (modified)}
}
\startdata
Mrk 110 &     7509.74 &       690.6 &     0.76 &    10.47 \\ 
Mrk 110 &     7520.75 &       683.9 &     0.94 &    10.38 \\ 
Mrk 110 &     7526.73 &       686.3 &     0.96 &    10.42 \\ 
Mrk 110 &     7527.73 &       678.7 &     0.93 &    10.30 \\ 
Mrk 110 &     7536.71 &       701.7 &     1.10 &    10.67 \\ 
Mrk 110 &     7540.71 &       695.0 &     1.05 &    10.56 \\ 
Mrk 110 &     7541.71 &       717.2 &     1.39 &    10.93 \\ 
Mrk 110 &     7543.72 &       665.7 &     1.34 &    10.15 \\ 
Mrk 110 &     7546.72 &       685.8 &     1.13 &    10.43 \\ 
Mrk 110 &     7549.69 &       683.1 &     1.35 &    10.41 
\enddata
\tablecomments{Dates are listed as HJD$-$2450000. Units for
  emission-line flux $f$ are in
  10$^{-15}$ erg s$^{-1}$ cm$^{-2}$. The original $\delta_f$ represents the photon-counting errors on the flux measurement, while the modified $\delta_f$ incorporates the spectral scaling uncertainties from \oiii~scatter. This table is
  published in its entirety in machine-readable format on the online
  version of the journal. Sample entries are shown here for guidance
  regarding the table's form and content.}
\end{deluxetable}

To account for the residual uncertainties from spectral scaling, we computed the excess variance in the \oiii~light curve for each AGN as the additional uncertainty term. We then combined this excess variance term with the statistical error in quadrature such that
\begin{equation}
    \delta_f^2 \mathrm{(modified)} = \delta^2_f \mathrm{(stat)} + (f \times \sigma_\mathrm{[O~\textsc{iii}]})^2 \quad .
\end{equation}
Both the statistical and the modified errors are listed in Table \ref{tbl:linelc}. Following~\cite{Bentz09b}, we compute the variability statistics $F_{\rm var}$ for each light curve and found that it ranges from 0.028 to 0.268, with a median of 0.069 (see Table \ref{tbl:windows}). Almost a third of our sample (3C 382, Ark 120, MCG $+$04$-$22$-$042, Mrk 1392, Mrk 315, and PG 2209$+$184) exhibit strong variability in the \hb~light curves with $F_{\rm var} > 0.1$. Among the rest of the sample, a few galaxies (such as Zw 535$-$012, Mrk 110, Mrk 841, RBS 1303, and RXJ2044.0$+$2833) still display discernible variability for plausible lag determination even if the signal is diluted by large error bars or outliers. 

\subsection{Integrated \hb~Lags} 
Following \cite{White94}, \cite{Peterson04} and others, 
cross-correlation was employed to determine the delay in
the integrated \hb~line signal 
relative to that from the $V$-band
continuum from the photometric campaign.  
We chose to adopt the interpolated cross-correlation function
~\cite[ICCF; e.g.,][]{Peterson98} method. Given a lag range that
represents reasonable bounds of the expected lag, the cross-correlation
function (CCF) between the two light curves is computed
at every time lag $\tau$ at 0.5\,day intervals via 
\begin{equation}
F_{\rm CCF}(\tau) = \frac{1}{N-1} \sum_{i=1}^{N}
\frac{[L(t_i)-\bar{L}][C(t_i-\tau)-\bar{C}]}{\sigma_C \sigma_L} \quad,
\end{equation}
where $\bar{L}$ ($\bar{C}$) and $\sigma_L$ ($\sigma_C$) are the
mean and standard deviation of the emission-line (continuum) time
series, respectively. 

\begin{figure*}[htbp]
   \centering
   \includegraphics[width=0.39\textwidth,angle=90,trim={1.3in 0 0 0},clip]{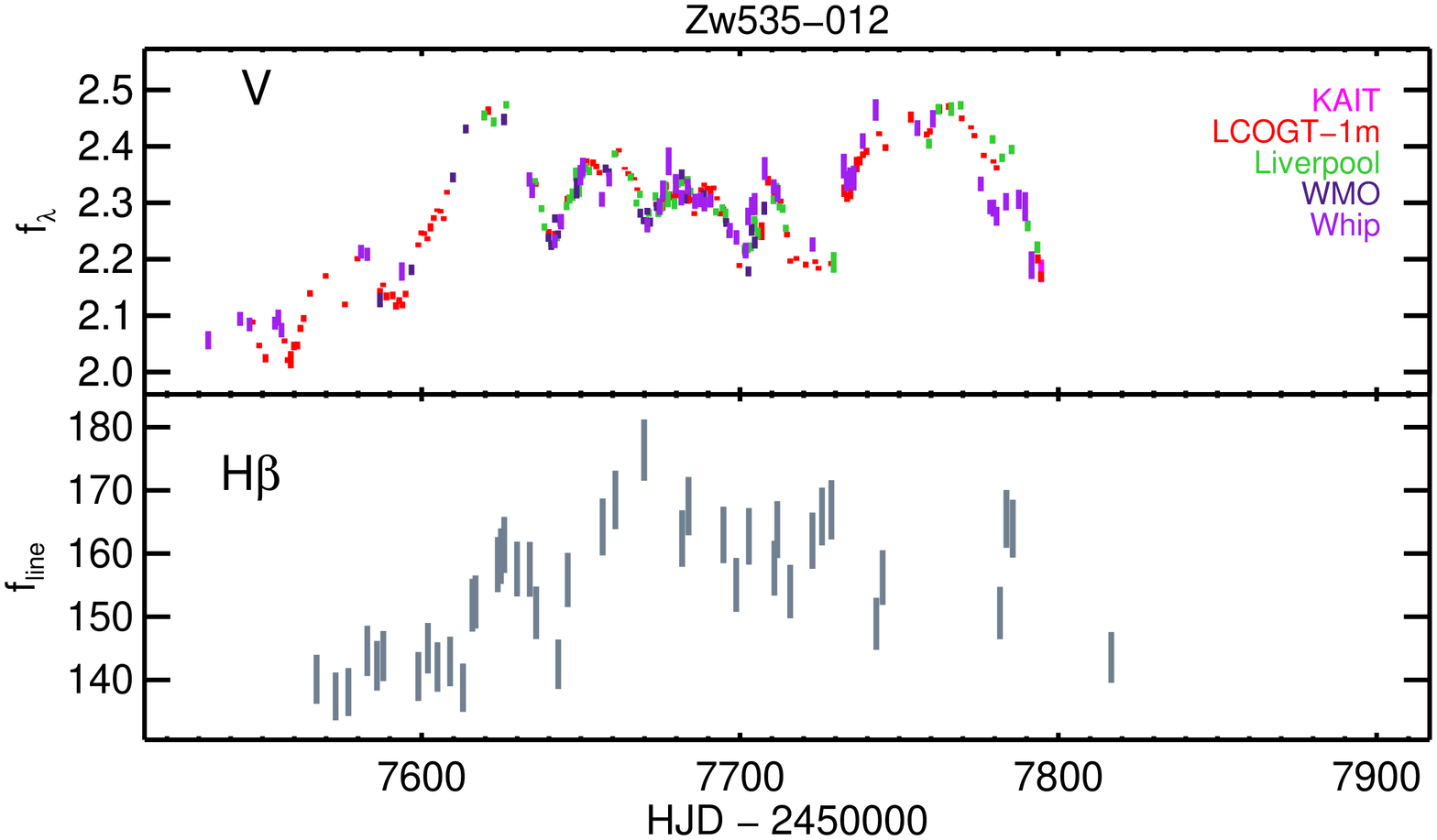}
   \hspace{0.1in}
   \includegraphics[width=0.26\textwidth,angle=90]{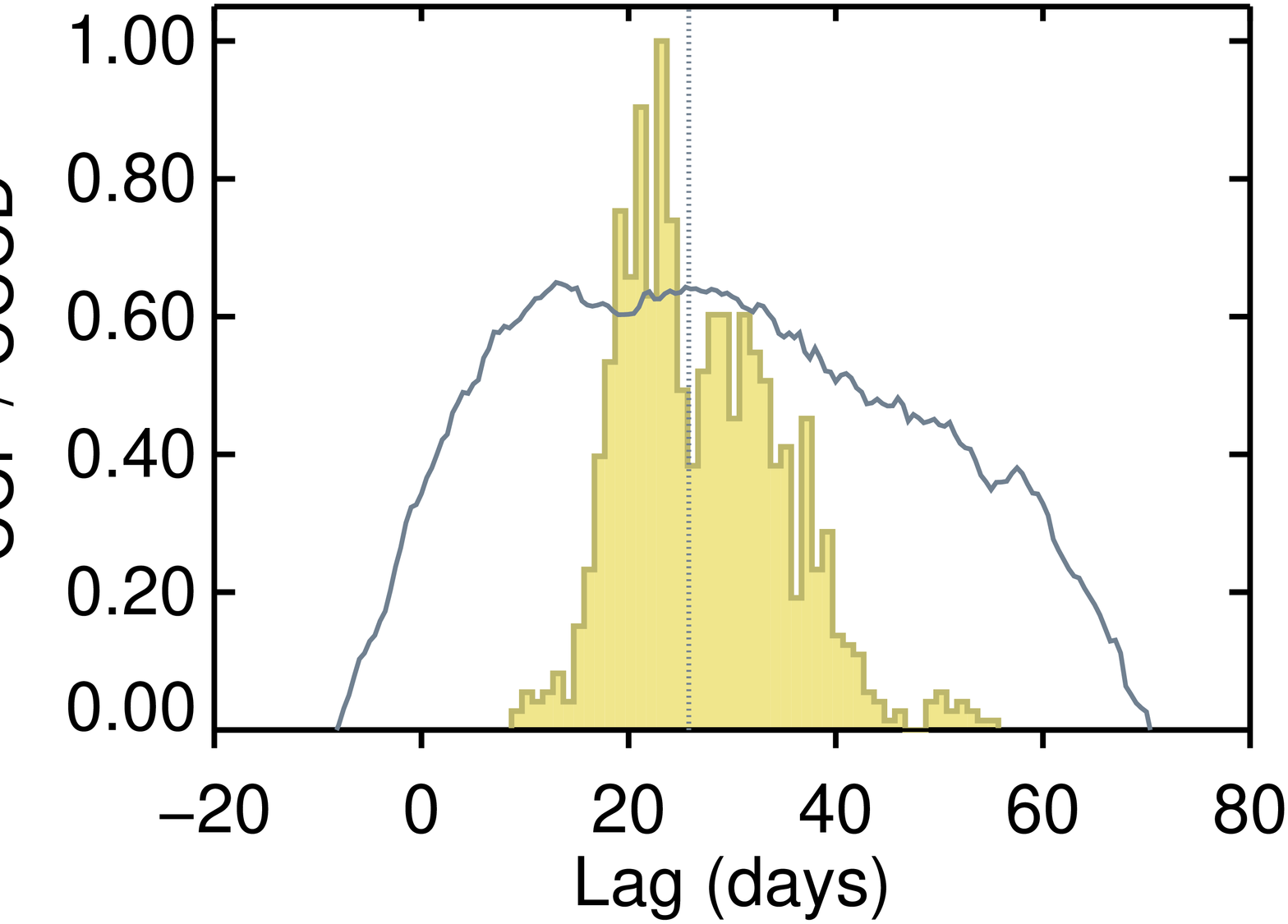}
   \includegraphics[width=0.39\textwidth,angle=90,trim={1.3in 0 0 0},clip]{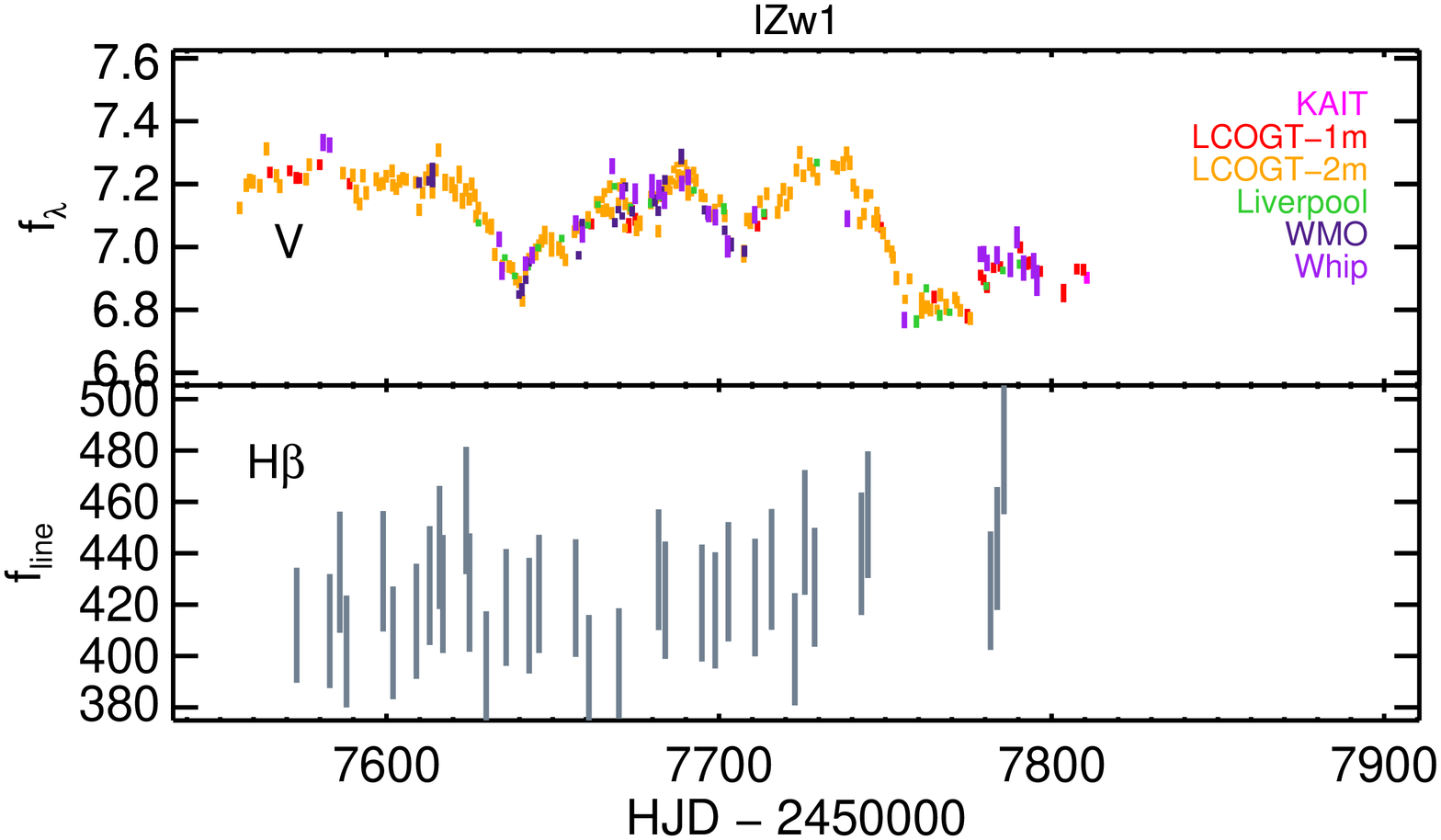}
   \hspace{0.1in}
   \includegraphics[width=0.26\textwidth,angle=90]{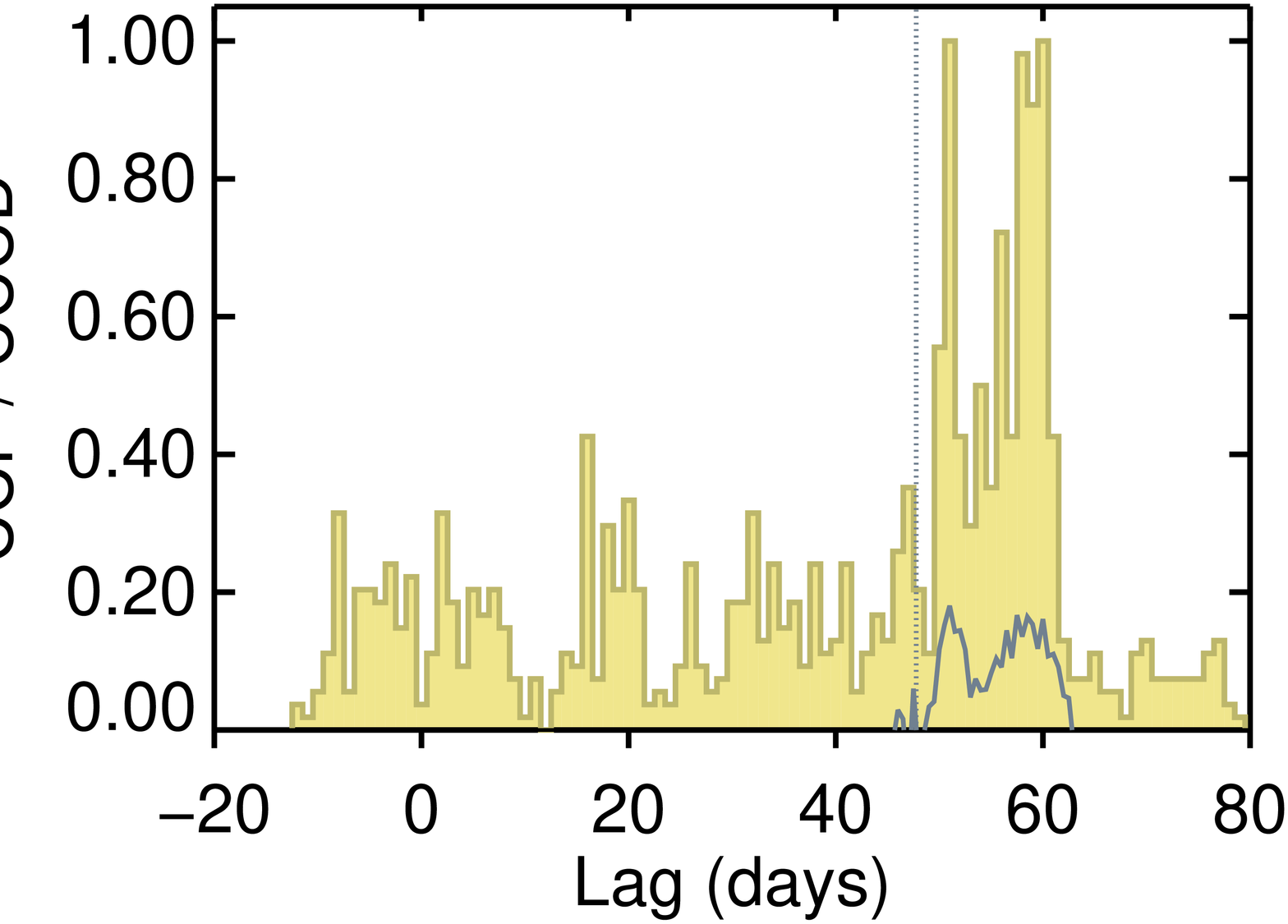}
   \includegraphics[width=0.39\textwidth,angle=90,trim={1.3in 0 0 0},clip]{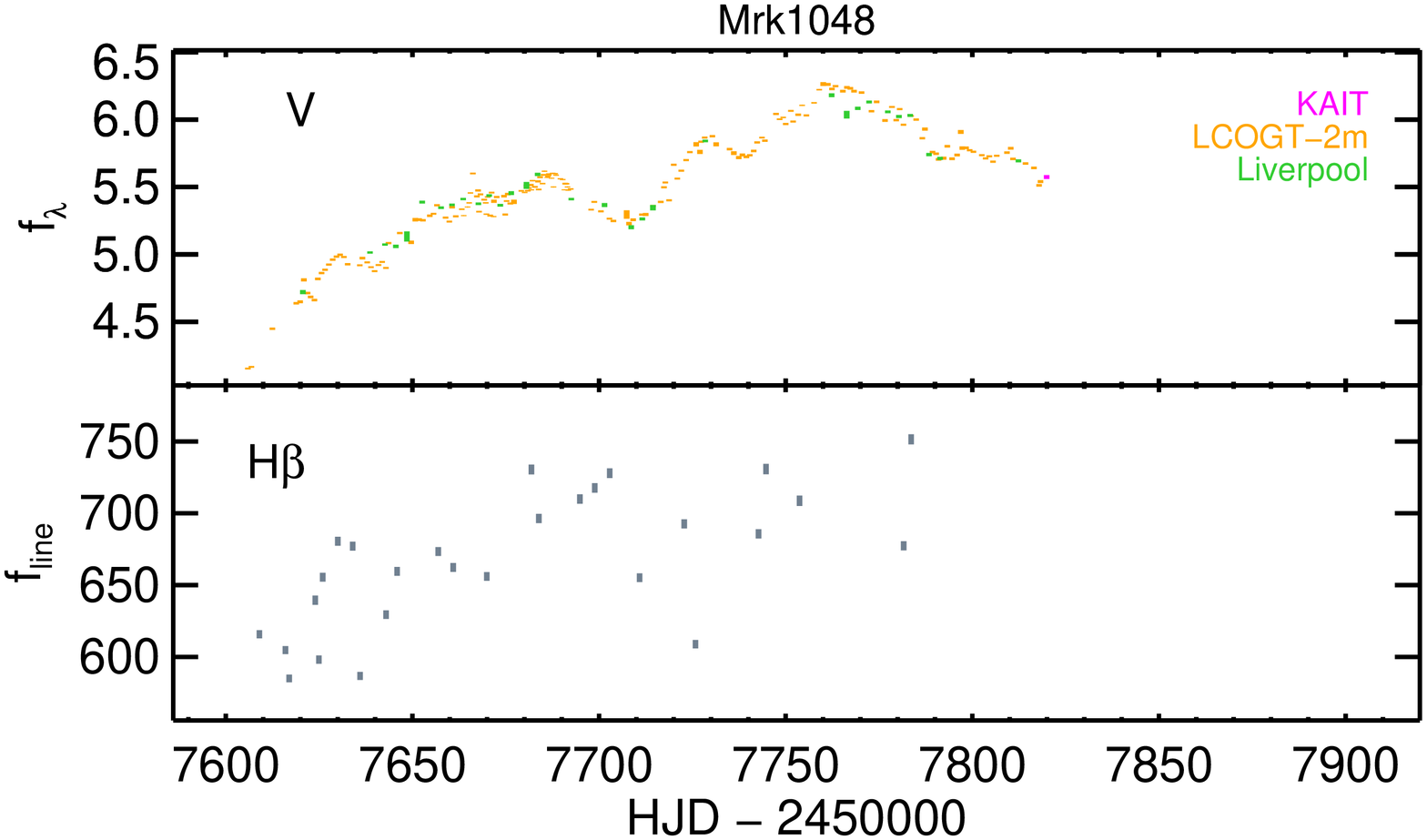}
   \hspace{0.1in}
   \includegraphics[width=0.26\textwidth,angle=90]{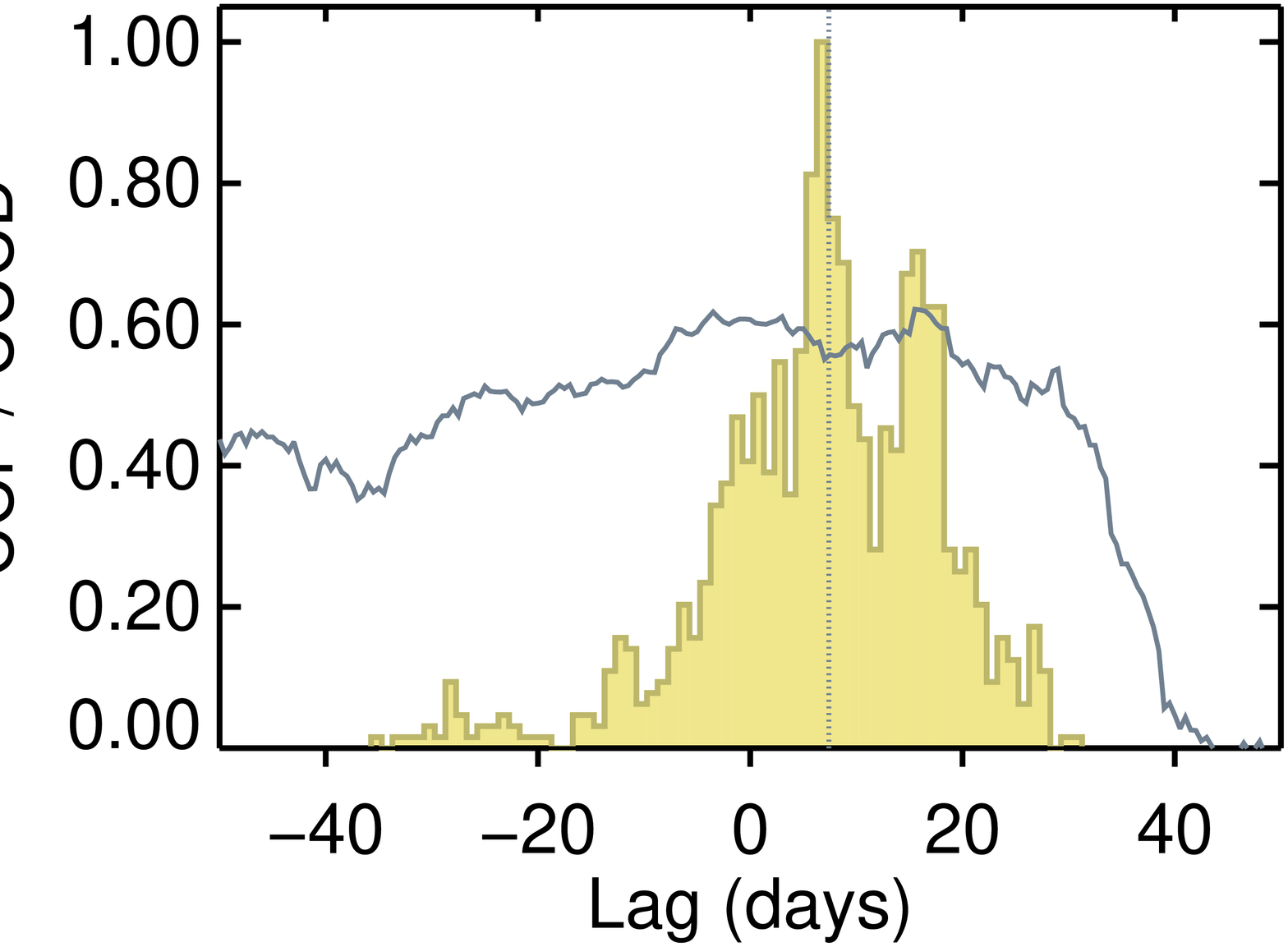}
\caption{Light curves for the LAMP2016 sample \edit1{(Zw 535$-$012, I Zw 1, Mrk 1048)}: $V$-band continuum flux density
  $f_{\lambda}$ in $10^{-15}$ erg cm$^{-2}$ s$^{-1}$ \AA$^{-1}$,
  color-coded by telescopes (top left); the \hb~emission line flux
  $f_{\rm line}$ in $10^{-15}$ erg cm$^{-2}$ s$^{-1}$ (bottom
  left). The error bars plotted here for the \hb~light curves incorporate the uncertainty term from the normalized excess scatter of \oiii. The cross-correlation function is shown in the right panel for each AGN, alongside the
  cross-correlation centroid distribution in yellow. The dotted vertical line indicates the median value of the CCF.}
  \label{fig:ccf}
\end{figure*}

\begin{figure*}[htbp]
   \centering
   \includegraphics[width=0.39\textwidth,angle=90,trim={1.3in 0 0 0},clip]{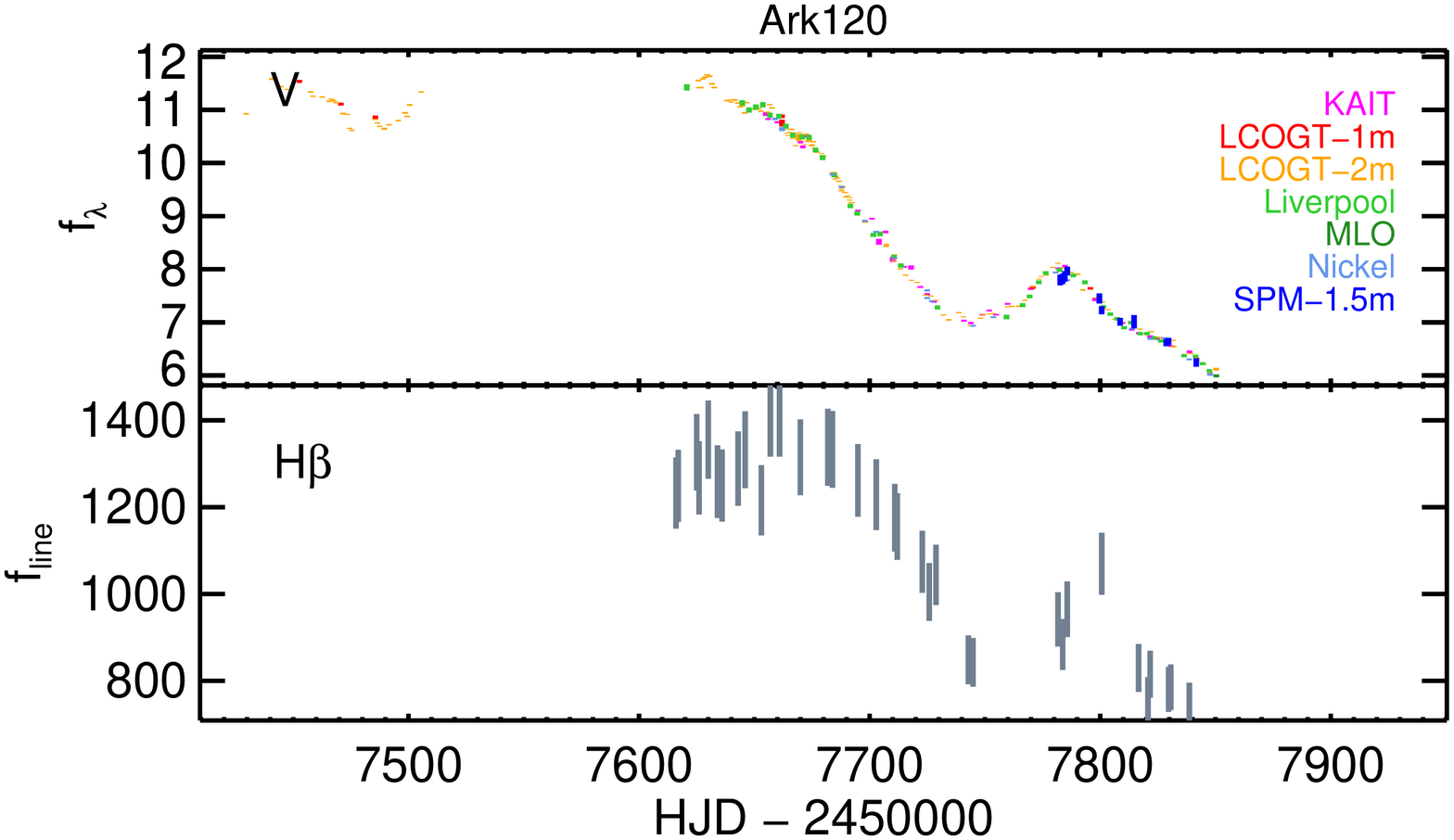}
   \hspace{0.1in}
   \includegraphics[width=0.26\textwidth,angle=90]{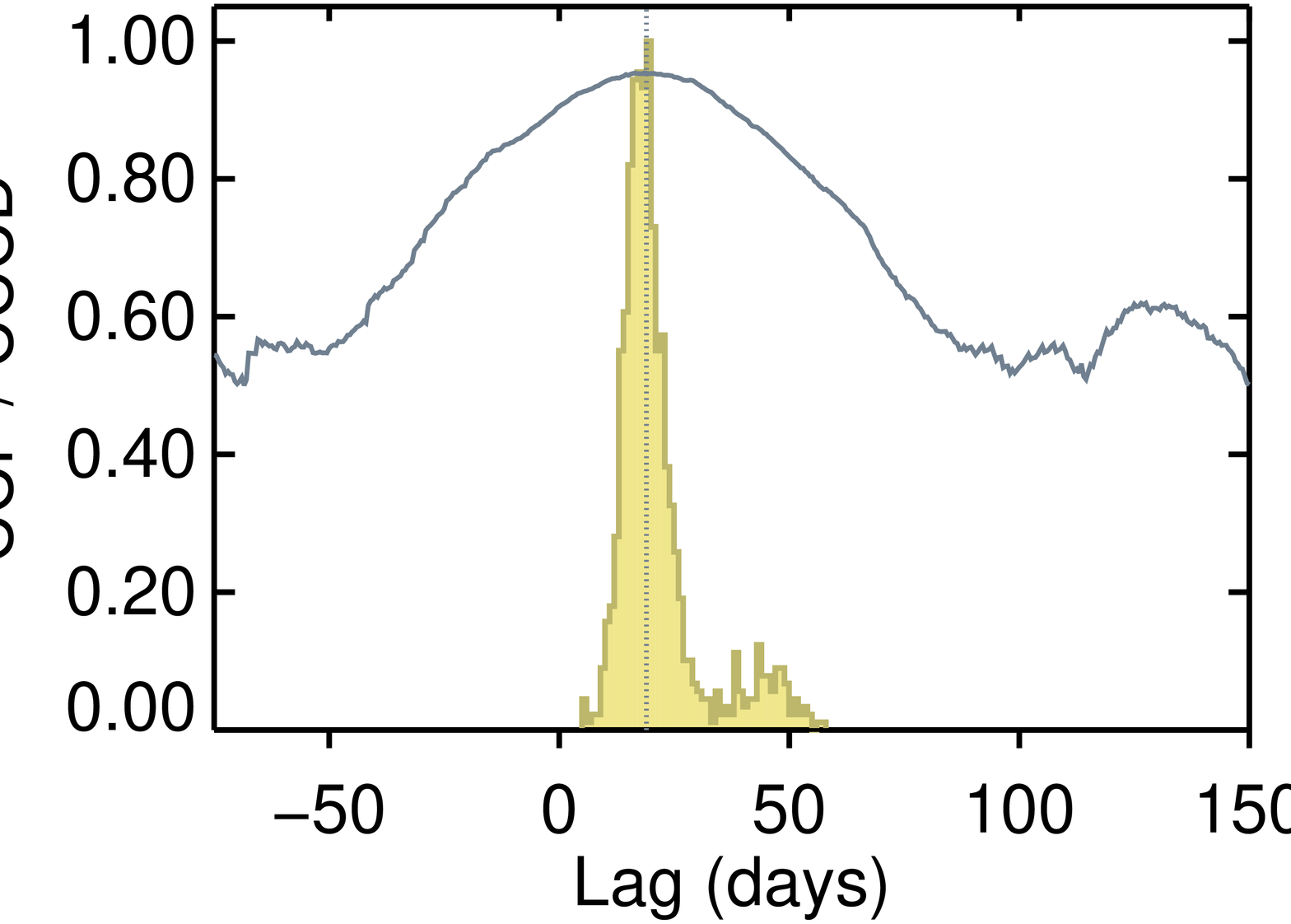}
   \includegraphics[width=0.39\textwidth,angle=90,trim={1.3in 0 0 0},clip]{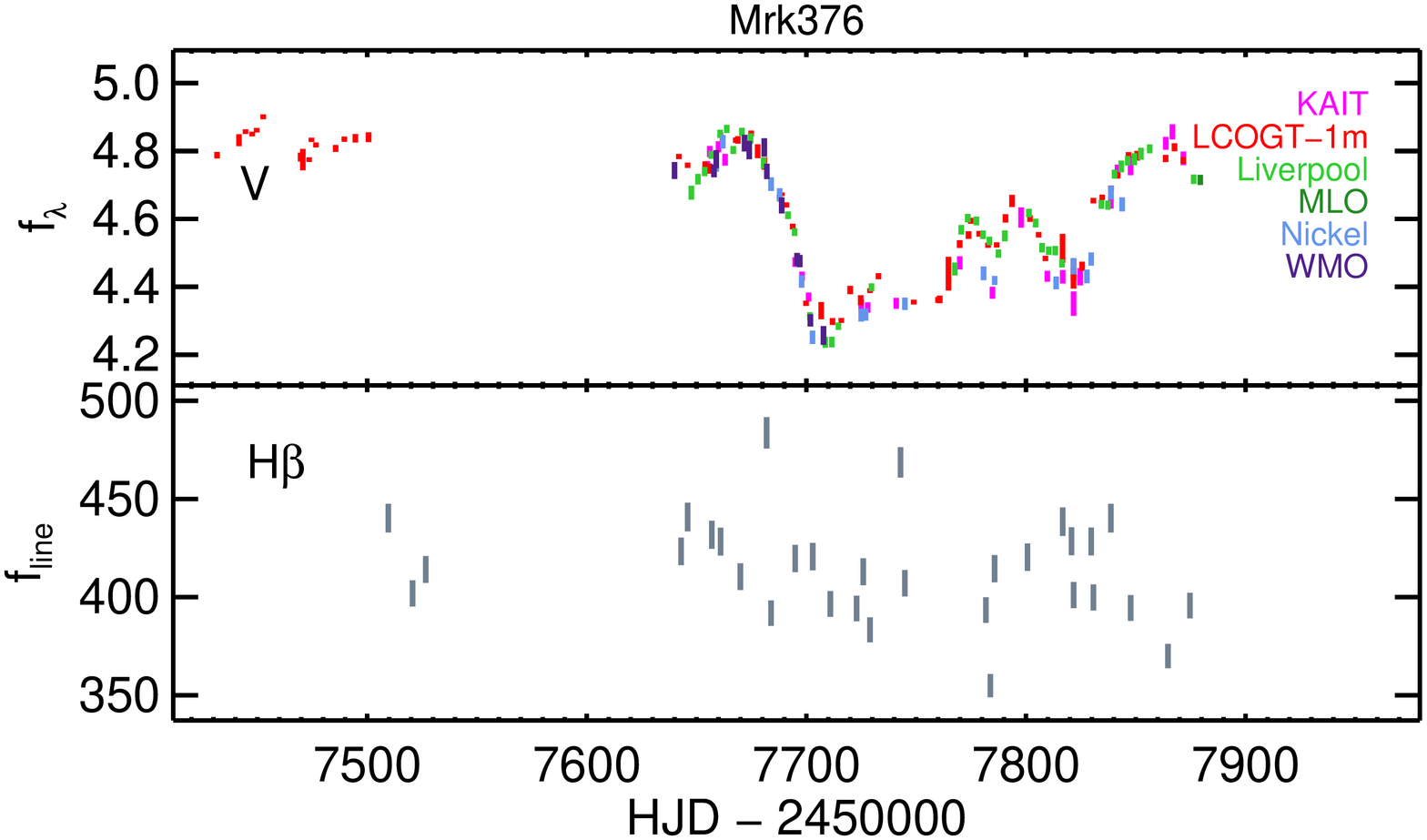}
   \hspace{0.1in}
   \includegraphics[width=0.26\textwidth,angle=90]{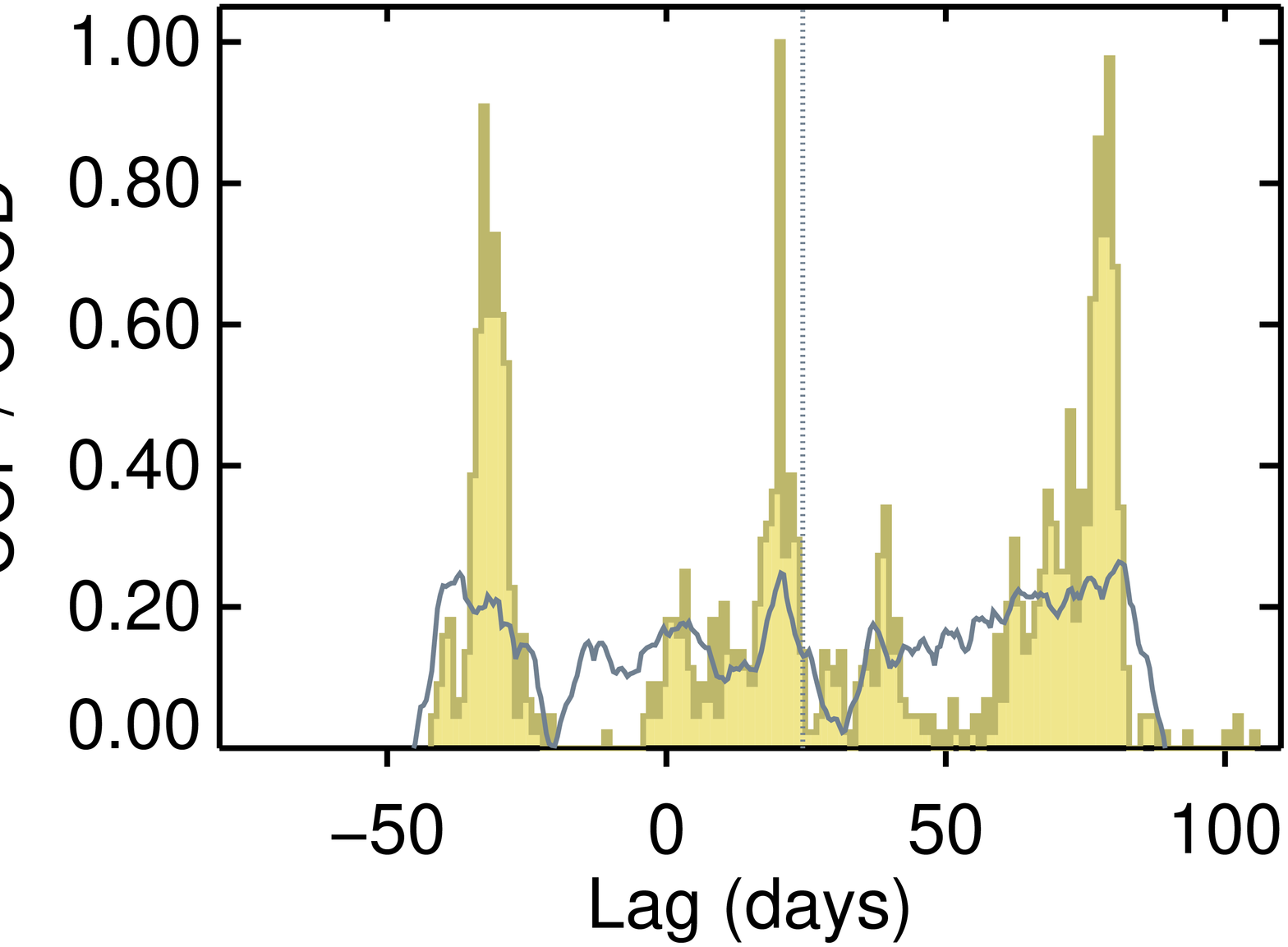}
   \includegraphics[width=0.39\textwidth,angle=90,trim={1.3in 0 0 0},clip]{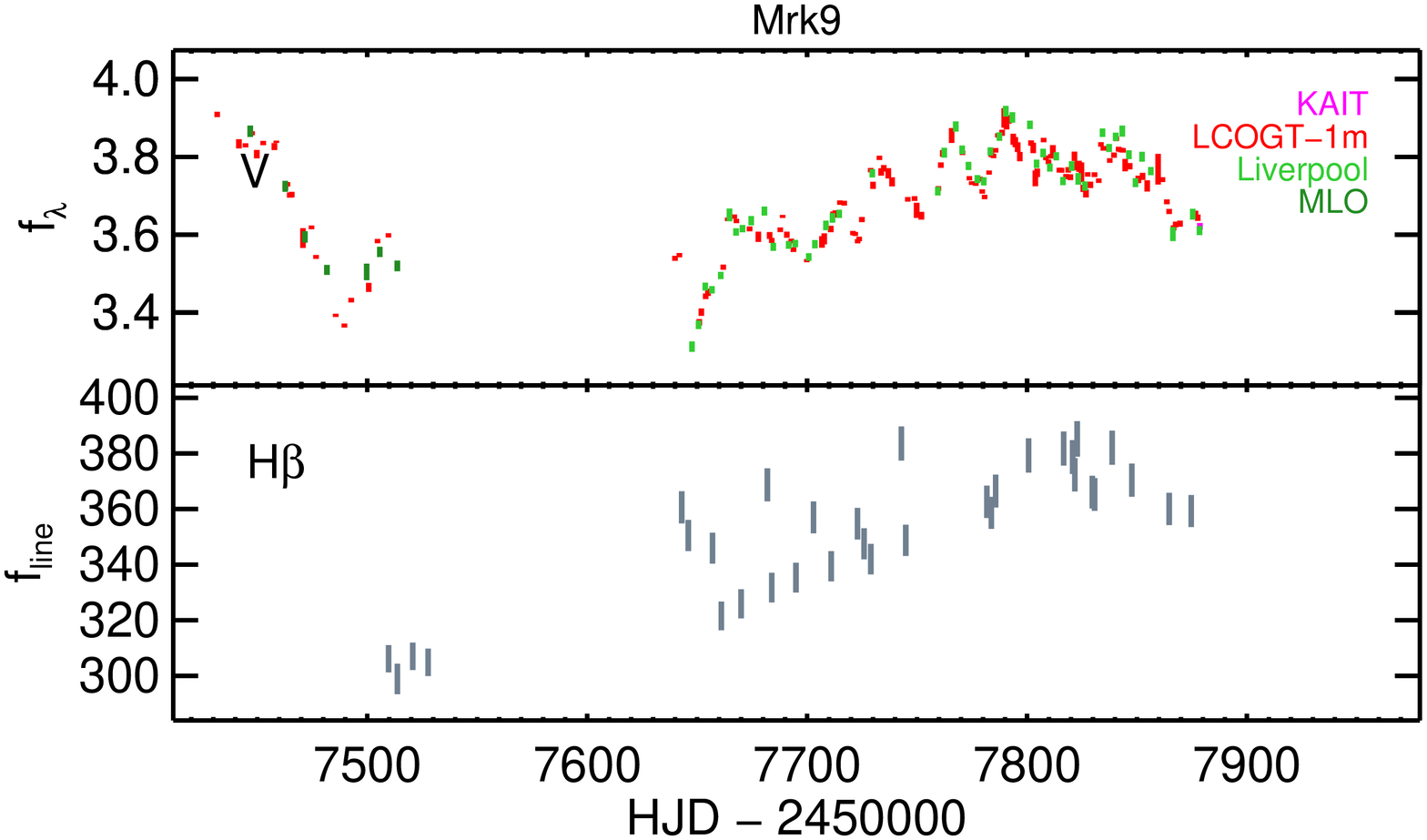}
   \hspace{0.1in}
   \includegraphics[width=0.26\textwidth,angle=90]{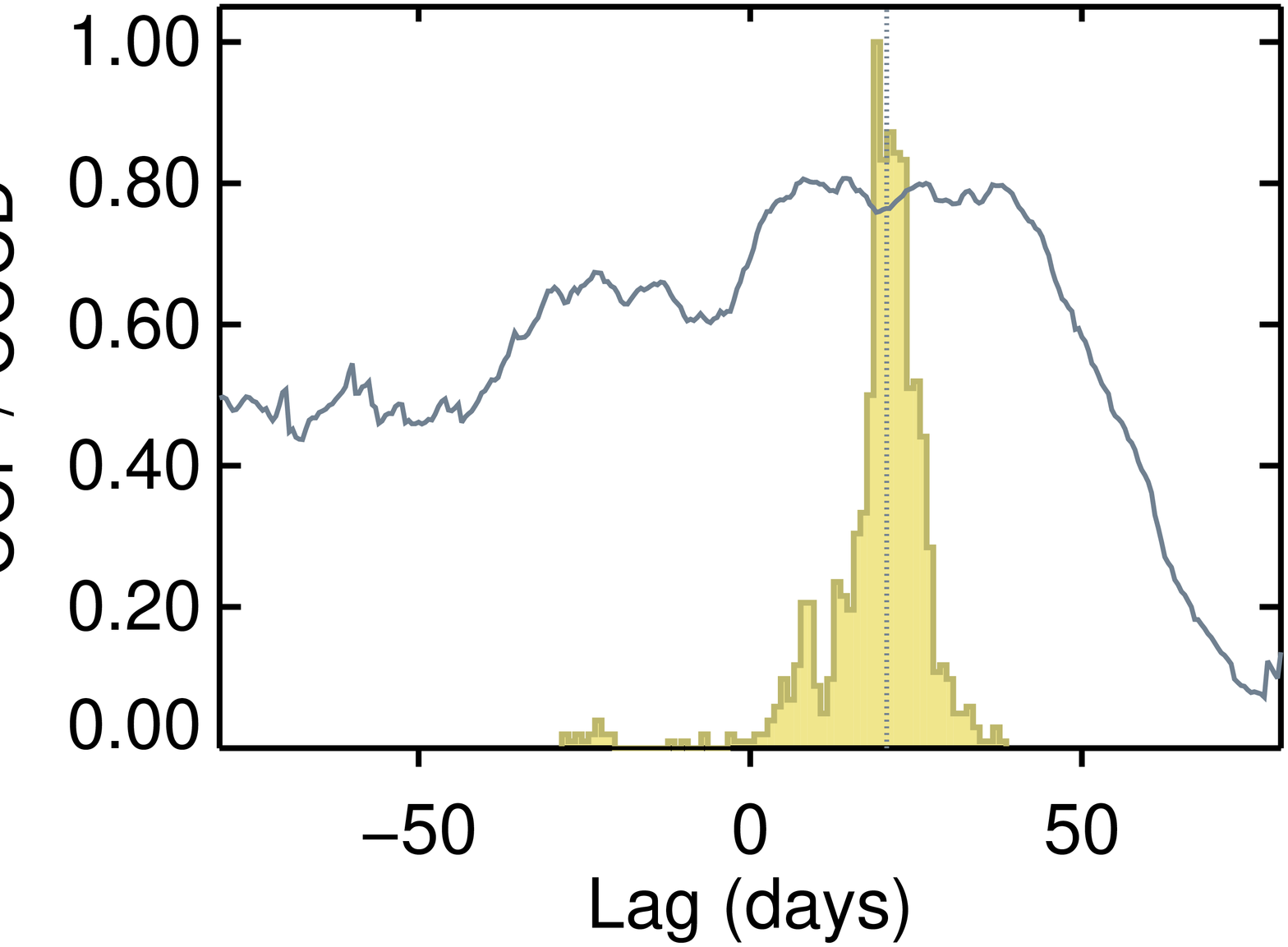}
    \caption{\edit1{Same as Figure \ref{fig:ccf} (but for Ark 120, Mrk 376, Mrk 9)}}
    \label{fig:ccf2}
\end{figure*}

\begin{figure*}[htbp]
   \centering
   \includegraphics[width=0.39\textwidth,angle=90,trim={1.3in 0 0 0},clip]{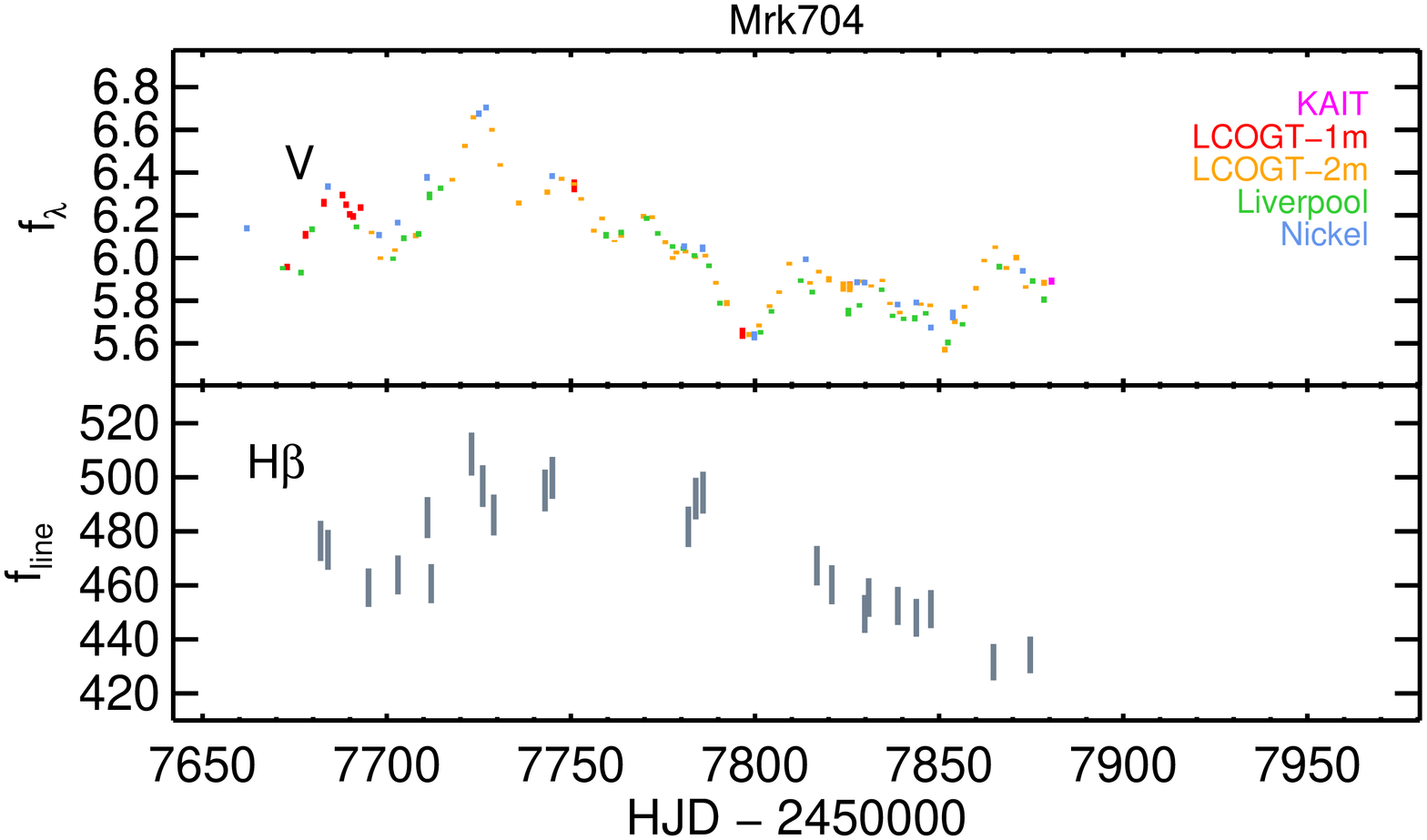}
   \hspace{0.1in}
   \includegraphics[width=0.26\textwidth,angle=90]{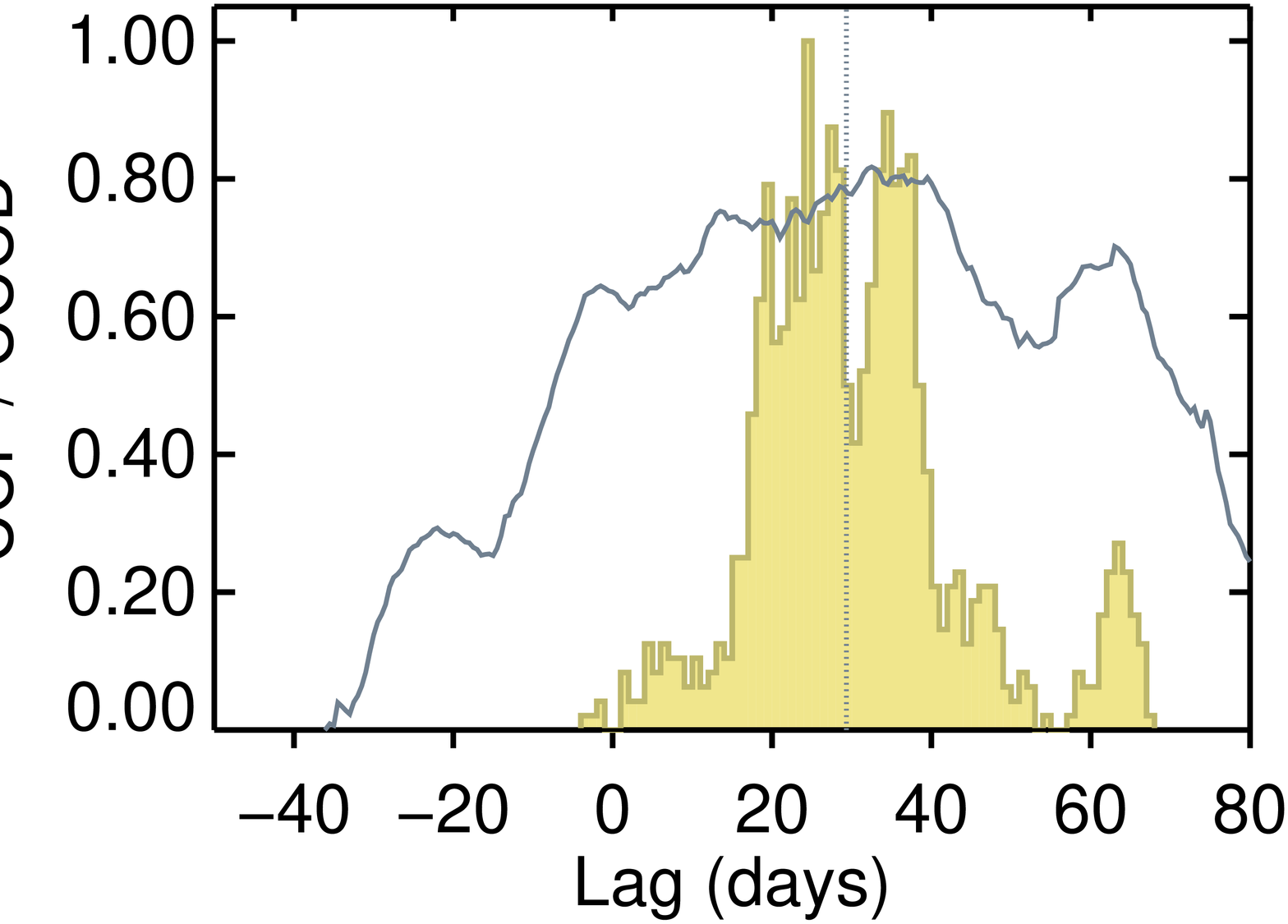}
   \includegraphics[width=0.39\textwidth,angle=90,trim={1.3in 0 0 0},clip]{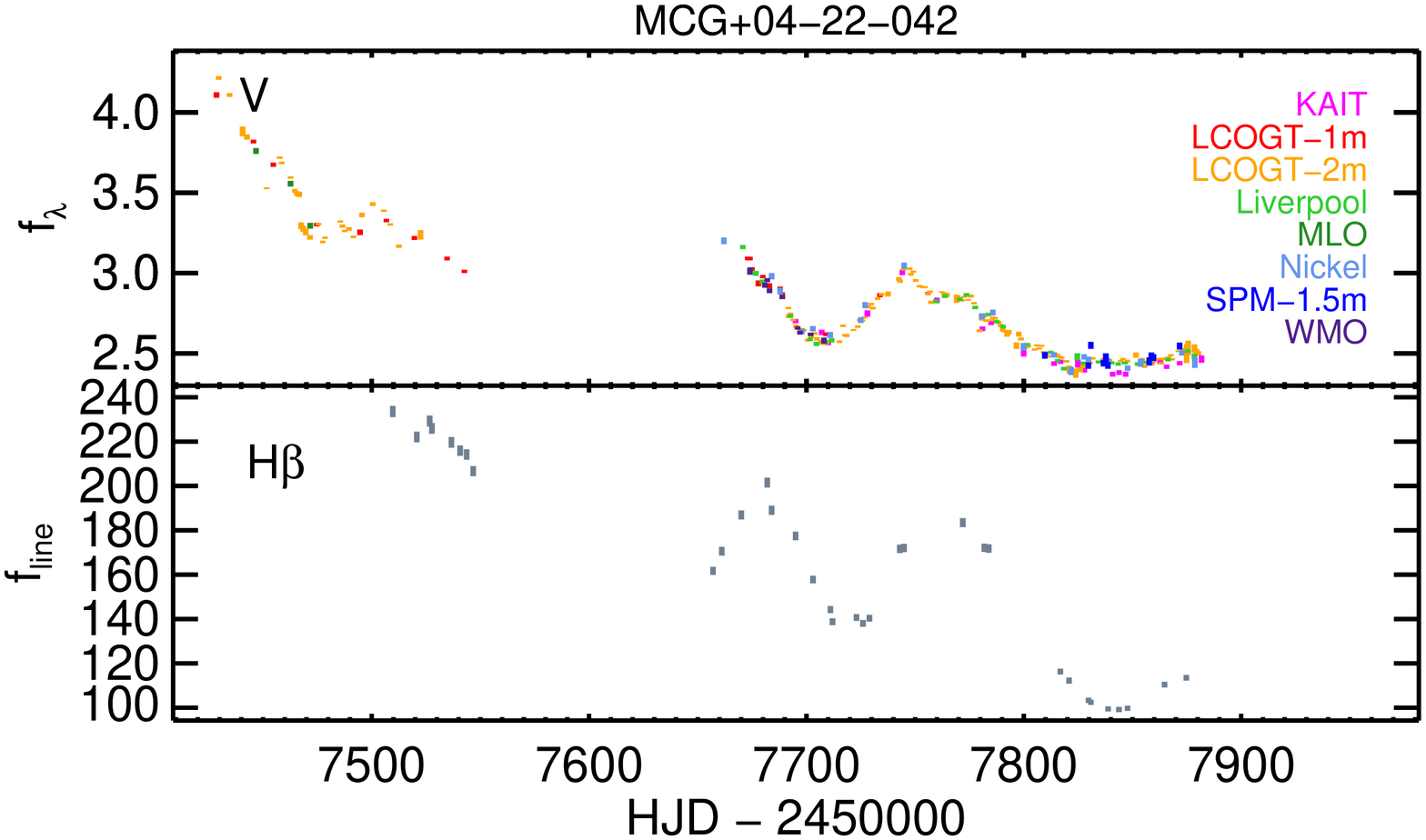}
   \includegraphics[width=0.26\textwidth,angle=90]{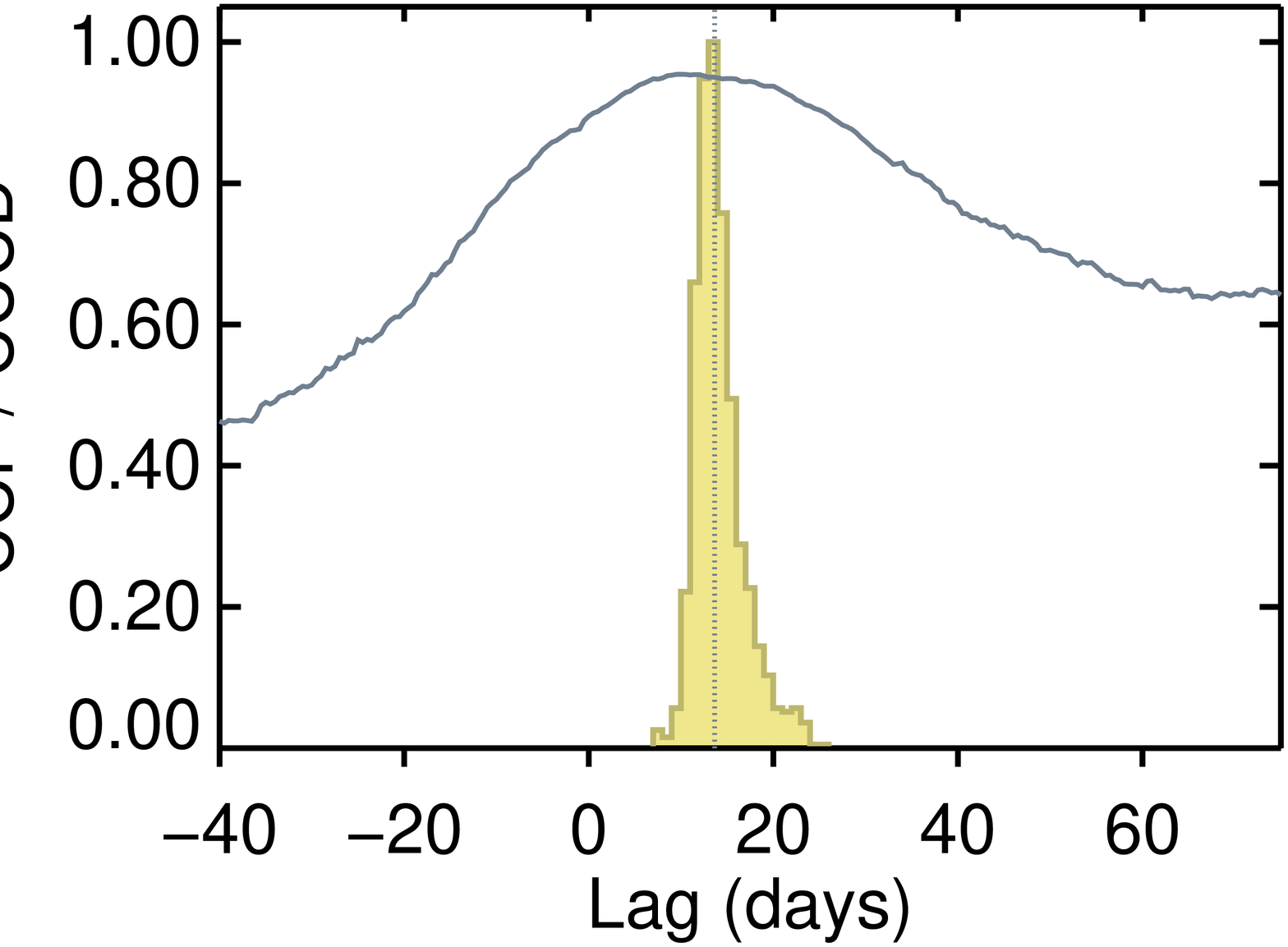}
   \includegraphics[width=0.39\textwidth,angle=90,trim={1.3in 0 0 0},clip]{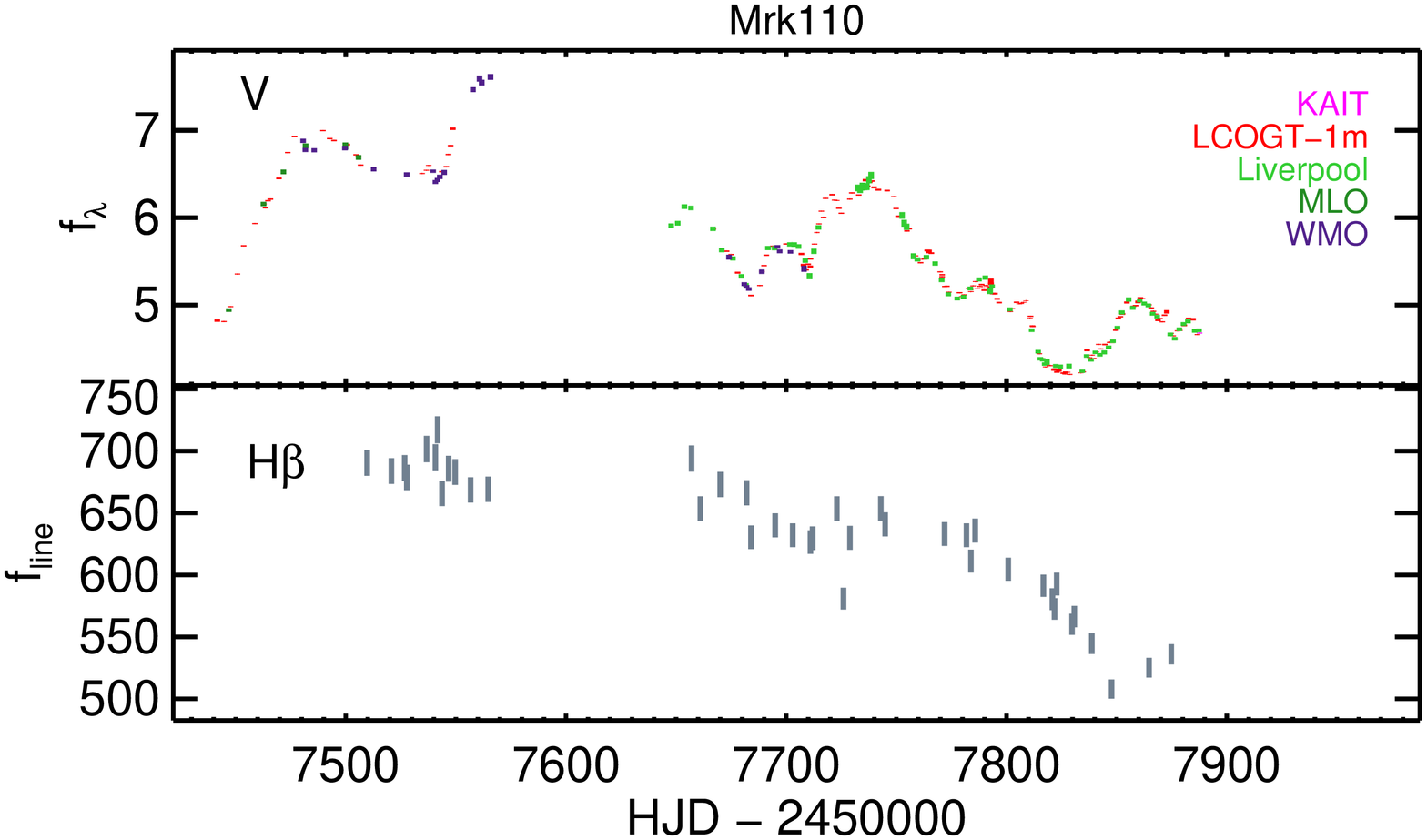}
   \hspace{0.1in}
   \includegraphics[width=0.26\textwidth,angle=90]{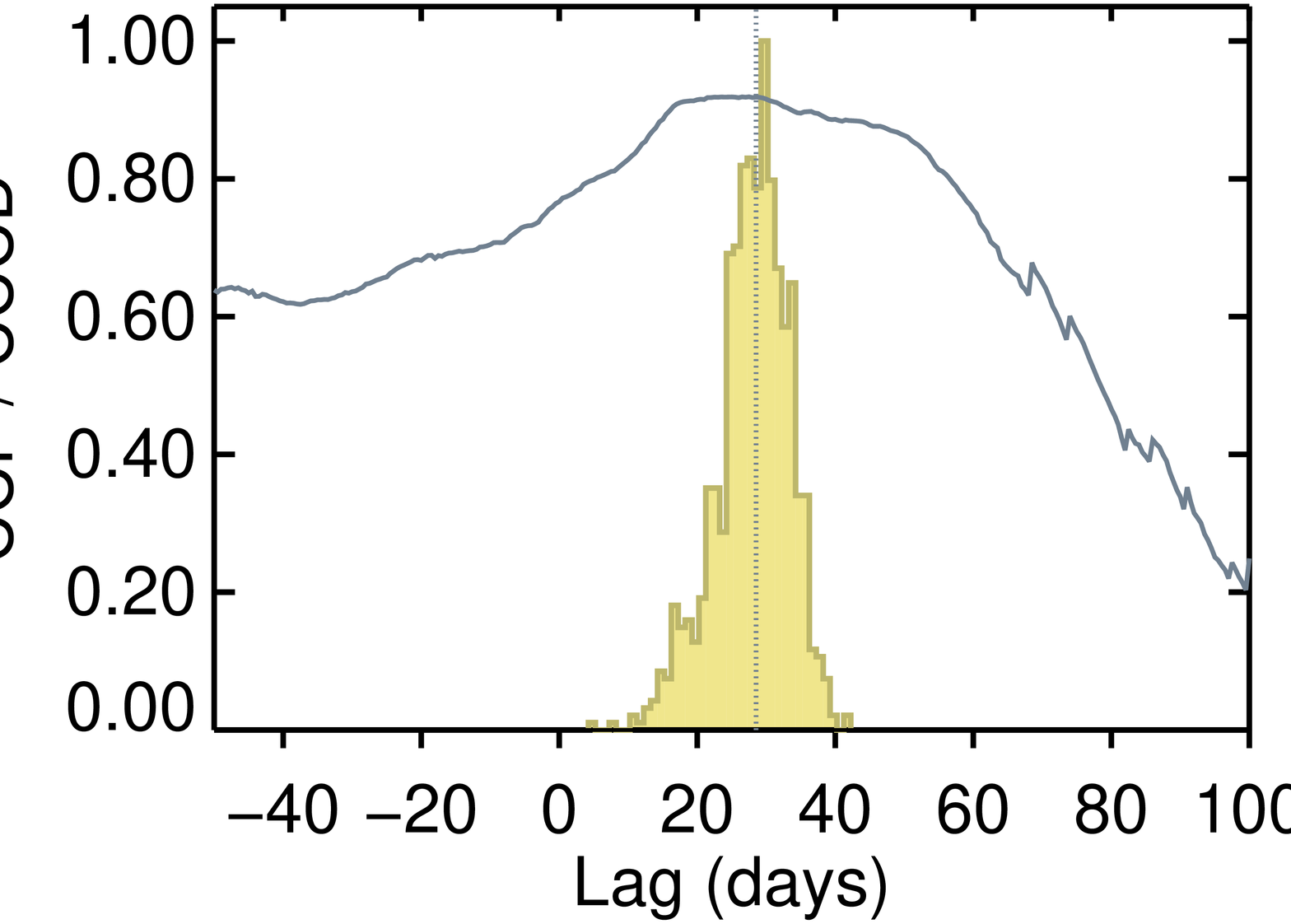}
    \caption{\edit1{Same as Figure \ref{fig:ccf} (but for Mrk 704, MCG $+$04$-$22$-$042, Mrk 110}}
    \label{fig:ccf3}
\end{figure*}

\begin{figure*}[htbp]
   \centering
   \includegraphics[width=0.39\textwidth,angle=90,trim={1.3in 0 0 0},clip]{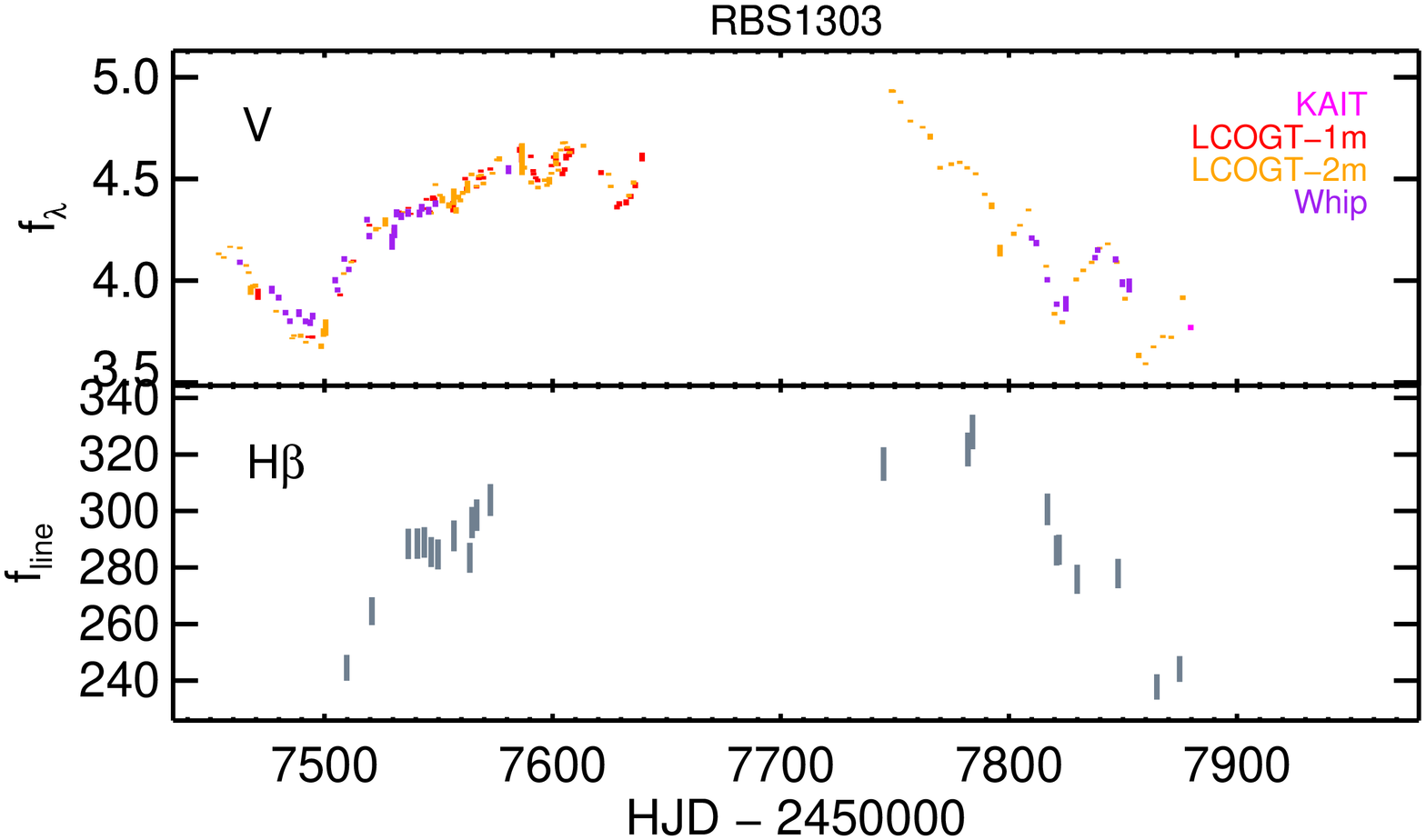}
   \includegraphics[width=0.26\textwidth,angle=90]{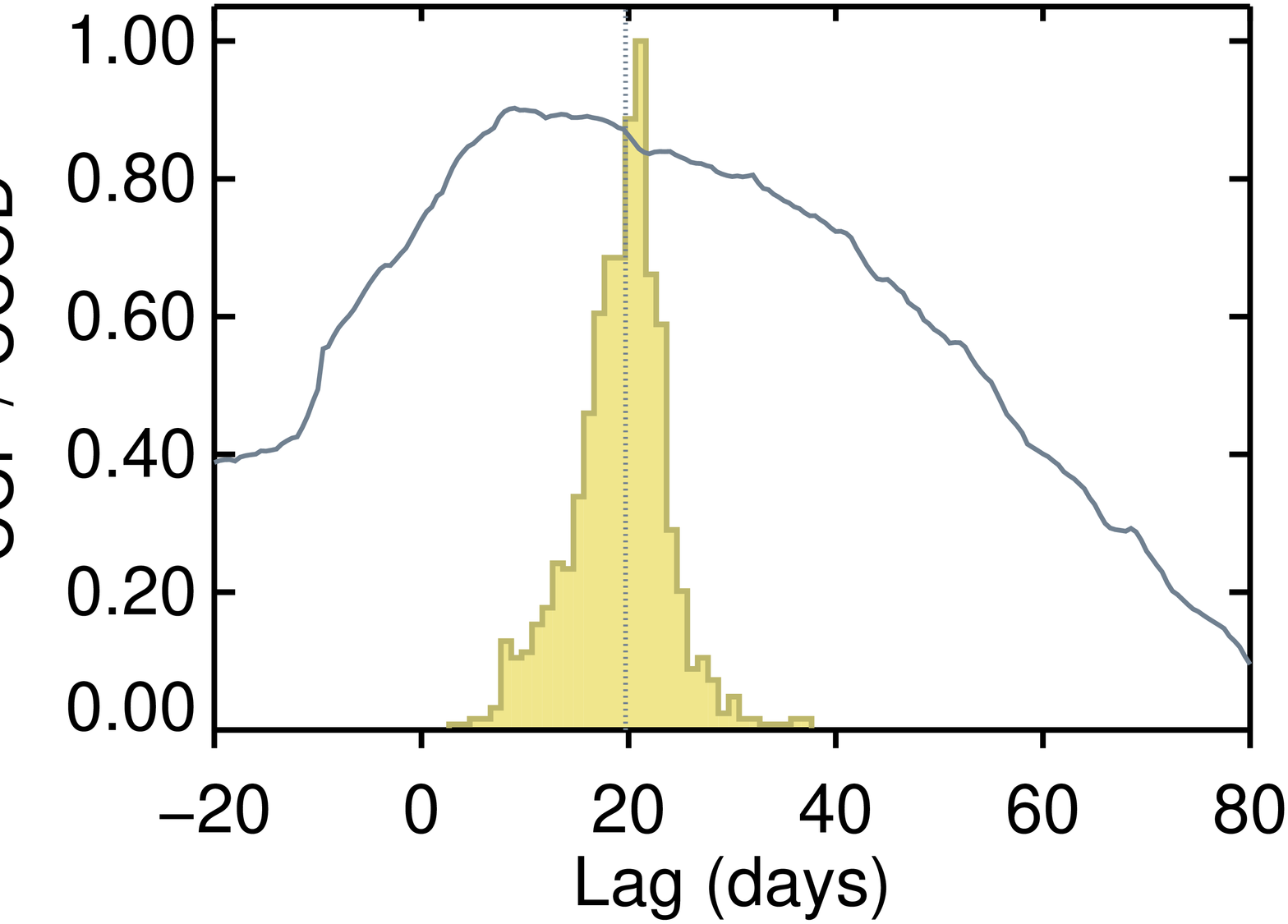}
   \includegraphics[width=0.39\textwidth,angle=90,trim={1.3in 0 0 0},clip]{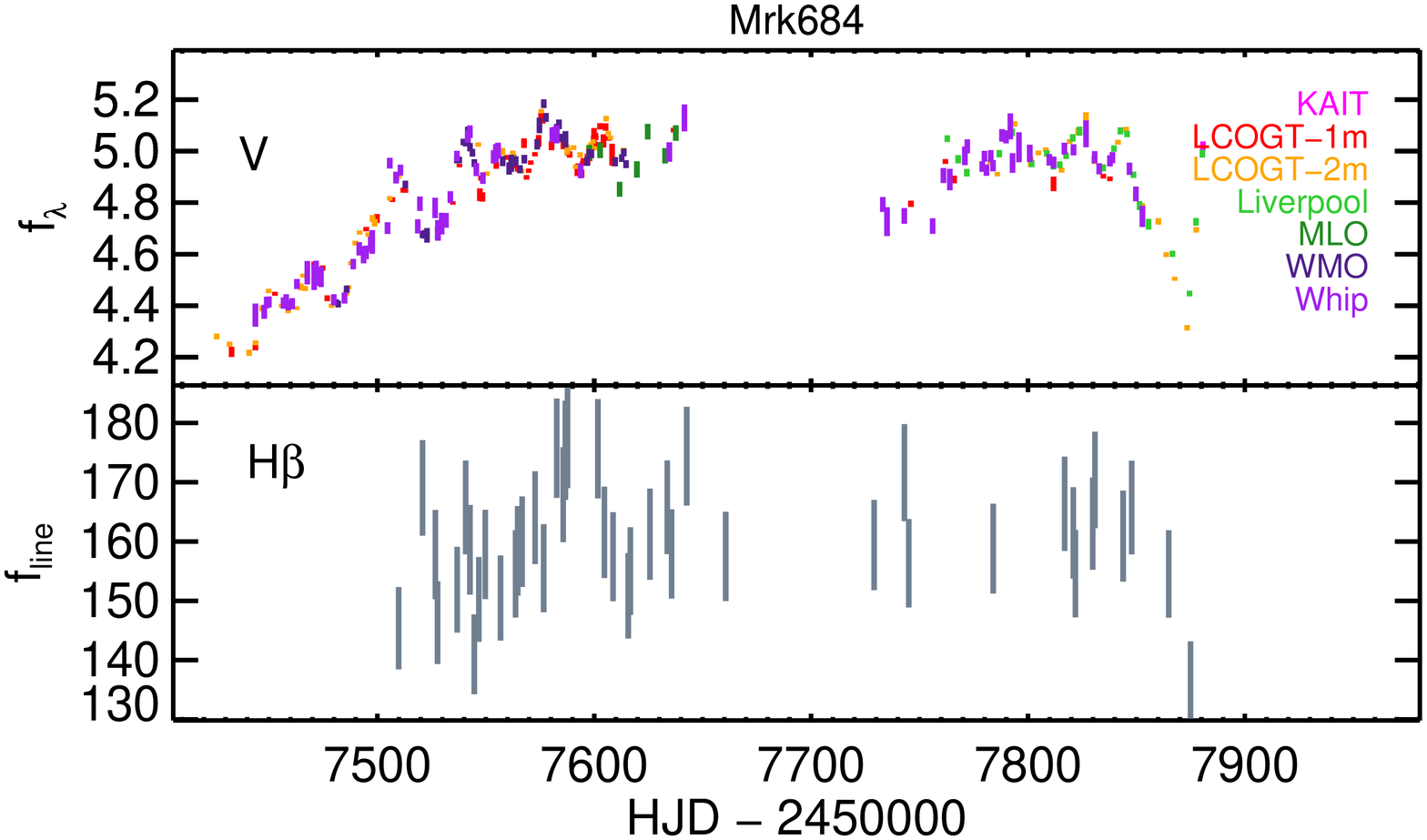}
   \hspace{0.1in}
   \includegraphics[width=0.26\textwidth,angle=90]{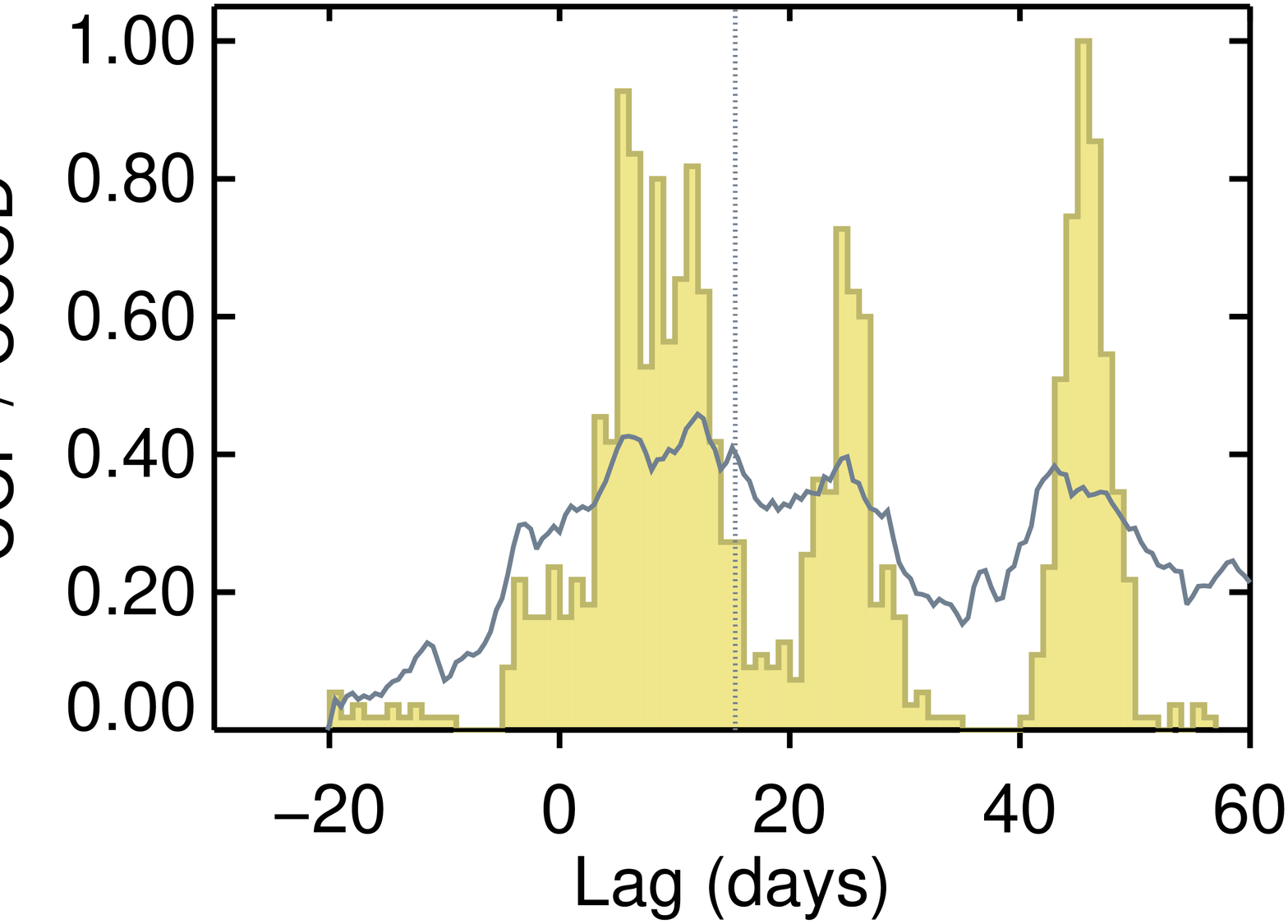}
   \includegraphics[width=0.39\textwidth,angle=90,trim={1.3in 0 0 0},clip]{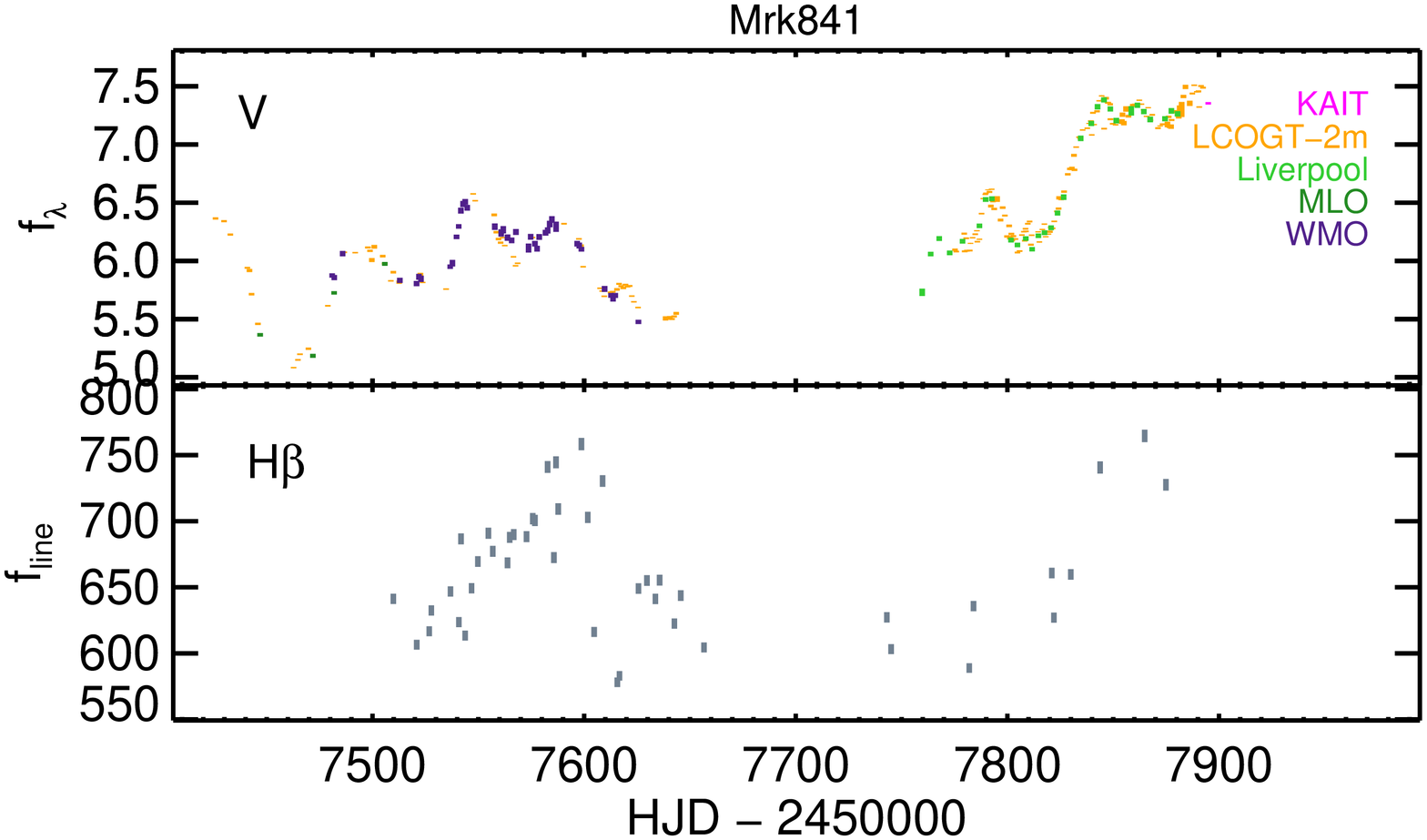}
   \hspace{0.1in}
   \includegraphics[width=0.26\textwidth,angle=90]{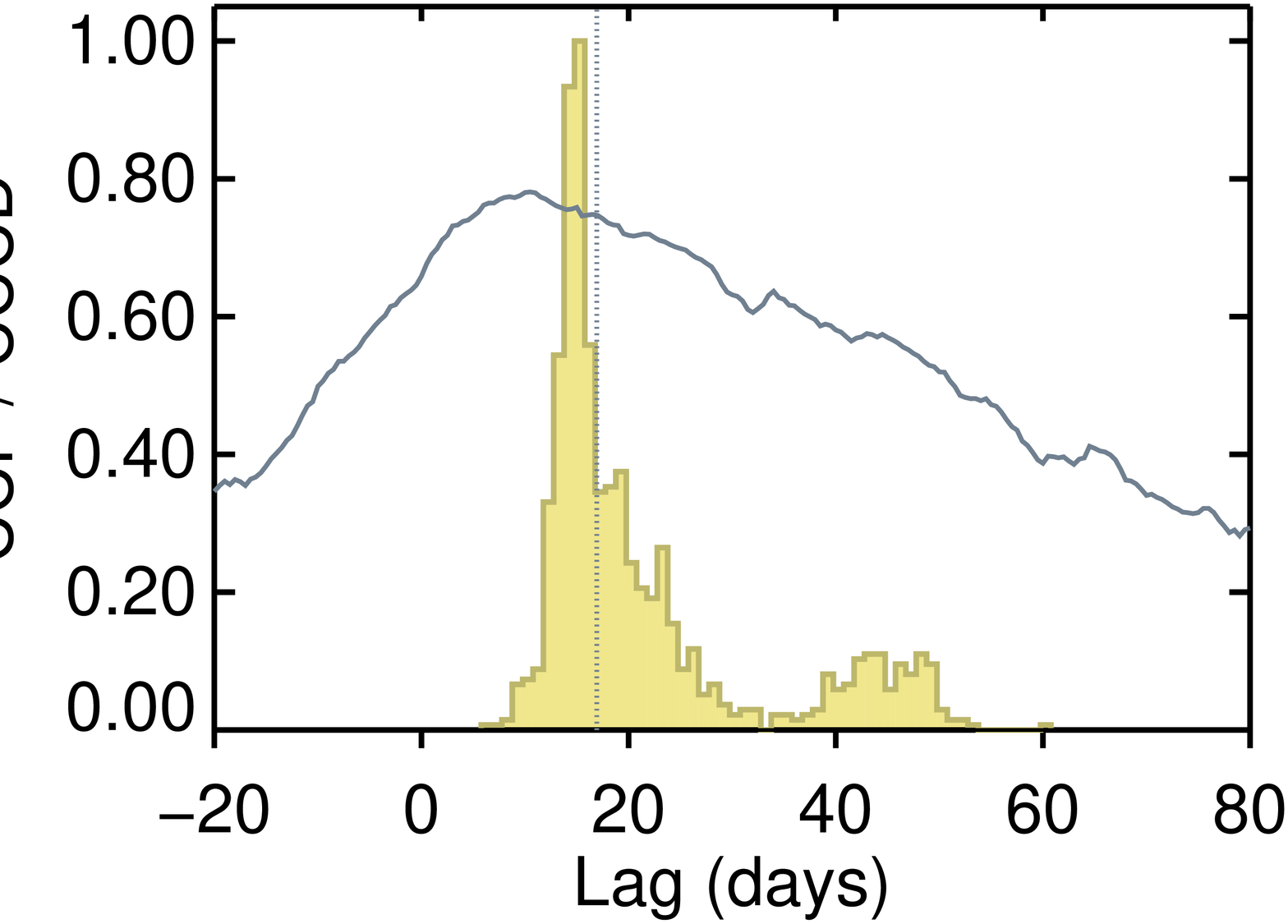}
    \caption{\edit1{Same as Figure \ref{fig:ccf} (but for RBS 1303, Mrk 684, Mrk 841)}}
    \label{fig:ccf4}
\end{figure*}

\begin{figure*}[htbp]
   \centering
   \includegraphics[width=0.39\textwidth,angle=90,trim={1.3in 0 0 0},clip]{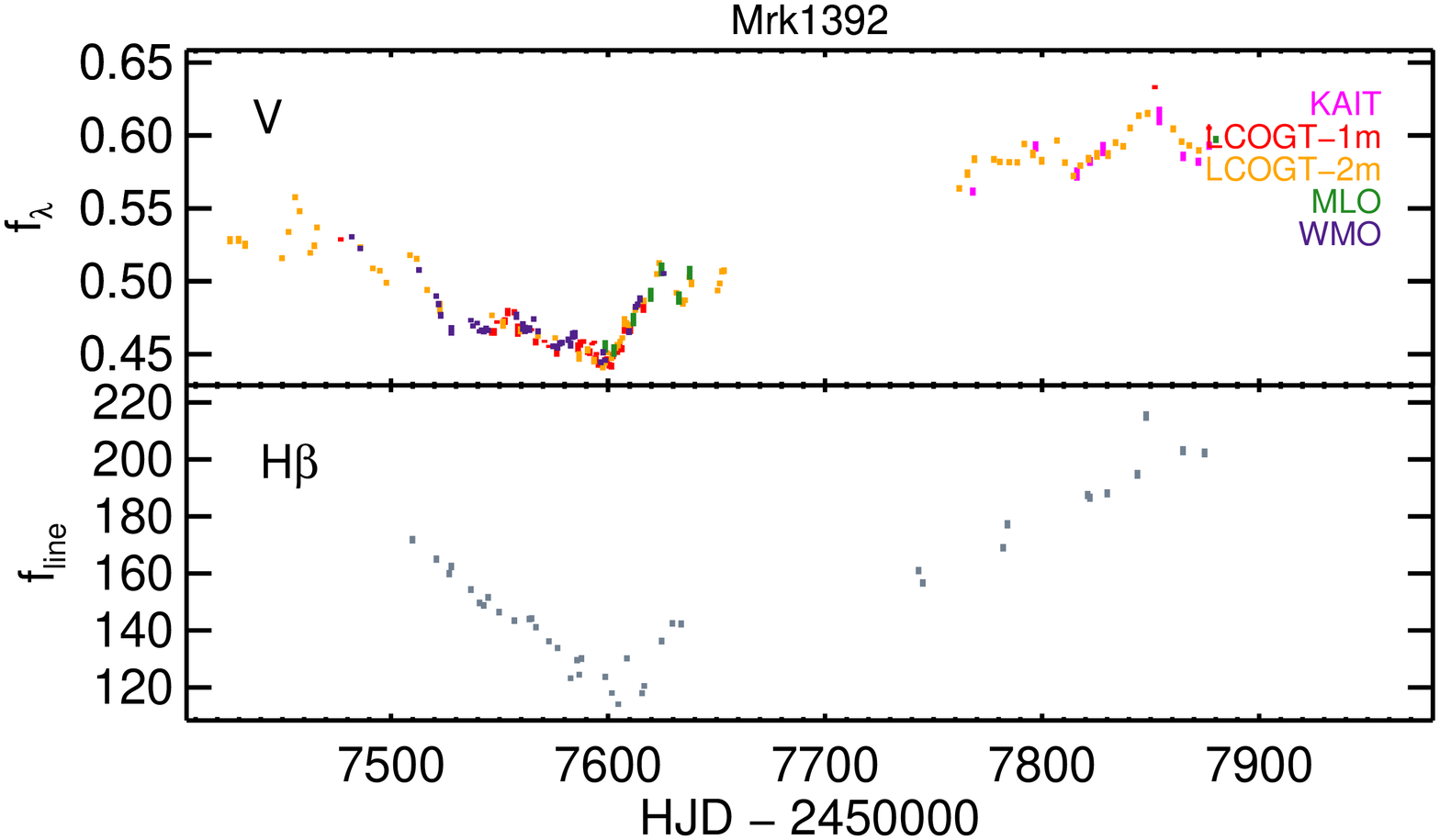}
   \hspace{0.1in}
   \includegraphics[width=0.26\textwidth,angle=90]{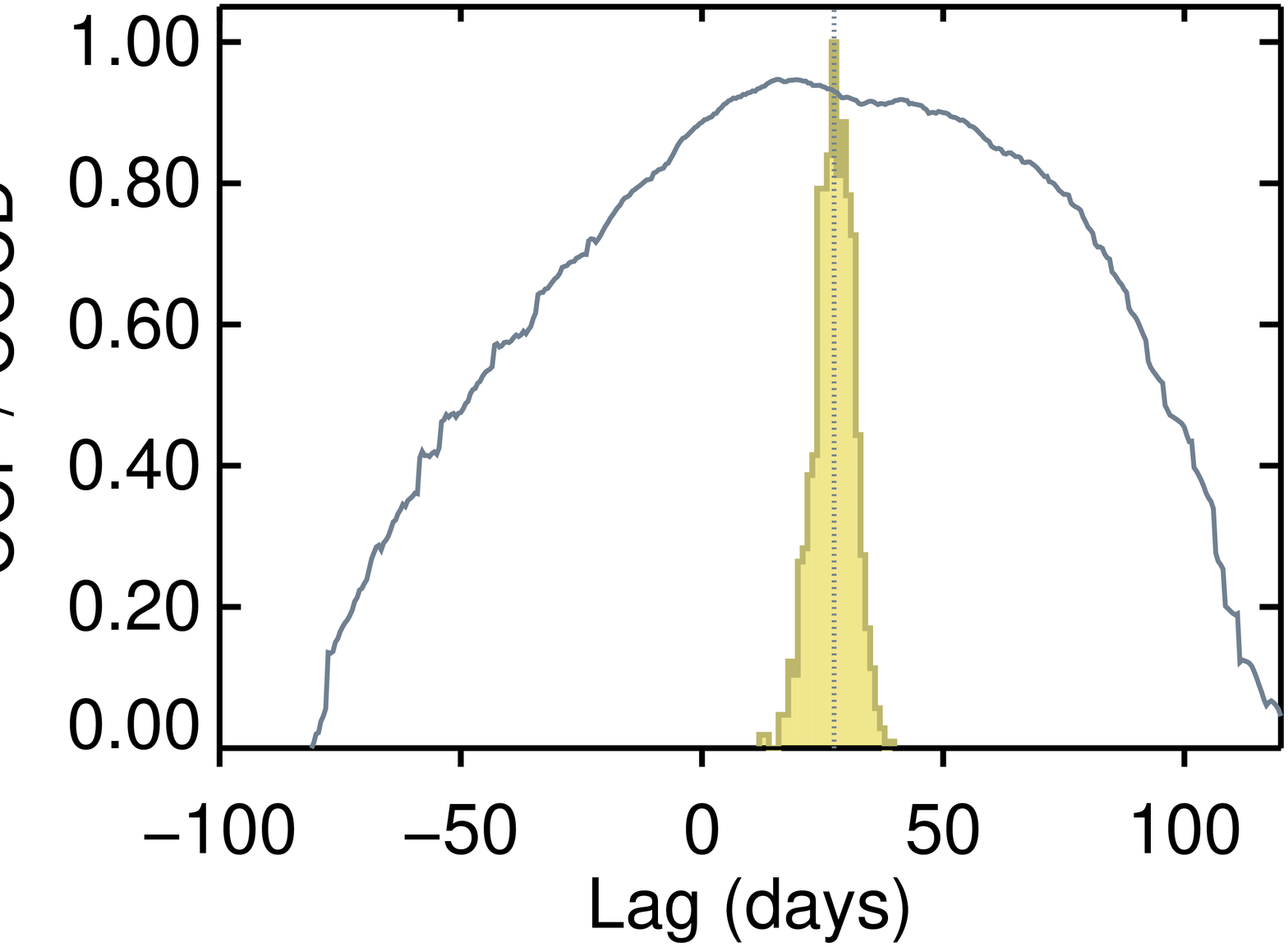}
   \includegraphics[width=0.39\textwidth,angle=90,trim={1.3in 0 0 0},clip]{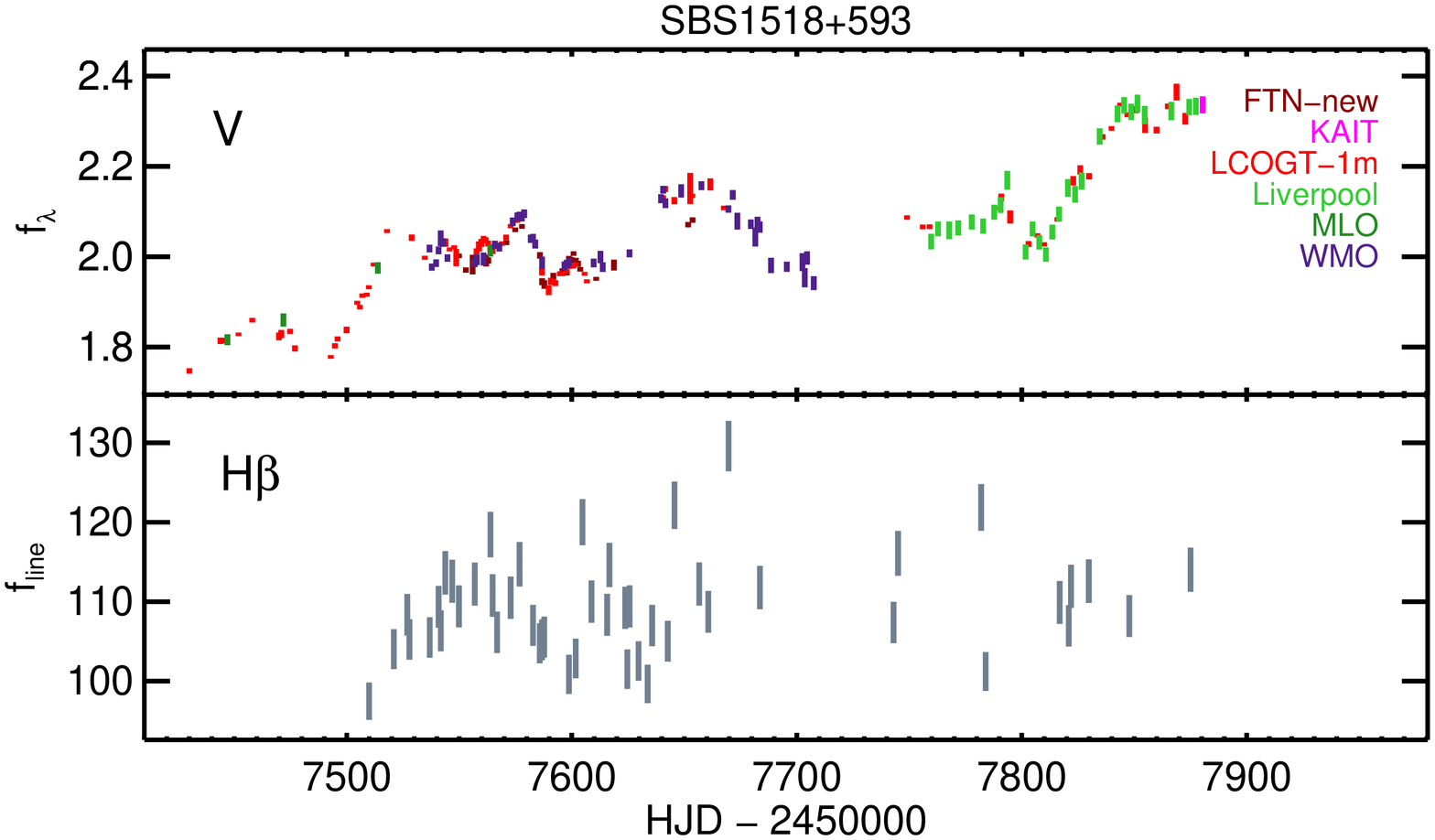}
   \hspace{0.1in}
   \includegraphics[width=0.26\textwidth,angle=90]{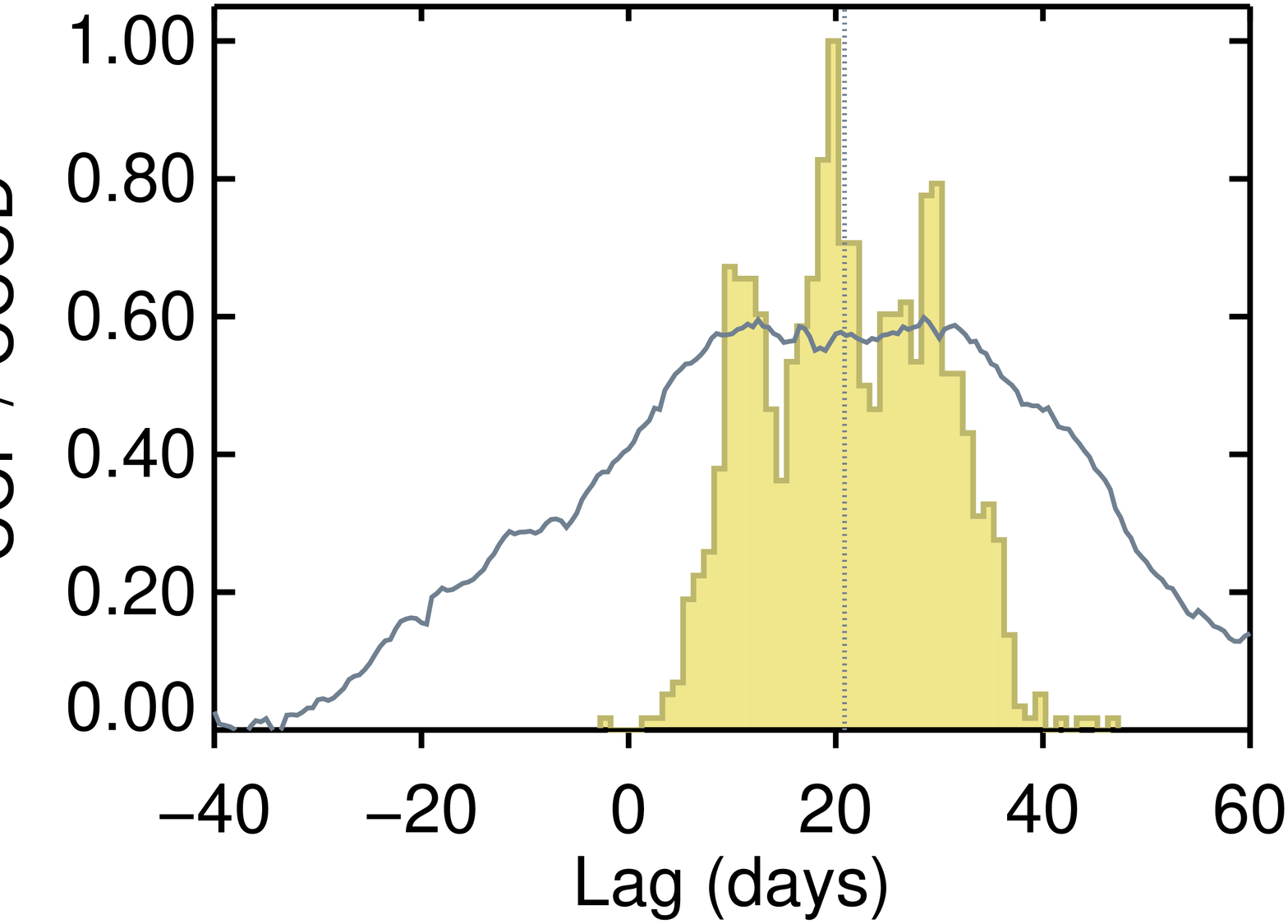}
   \includegraphics[width=0.39\textwidth,angle=90,trim={1.3in 0 0 0},clip]{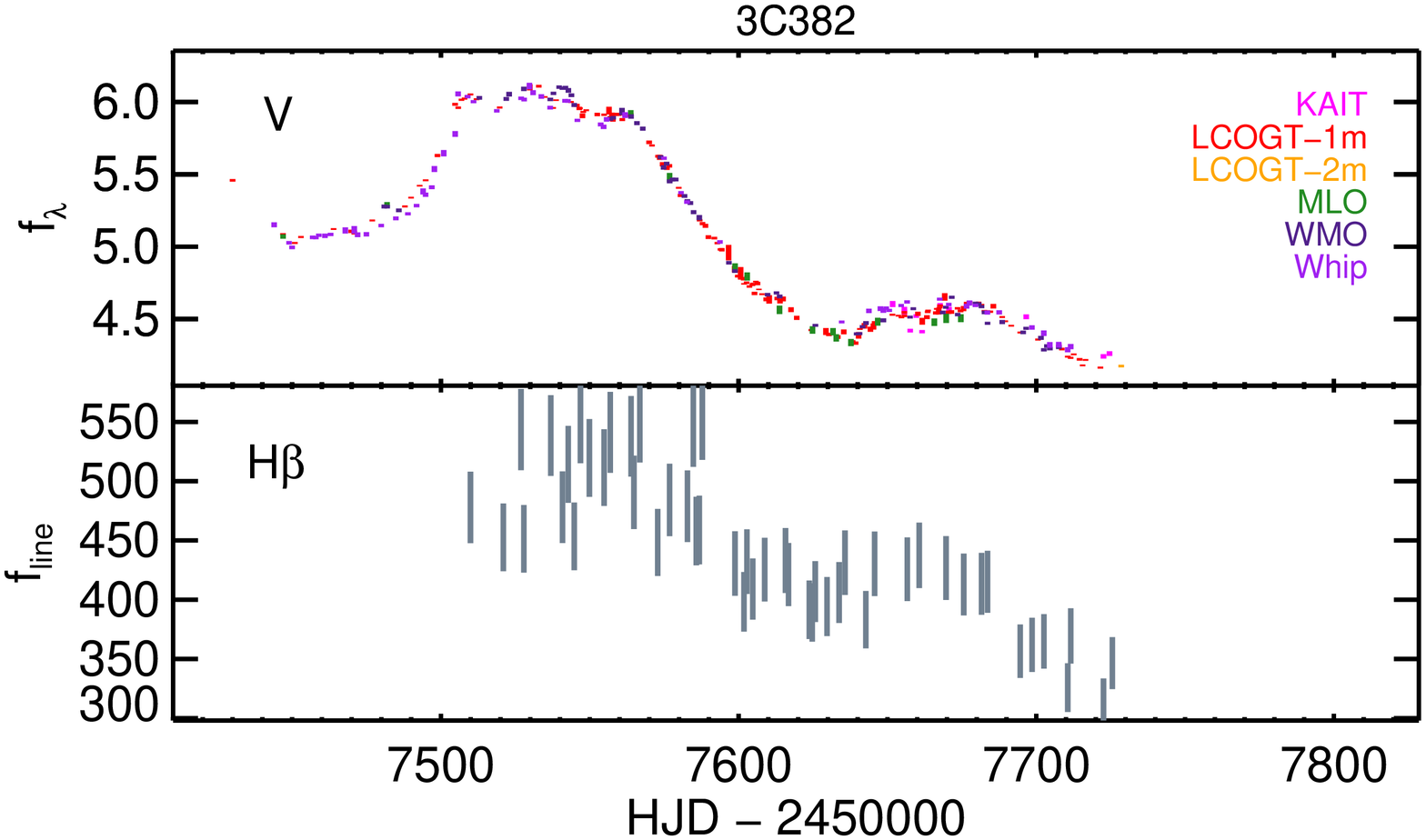}
   \hspace{0.1in}
   \includegraphics[width=0.26\textwidth,angle=90]{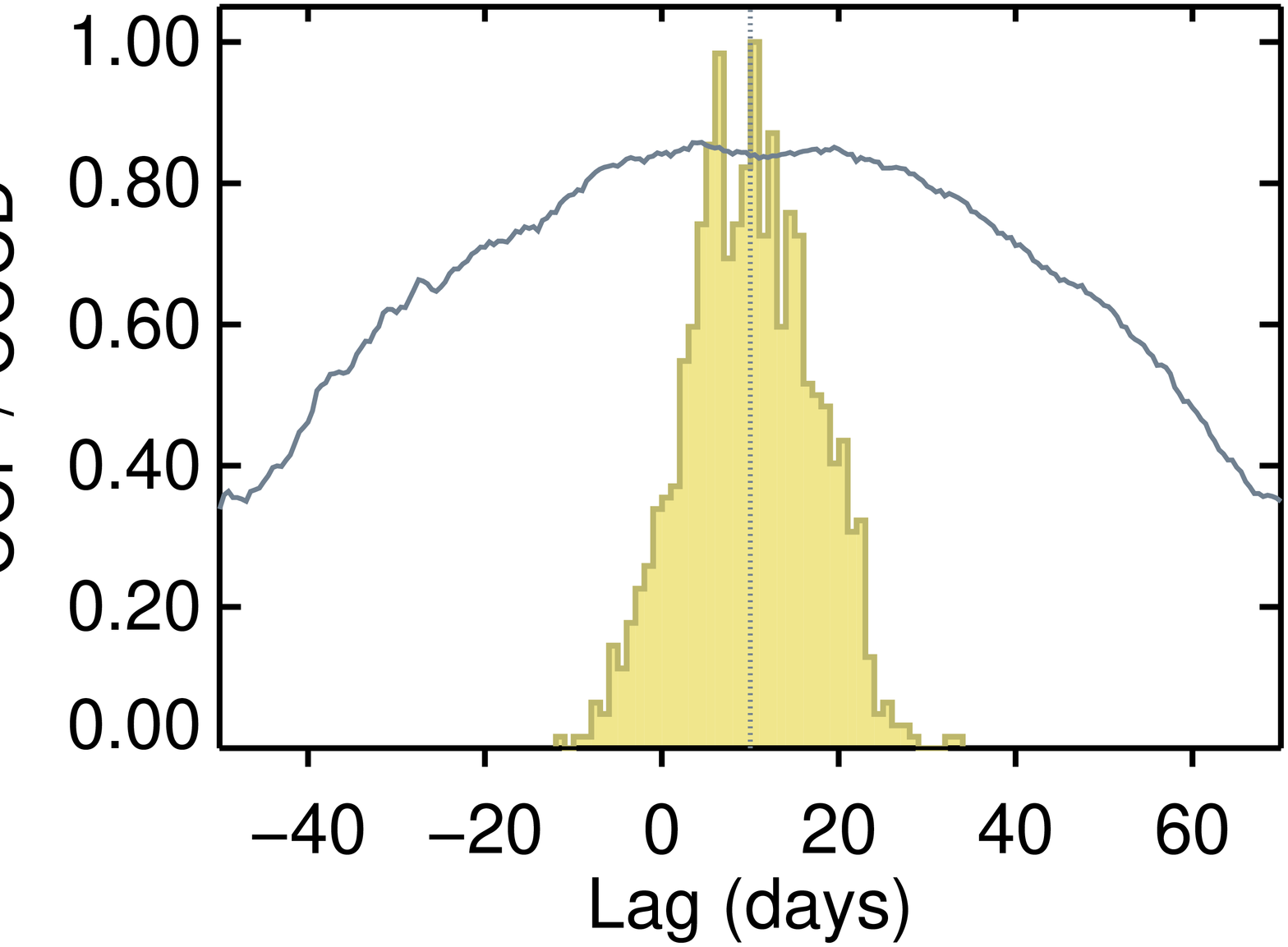}
    \caption{\edit1{Same as Figure \ref{fig:ccf} (but for Mrk 1392, SBS 1518$+$593, 3C 382)}}
    \label{fig:ccf5}
\end{figure*}

\begin{figure*}[htbp]
   \centering
   \includegraphics[width=0.39\textwidth,angle=90,trim={1.3in 0 0 0},clip]{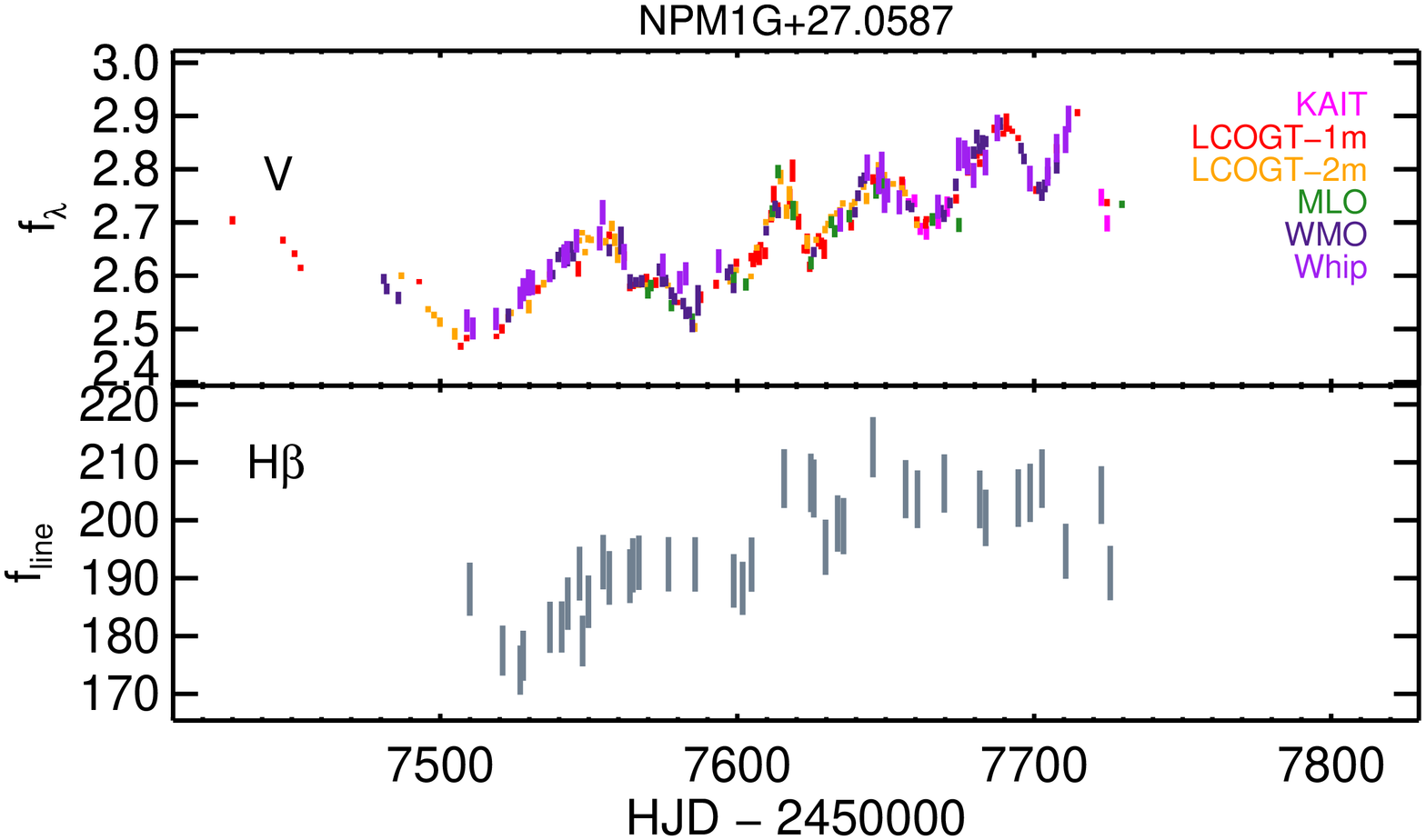}
   \hspace{0.1in}
   \includegraphics[width=0.26\textwidth,angle=90]{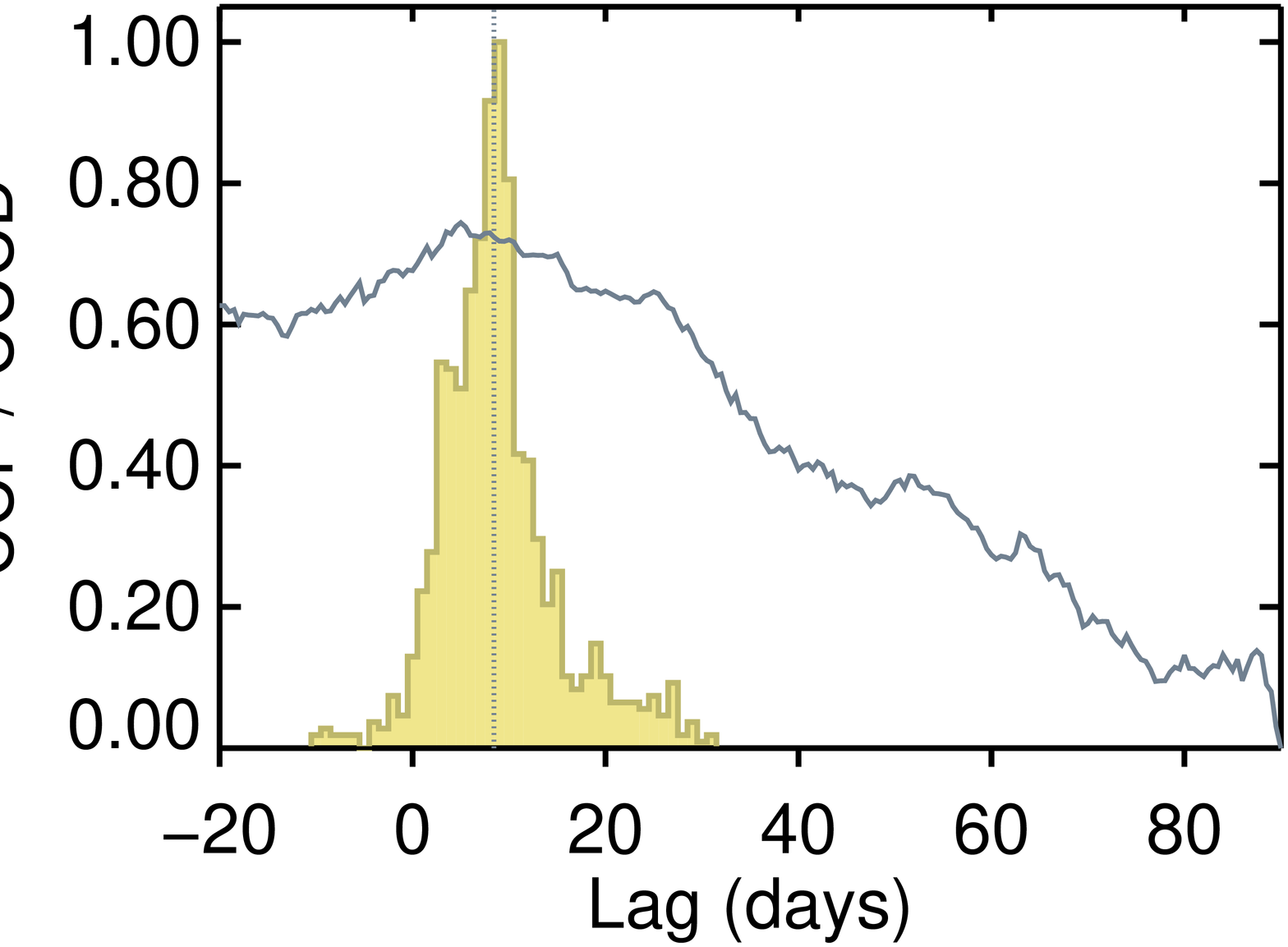}
   \includegraphics[width=0.39\textwidth,angle=90,trim={1.3in 0 0 0},clip]{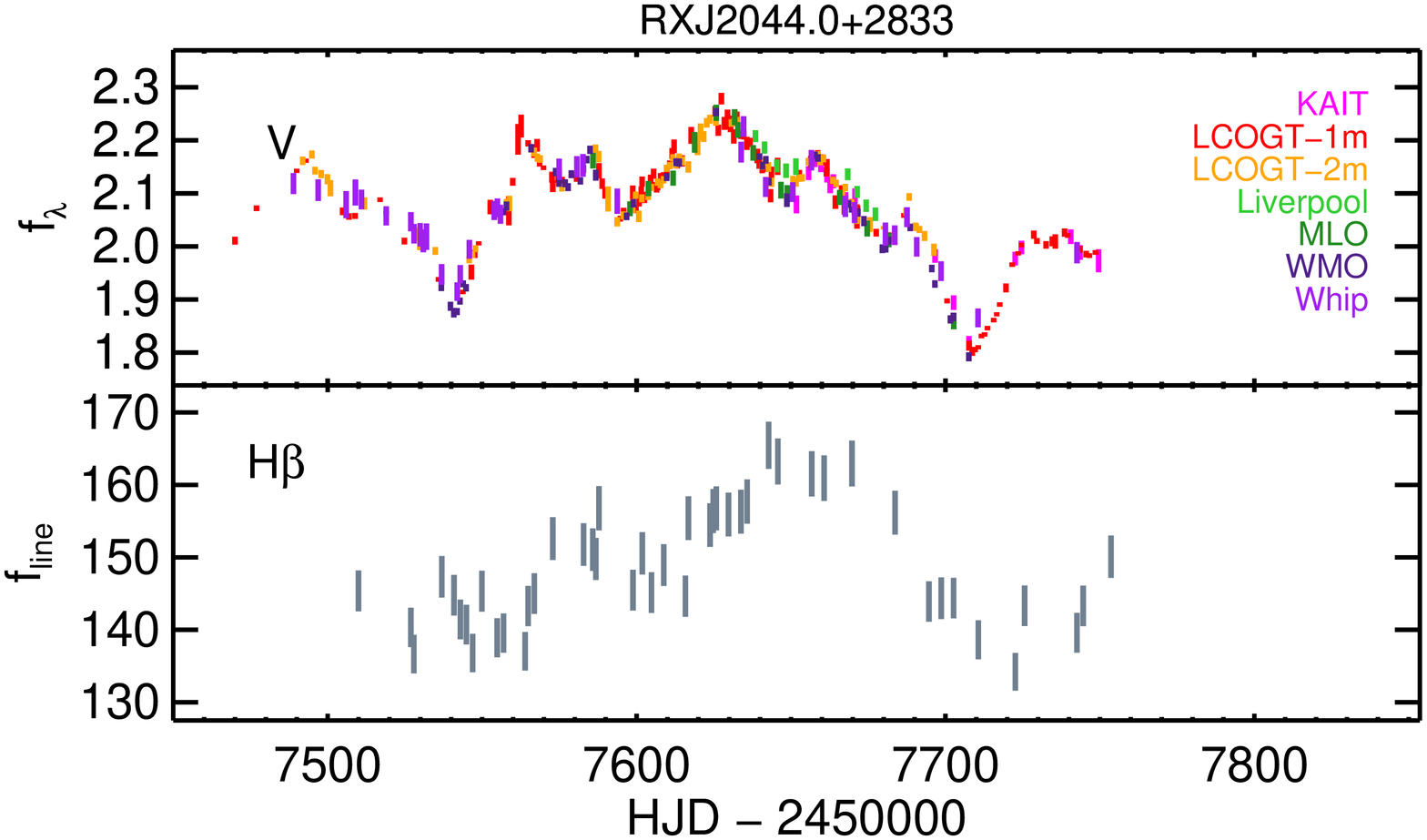}
   \hspace{0.1in}
   \includegraphics[width=0.26\textwidth,angle=90]{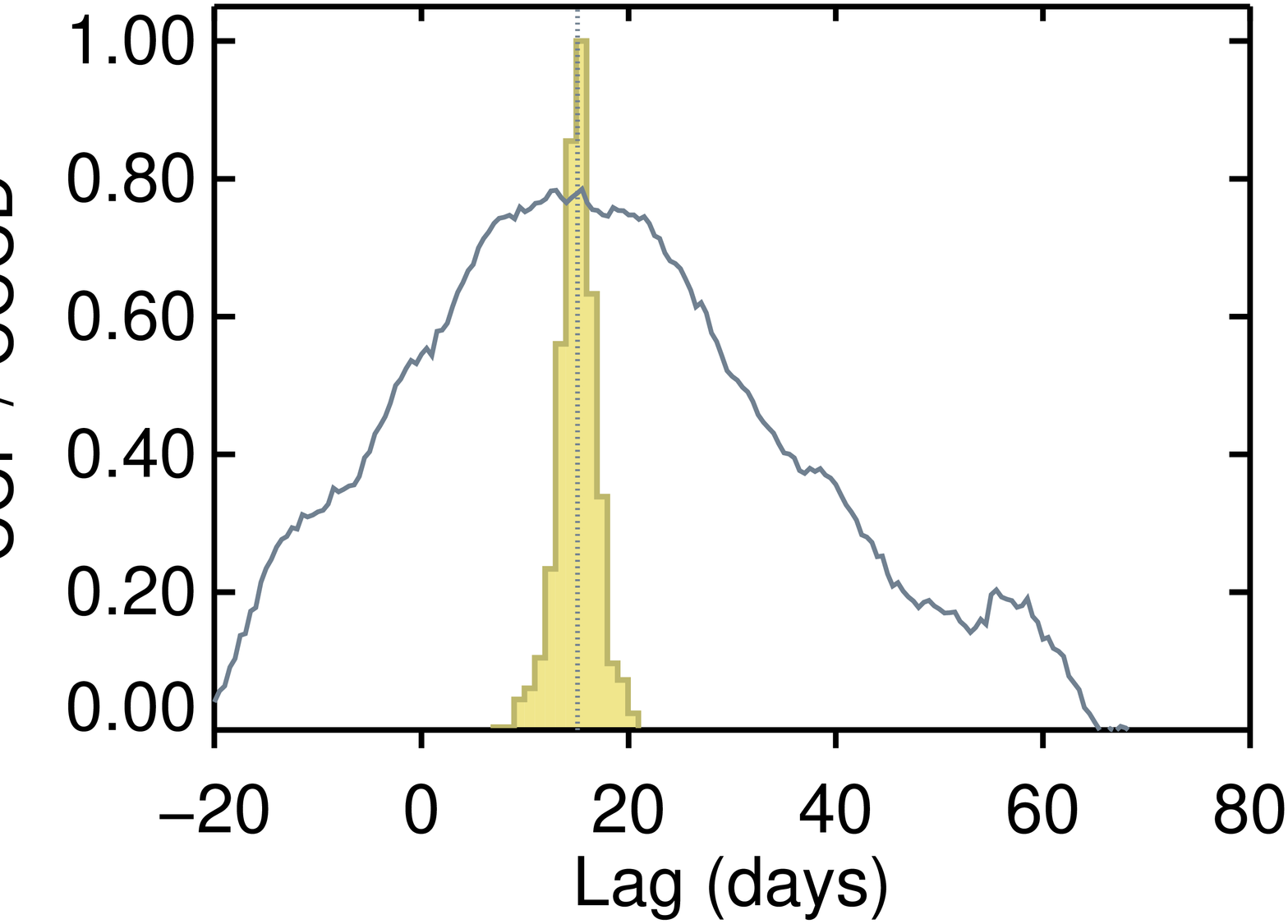}
   \includegraphics[width=0.39\textwidth,angle=90,trim={1.3in 0 0 0},clip]{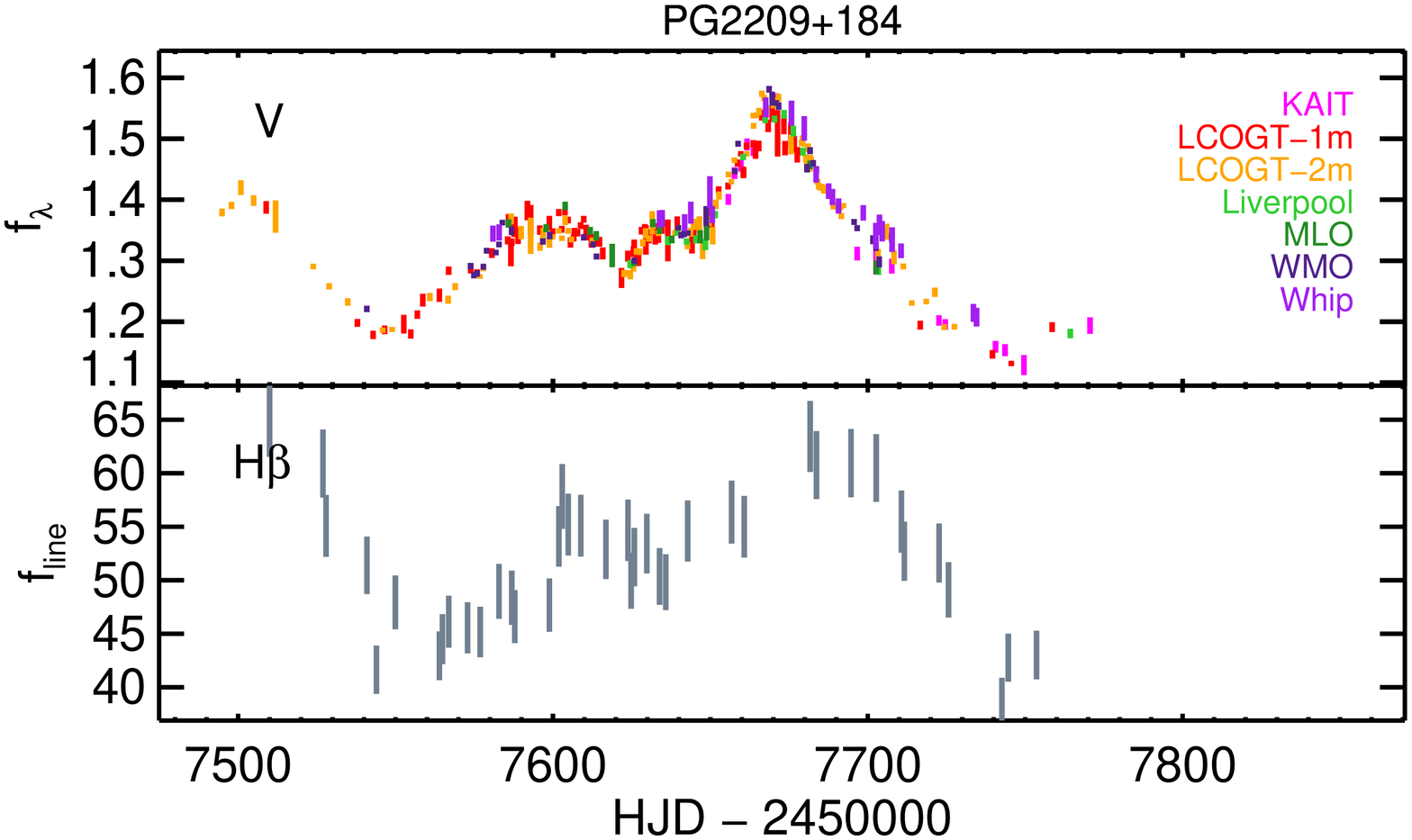}
   \hspace{0.1in}
   \includegraphics[width=0.26\textwidth,angle=90]{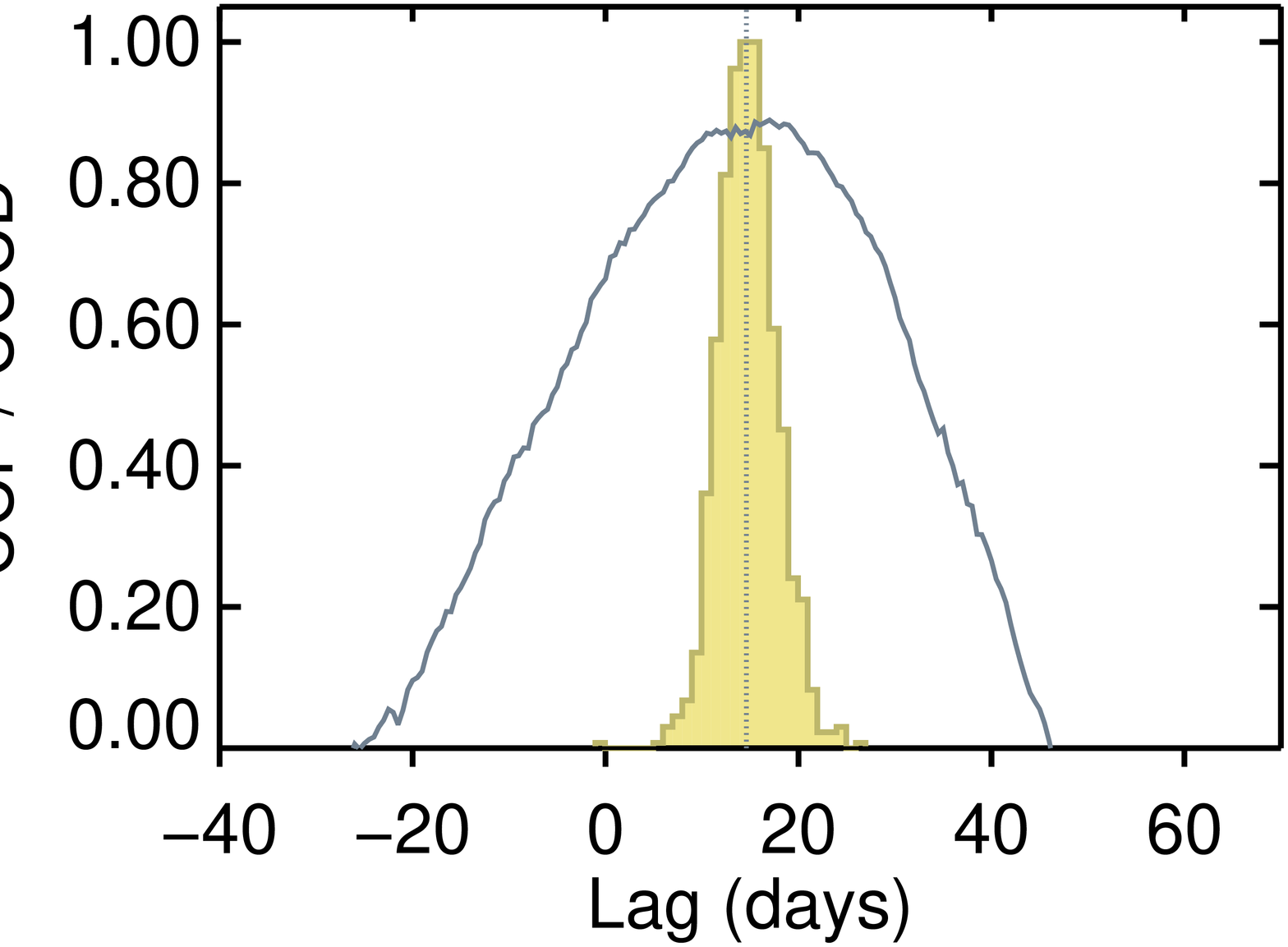}
    \caption{\edit1{Same as Figure \ref{fig:ccf} (but for NPM1G$+$27.0587, RXJ 2044.0$+$2833, PG 2209$+$184)}}
    \label{fig:ccf6}
\end{figure*}

\begin{figure*}[htbp]
   \centering
   \includegraphics[width=0.39\textwidth,angle=90,trim={1.3in 0 0 0},clip]{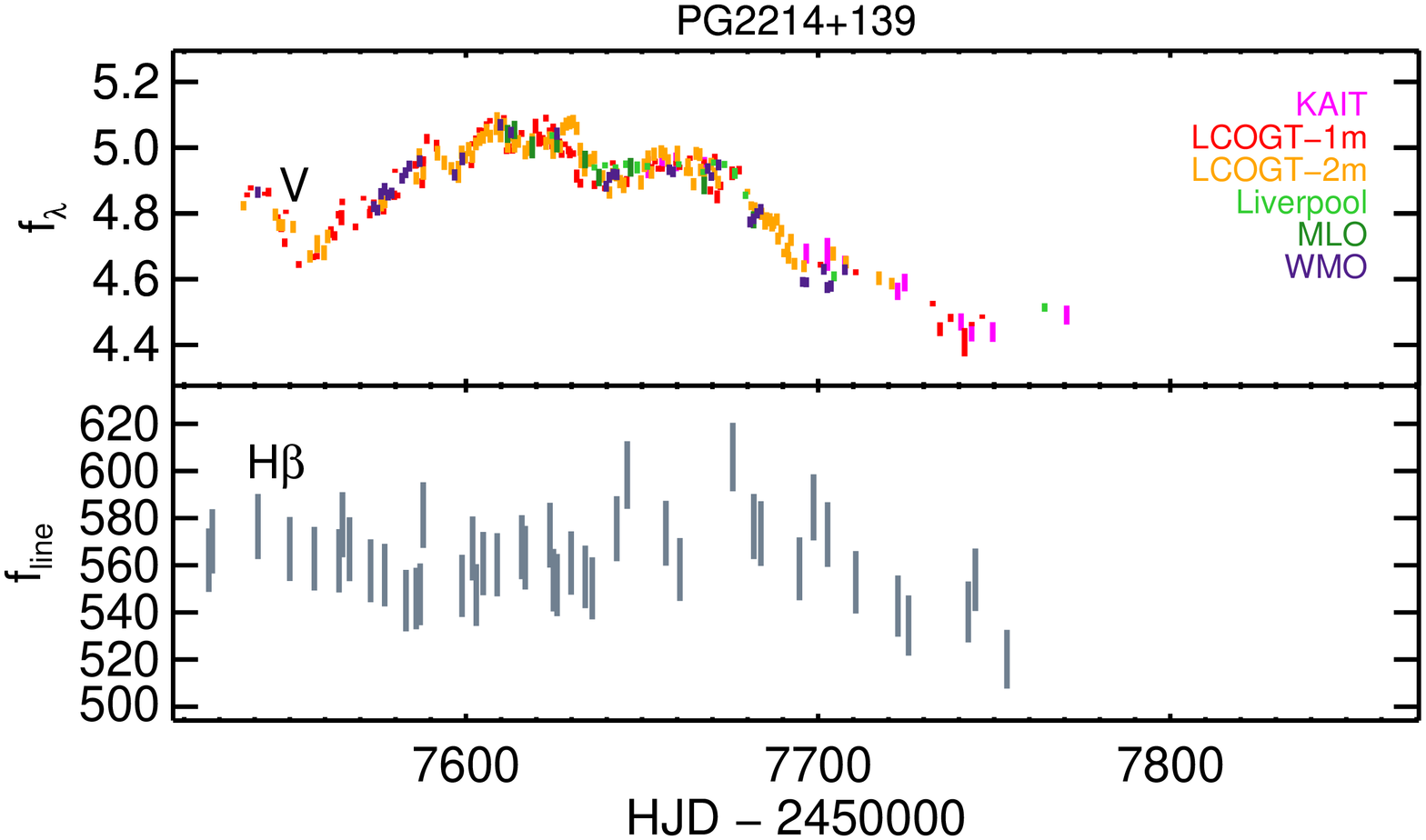}
   \hspace{0.1in}
   \includegraphics[width=0.26\textwidth,angle=90]{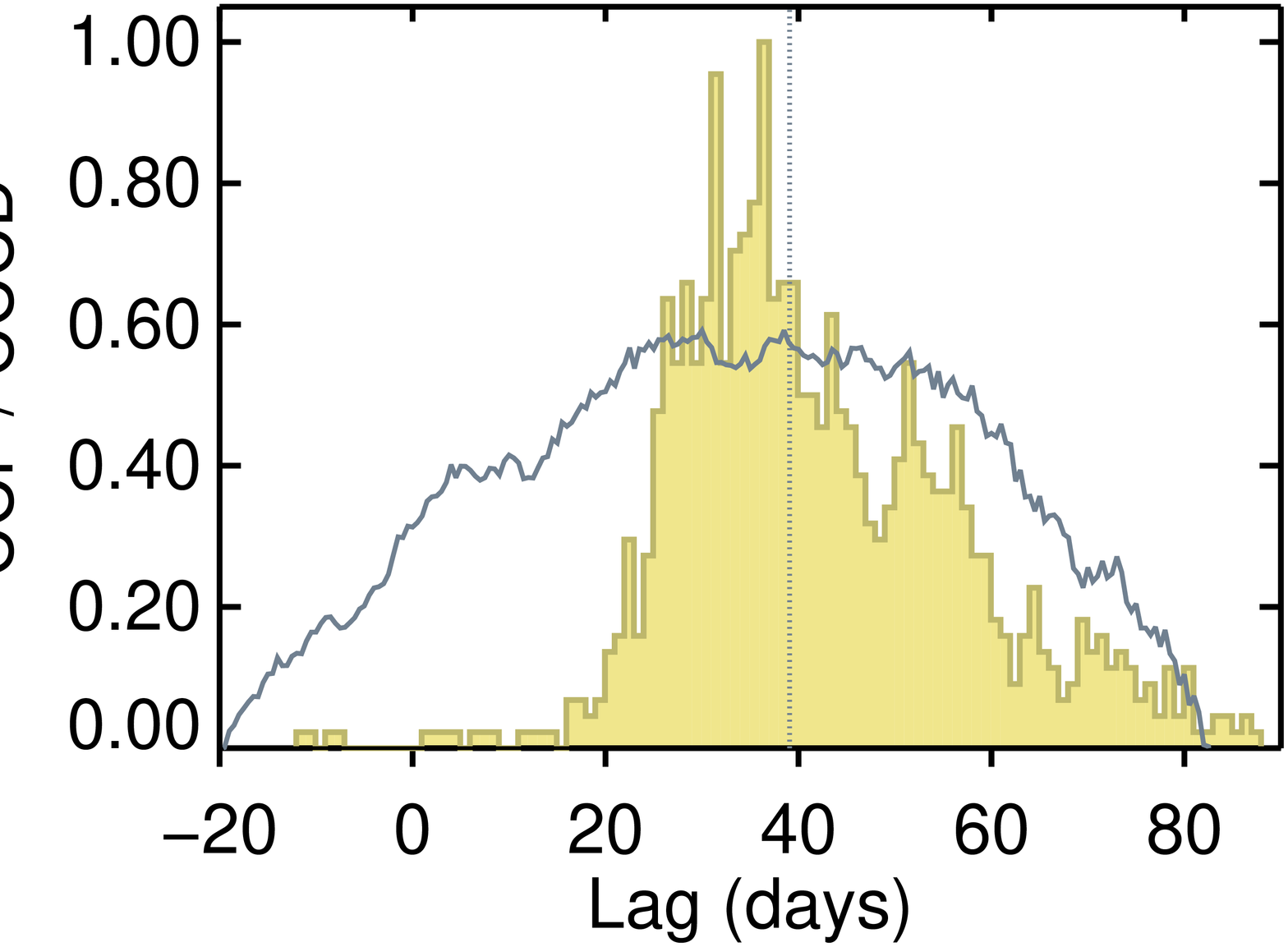}
   \includegraphics[width=0.39\textwidth,angle=90,trim={1.3in 0 0 0},clip]{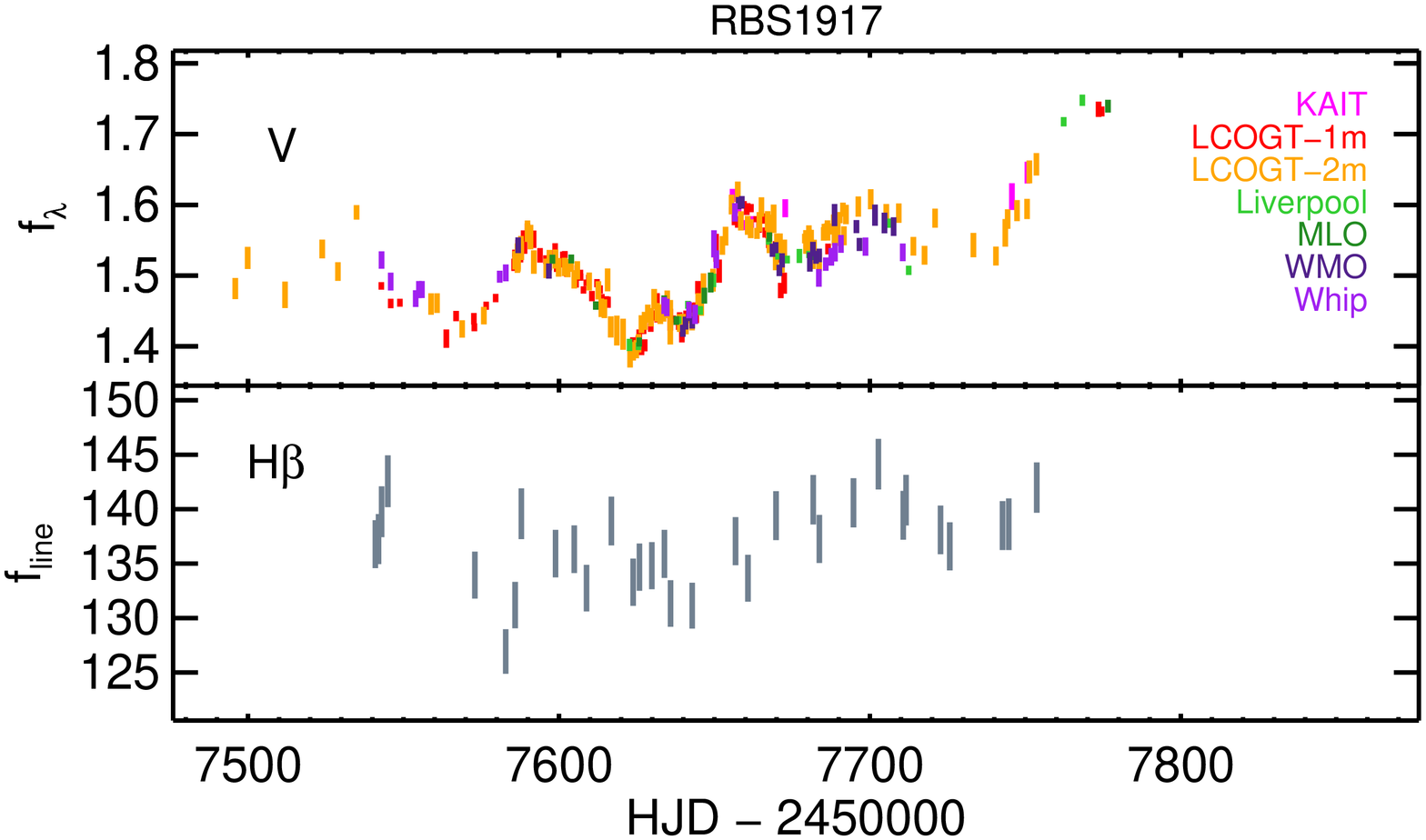}
   \hspace{0.1in}
   \includegraphics[width=0.26\textwidth,angle=90]{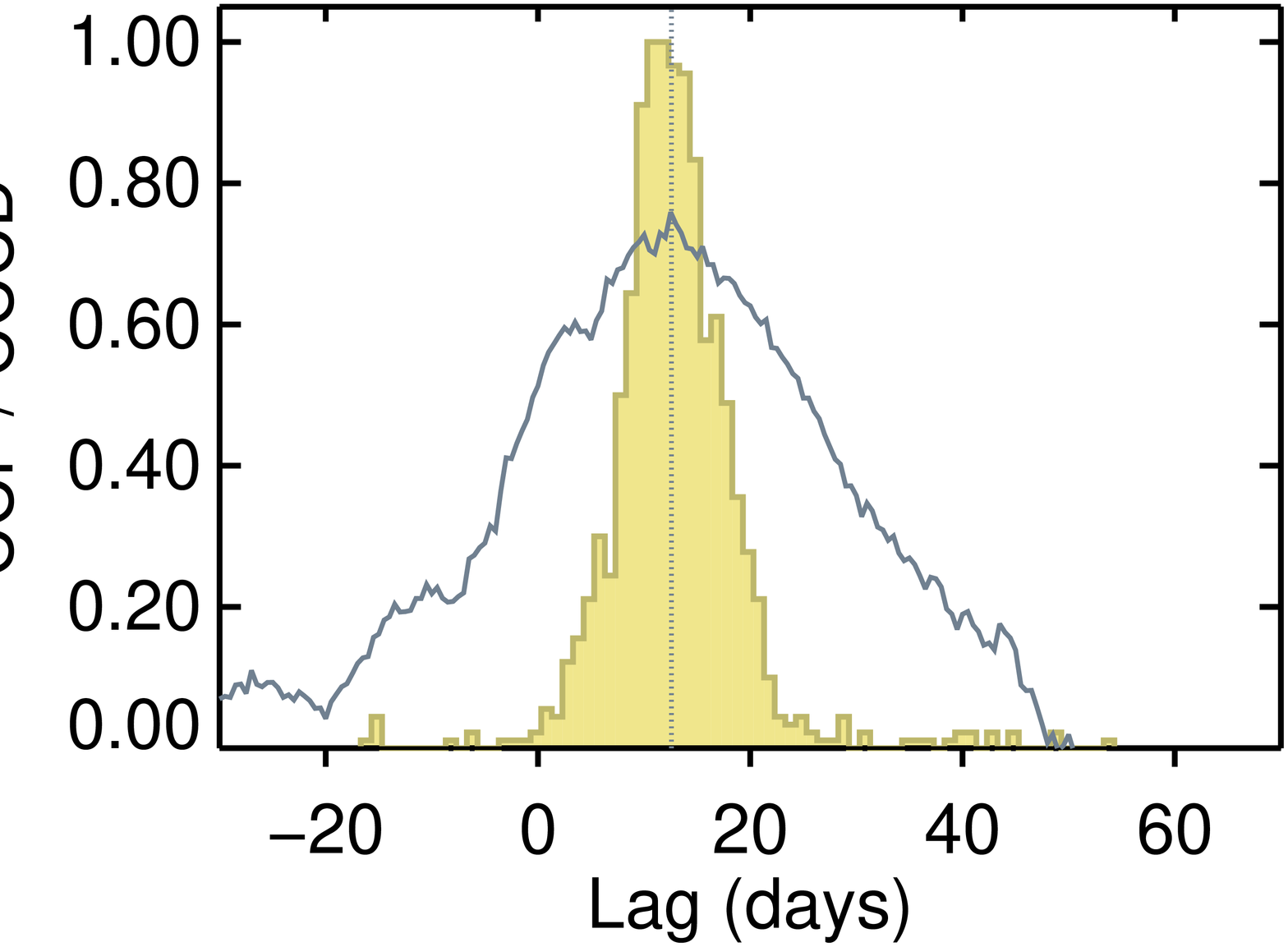}
   \includegraphics[width=0.39\textwidth,angle=90,trim={1.3in 0 0 0},clip]{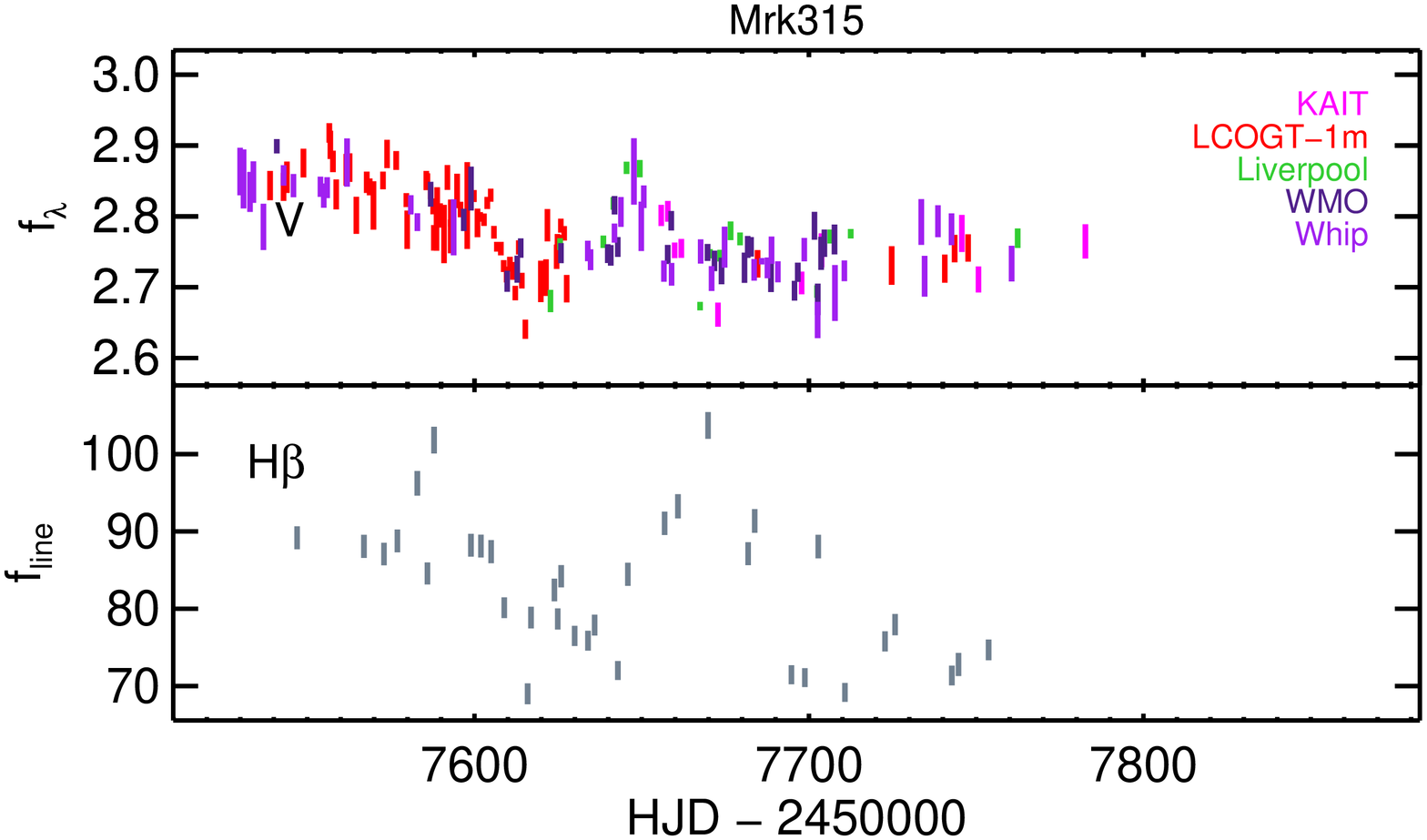}
   \hspace{0.1in}
   \includegraphics[width=0.26\textwidth,angle=90]{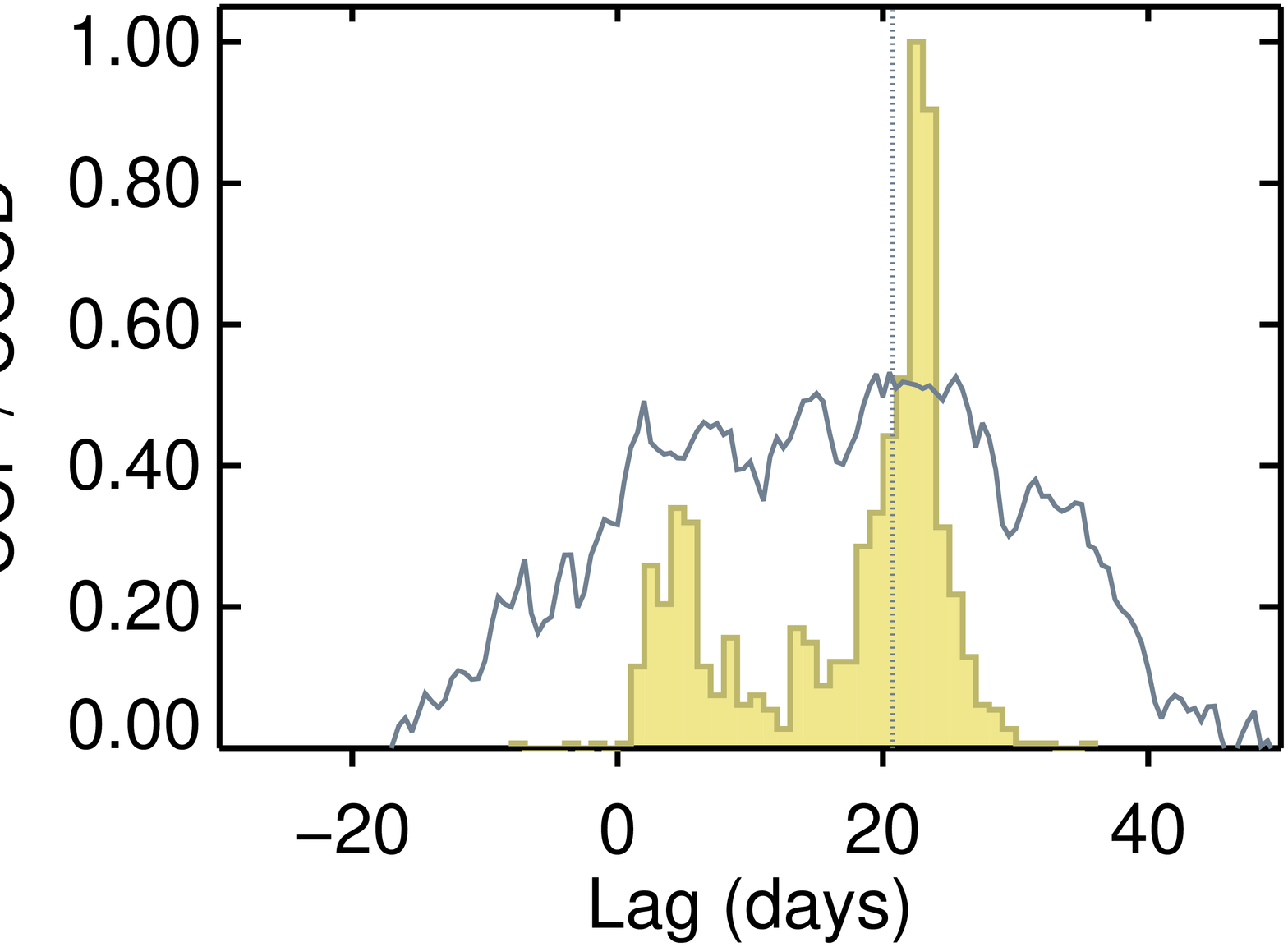}
    \caption{\edit1{Same as Figure \ref{fig:ccf} (but for PG 2214$+$139, RBS 1917, Mrk 315)}}
    \label{fig:ccf7}
\end{figure*}

Two lag measures are computed: $\tau_{\rm peak}$, the lag at the peak
of the CCF, and $\tau_{\rm cen}$, the centroid of the CCF 
for all points in the CCF above a predetermined threshold value
(80\% of the peak CCF value). We adopt the latter for the subsequent 
analysis. For error analysis, we employed the Monte Carlo flux randomization method~\edit1{\citep{Peterson98}} and
repeated the CCF calculation for 10$^3$ realizations, building up
distributions of correlation measurements. The median and $\pm$1$\sigma$
widths of the cross-correlation centroid distributions (CCCD) were then
adopted as the final lag measurements, and their associated
uncertainties as shown in \edit1{Figures~\ref{fig:ccf}--\ref{fig:ccf7}}. 

\subsection{Assessment of Cross-Correlation Reliability}
In any RM campaign, it is expected that some AGN will yield highly robust measurements of reverberation lag, thanks to strong variability and high-quality data, while other objects may exhibit little or no evidence for correlated variability between the continuum and emission-line light curves. This can occur as a result of observational factors including low S/N or poor temporal sampling, or factors  intrinsic to the AGN including low variability amplitude or low responsivity of the emission line to variations in the ionizing continuum. There is not necessarily a clear demarcation between reliable and unreliable lag measurements, and a variety of methods have been used to assess the significance or quality of lag detections \citep[e.g.,][]{Grier17}. In some cases, a simple threshold value of the correlation strength $r_\mathrm{max}$ is used, but $r_\mathrm{max}$ alone is not necessarily a good indicator of correlation significance for red-noise light curves. An alternative method is to employ null-hypothesis testing to determine the probability that two uncorrelated red-noise light curves having the same S/N and cadence as the data would yield an $r_\mathrm{max}$ at least as strong as the observed value. Such methods have been employed for analysis of multiwavelength continuum correlations in AGN \citep[e.g.,][]{Uttley03,Arevalo08,Chatterjee08}, but until recently have rarely been employed to assess broad-line reverberation lags \citep{Penton2021,Li21}. Our method will be presented in detail by Guo et al.\ (in prep.), and we provide a brief description here.

We first generated light curves according to the damped random walk (DRW) model with the Python software {\tt CARMA}\footnote{\url{https://github.com/brandonckelly/carma_pack}} ~\citep{Kelly09,Kelly14}. The DRW model provides an adequate description of ultraviolet/optical variabilities in AGN, though with plausible deviations on short~\citep{McHardy06,Mushotzky11,Kasliwal15,Smith18} or long \citep{MacLeod10,Guo17} timescales. Each simulated light curve represents a segment randomly selected from a 100-times longer light curve predicted from the same DRW model fitted to the observed light curve. We resampled each mock light curve to have the same cadence as the real observations and added Gaussian random noise based on the S/N of the data, creating $10^{3}$ realizations of each light curve. Cross-correlation measurements were then carried out using the simulated data to determine the distribution of $r_\mathrm{max}$ values that would be obtained for uncorrelated light curves. A two-way simulation was performed --- that is, we calculated the CCF between the real continuum and each simulated emission-line light curve for the first 500 simulations, and then the opposite way for the rest of the realizations.

We measured the CCF for all the simulations, using the same lag search range employed for each AGN, and counted, out of 10$^3$,
the number of positive lags ($\tau > 0$) with peak values $r_{\rm max}$ higher
than our observed $r_{\rm max}$. The resulting fraction represents the
probability that a correlation signal found between two uncorrelated light curves would exceed the correlation signal of the data.  
This derived $p$-value, denoted $p(r_\mathrm{max})$, thus provides an indicator of the robustness of our lag detections given the observed $r_\mathrm{max}$ and the observed properties of the light curves including S/N, cadence, and duration. 

We emphasize that the $p$-values derived from this method are not false-positive probabilities, and they do not give a measure of the absolute ``significance'' of a lag detection. Strictly speaking, our $p$-values simply give the probability that uncorrelated light curves having the same statistical properties as the data would give a cross-correlation signal as strong as that seen in the data. Consequently, a smaller $p$-value denotes a more robust and reliable detection of the correlation signal between two light curves. In general, for broad-line RM we have a strong prior that a reverberation signal is very likely to be present in the data, thus our null hypothesis of intrinsically uncorrelated light curves is an extreme and fairly unlikely scenario. We further emphasize that there is no strict cutoff between significant and insignificant lag detections by this method; the $p$-value merely gives an indication of the relative degree of reliability between different measurements.

\begin{figure}[htbp]
   \centering
   \includegraphics[width=0.39\textwidth,angle=90]{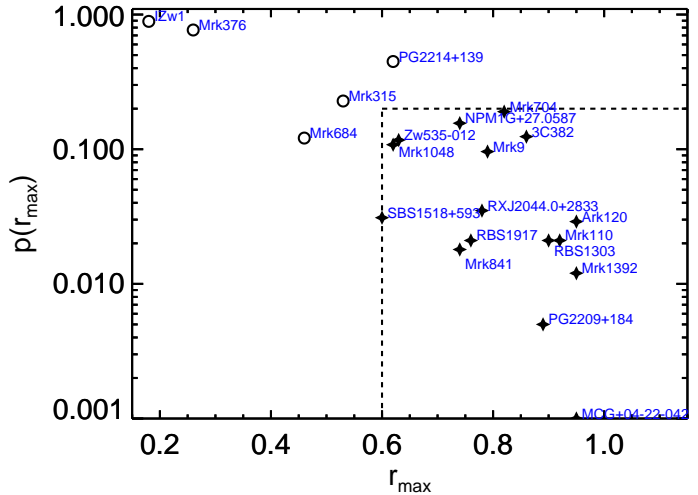}
\caption{$p(r_\mathrm{max})$ versus $r_{\rm max}$ as two objective lag-assessment parameters for our sample. Sources with  $p(r_\mathrm{max})$ above 0.2 or $r_{\rm max}$ less than 0.6 are deemed to have unreliable lags as represented by open circles.}
  \label{fig:lagsig}
\end{figure}

We assess the quality of our resulting lags based on these
quantitative assessment indicators in Figure~\ref{fig:lagsig}. As expected, there is an anticorrelation between $r_\mathrm{max}$ and $p(r_\mathrm{max})$, but it is not a tight one-to-one relationship owing to the differences in S/N, sampling cadence, and variability amplitude among our sample. 
The results are consistent with the general
expectation that the highest-quality light curves that exhibit clear correlated variability have high $r_{\rm max}$ 
and small $p(r_\mathrm{max})$ values. 
Considering these lag-assessment results, we conclude that 16 of the 21 AGN in our sample falling in the region $p(r_\mathrm{max}) < 0.2$ and $r_\mathrm{max} > 0.6$ have correlations between continuum and \hb\ light curves that are sufficiently robust for the ensuing analyses (noting that they display a broad range of $p$-values and thus range from highly robust to relatively weak detections of correlated variability), while we discard the remaining five objects from further analysis based on their low $r_\mathrm{max}$ and high $p$-values.
Based partly on visual inspection of the light-curve quality, our chosen $r_\mathrm{max}$ threshold is relatively conservative compare to those selected by other large surveys~\cite[e.g., minimum $r_\mathrm{max}$ = 0.45 by SDSS-RM;][]{Grier17}.
These $r_{\rm max}$ and $p(r_\mathrm{max})$ values, along with AGN luminosities and lag measurements $\tau_{\rm cen}$
in both observed and rest frames, are reported in Table \ref{tbl:lags}. 
[Only the $r_{\rm max}$ and $p(r_\mathrm{max})$ are reported for the subsample of sources with unreliable lag measurements.]

\begin{deluxetable*}{@{\extracolsep{4pt}}lcccccccch@{}}[htb]
\tablecaption{AGN Luminosity and \hb~Cross-Correlation Lag Results \label{tbl:lags}}
\tablecolumns{10}
\tablewidth{0pt}
\tablehead{
  \colhead{Object} &
  \colhead{$\lambda L_{\lambda}$(5100\,\AA)} &
  \multicolumn{2}{c}{Observed Frame} &
  \multicolumn{2}{c}{Rest Frame} &
  \colhead{$r_{\rm max}$} &
  \colhead{$p(r_\mathrm{max})$} &
  \colhead{BLR Kinematics} \\ 
  \cline{3-4} \cline{5-6}
\colhead{} &
\colhead{} &
\colhead{$\tau_{\rm cen}$} & 
\colhead{$\tau_{\rm peak}$} & 
\colhead{$\tau_{\rm cen}$} & 
\colhead{$\tau_{\rm peak}$} & 
\colhead{} &
\colhead{} &
\colhead{} &
\colhead{} \\
\colhead{} &
\colhead{(10$^{43}$ erg s$^{-1}$)} &
\colhead{(days)} &
\colhead{(days)} &
\colhead{(days)} &
\colhead{(days)} &
\colhead{} &
\colhead{} &
\colhead{} 
}
\startdata
        Zw 535$-$012 & 4.9 $\pm$  0.7 & 21.3$_{-  4.8}^{+  8.5}$ &   22.0$_{-  8.0}^{+  6.5}$ &   20.3$_{-  4.6}^{+  8.1}$ &   21.0$_{-  7.6}^{+  6.2}$ &  0.63 & 0.12 & Infalling \\ 
            Mrk 1048 & 9.5 $\pm$  1.8 &  7.8$_{-  9.8}^{+ 10.1}$ &   10.0$_{- 13.0}^{+  9.0}$ &    7.4$_{-  9.4}^{+  9.7}$ &    9.6$_{- 12.5}^{+  8.6}$ &  0.62 & 0.11 & Infalling \\ 
             Ark 120 & 9.2 $\pm$  3.4 &  19.3$_{-  4.6}^{+  6.1}$ &   20.5$_{-  4.5}^{+  5.5}$ &   18.7$_{-  4.5}^{+  5.9}$ &   19.9$_{-  4.4}^{+  5.3}$ &  0.95 & 0.03 & Ambiguous \\ 
               Mrk 9 & 5.6 $\pm$  0.4 & 20.3$_{-  7.6}^{+  4.2}$ &   15.0$_{-  7.0}^{+ 12.0}$ &   19.5$_{-  7.3}^{+  4.1}$ &   14.4$_{-  6.7}^{+ 11.5}$ &  0.79 & 0.10 & Ambiguous \\ 
             Mrk 704 & 6.2 $\pm$  0.6 &   29.8$_{- 10.3}^{+ 10.4}$ &   34.5$_{- 19.5}^{+  5.5}$ &   28.9$_{- 10.0}^{+ 10.2}$ &   33.5$_{- 18.9}^{+  5.3}$ &  0.82 & 0.19 & Infalling \\ 
MCG $+$04$-$22$-$042 & 1.6 $\pm$  0.4 &  13.7$_{-  1.9}^{+  2.5}$ &   10.5$_{-  1.0}^{+  3.0}$ &   13.3$_{-  1.8}^{+  2.4}$ &   10.2$_{-  1.0}^{+  2.9}$ &  0.95 & 0.00 & Symmetric \\ 
             Mrk 110 & 7.2 $\pm$  1.7 &  28.8$_{-  5.2}^{+  4.4}$ &   22.5$_{-  4.0}^{+  7.5}$ &   27.8$_{-  5.1}^{+  4.3}$ &   21.7$_{-  3.9}^{+  7.2}$ &  0.92 & 0.02 & Symmetric \\ 
            RBS 1303 & 2.4 $\pm$  0.4 &  19.4$_{-  4.5}^{+  3.6}$ &   13.5$_{-  4.5}^{+  7.5}$ &   18.7$_{-  4.3}^{+  3.4}$ &   13.0$_{-  4.3}^{+  7.2}$ &  0.90 & 0.02 & Outflowing \\ 
             Mrk 841 & 6.7 $\pm$  0.9 &  11.7$_{-  3.7}^{+  5.0}$ &    9.0$_{-  2.0}^{+  5.0}$ &   11.2$_{-  3.5}^{+  4.8}$ &    8.7$_{-  1.9}^{+  4.8}$ &  0.74 & 0.02 & Infalling \\ 
            Mrk 1392 &  1.6 $\pm$  0.6 &  27.6$_{-  4.0}^{+  3.6}$ &   17.5$_{-  2.0}^{+  3.0}$ &   26.7$_{-  3.9}^{+  3.5}$ &   16.9$_{-  1.9}^{+  2.9}$ &  0.95 & 0.01 & Symmetric \\ 
      SBS 1518$+$593 & 10.6 $\pm$  1.0 &  21.7$_{-  9.5}^{+  9.0}$ &   22.0$_{- 11.0}^{+  9.5}$ &   20.1$_{-  8.9}^{+  8.4}$ &   20.4$_{- 10.2}^{+  8.8}$ &  0.60 & 0.03 & Infalling \\ 
              3C 382 & 15.0 $\pm$  3.1 &  10.0$_{-  7.1}^{+  7.2}$ &    6.5$_{-  7.0}^{+ 13.5}$ &    9.5$_{-  6.7}^{+  6.8}$ &    6.1$_{-  6.6}^{+ 12.8}$ &  0.86 & 0.12 & Symmetric \\ 
    NPM1G $+$27.0587 & 10.9 $\pm$  1.0 &   8.5$_{-  4.7}^{+  5.0}$ &    7.5$_{-  4.0}^{+  6.5}$ &    8.0$_{-  4.5}^{+  4.7}$ &    7.1$_{-  3.8}^{+  6.1}$ &  0.74 & 0.16 & Infalling \\ 
   RXJ 2044.0$+$2833 &  5.3 $\pm$  0.5 &  15.1$_{-  2.0}^{+  1.7}$ &   15.0$_{-  3.5}^{+  4.0}$ &   14.4$_{-  1.9}^{+  1.6}$ &   14.3$_{-  3.3}^{+  3.8}$ &  0.78 & 0.04 & Infalling \\ 
       PG 2209$+$184 & 2.3 $\pm$  0.6 &   14.6$_{-  3.1}^{+  3.0}$ &   16.0$_{-  5.0}^{+  3.0}$ &   13.7$_{-  2.9}^{+  2.8}$ &   15.0$_{-  4.7}^{+  2.8}$ &  0.89 & 0.01 & Ambiguous \\ 
            RBS 1917 & 4.9 $\pm$  0.5 &   12.7$_{-  4.1}^{+  4.6}$ &   12.5$_{-  3.5}^{+  4.5}$ &   11.9$_{-  3.9}^{+  4.3}$ &   11.7$_{-  3.3}^{+  4.2}$ &  0.76 & 0.02 & Outflowing \\ 
\hline
             I Zw 1 & 28.5 $\pm$  4.0 & \nodata & \nodata & \nodata & \nodata & 0.18 & 0.89 & \nodata \\ 
             Mrk 376 & 17.6 $\pm$  1.8 & \nodata & \nodata & \nodata & \nodata & 0.26 & 0.77 & \nodata \\ 
             Mrk 684 & 8.3 $\pm$  0.9 & \nodata & \nodata & \nodata & \nodata & 0.46 & 0.12 & \nodata \\
             PG 2214$+$139 & 21.2 $\pm$  1.3 & \nodata & \nodata & \nodata & \nodata &  0.62 & 0.45 & \nodata \\ 
             Mrk 315 & 3.2 $\pm$  0.4 & \nodata & \nodata & \nodata & \nodata & 0.53 & 0.23 & \nodata \\
\enddata
\end{deluxetable*}

\subsection{Velocity-Resolved \hb~Lags}
To obtain velocity-resolved reverberation results as a means to
probe the kinematics of the BLR gas, we used CCF to compute lag measurements
for individual velocity segments of the
emission line. Our procedure follows that described
by \citet{Bentz09c}, \citet{Denney09}, and \citet{Grier13a}, and is detailed below.  

To verify the effect of binning on the apparent BLR kinematics, we
divided the broad \hb~component into eight bins via two different
schemes: (i) bins of equal rms flux, and (ii) bins of equal velocity
width. In the first approach, we initially determined the total rms flux
from integrating the \hb~line in the e-rms spectrum. Then we
established velocity bins such that the rms flux within each bin
would equal one-eighth of the total flux. There are some exceptions in the
cases where the red wing of the rms profile suffered from significant
noise and the last bin was thus truncated at the red line wing.
In the second approach, bins are divided evenly across the width
of the line in velocity space. In either binning scheme, light curves were
computed from each bin of the continuum-subtracted \hb~profile and
lags were measured against the $V$-band continuum light curve. Resulting lags from the two binning schemes generally agree
in the line's core where the S/N is high; minor
disagreement is found in the noisy line wings of some sources.
For clarity, we adopt the first scheme (equal-rms-flux binning) and show the
resulting \edit1{velocity-resolved lag spectra} for each AGN with robust lag in Figure 
\ref{fig:velresolve} and in Table \ref{tbl:velresolve}.

\begin{figure*}[htbp]
   \centering
   \includegraphics[width=.32\textwidth]{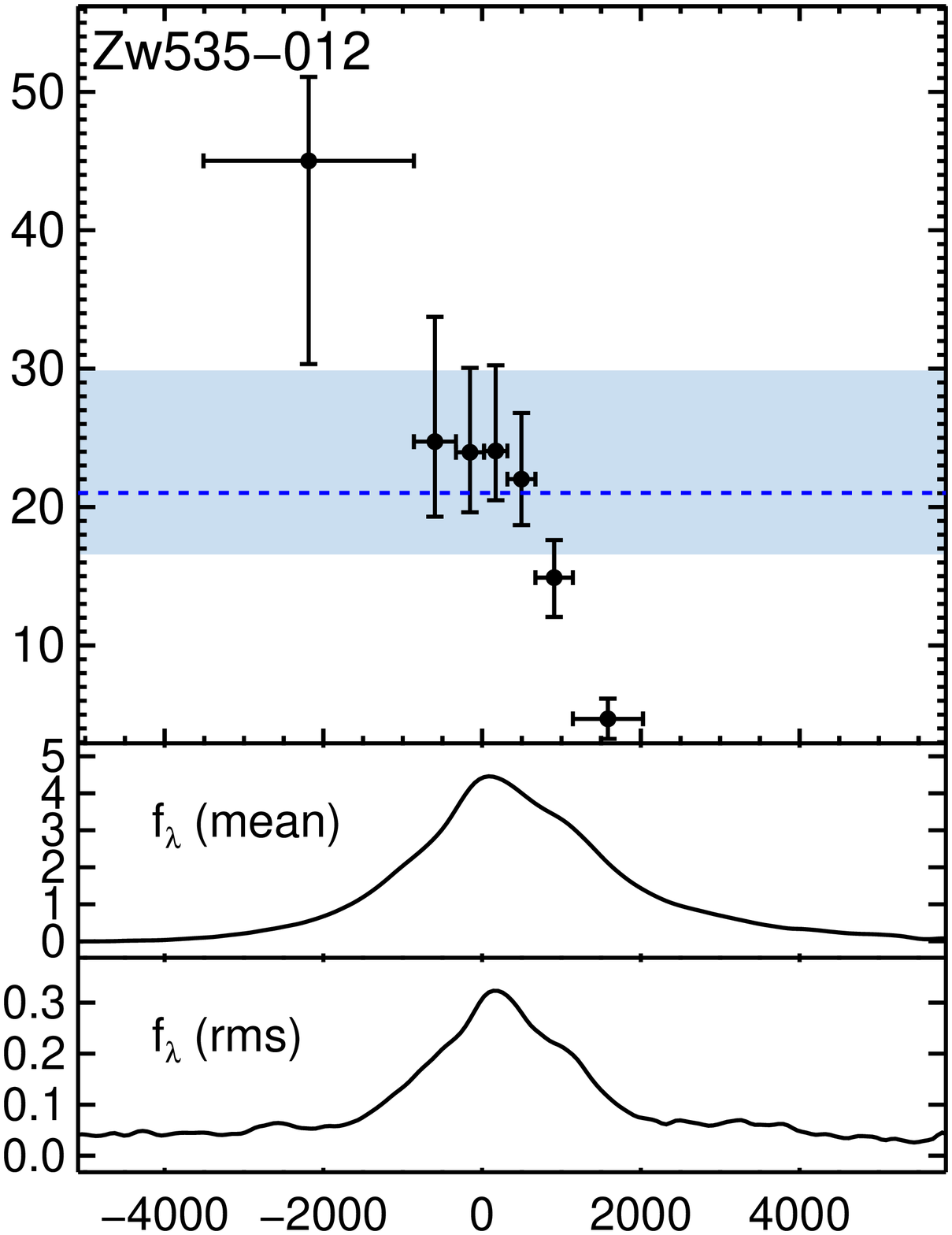}
   \includegraphics[width=.32\textwidth]{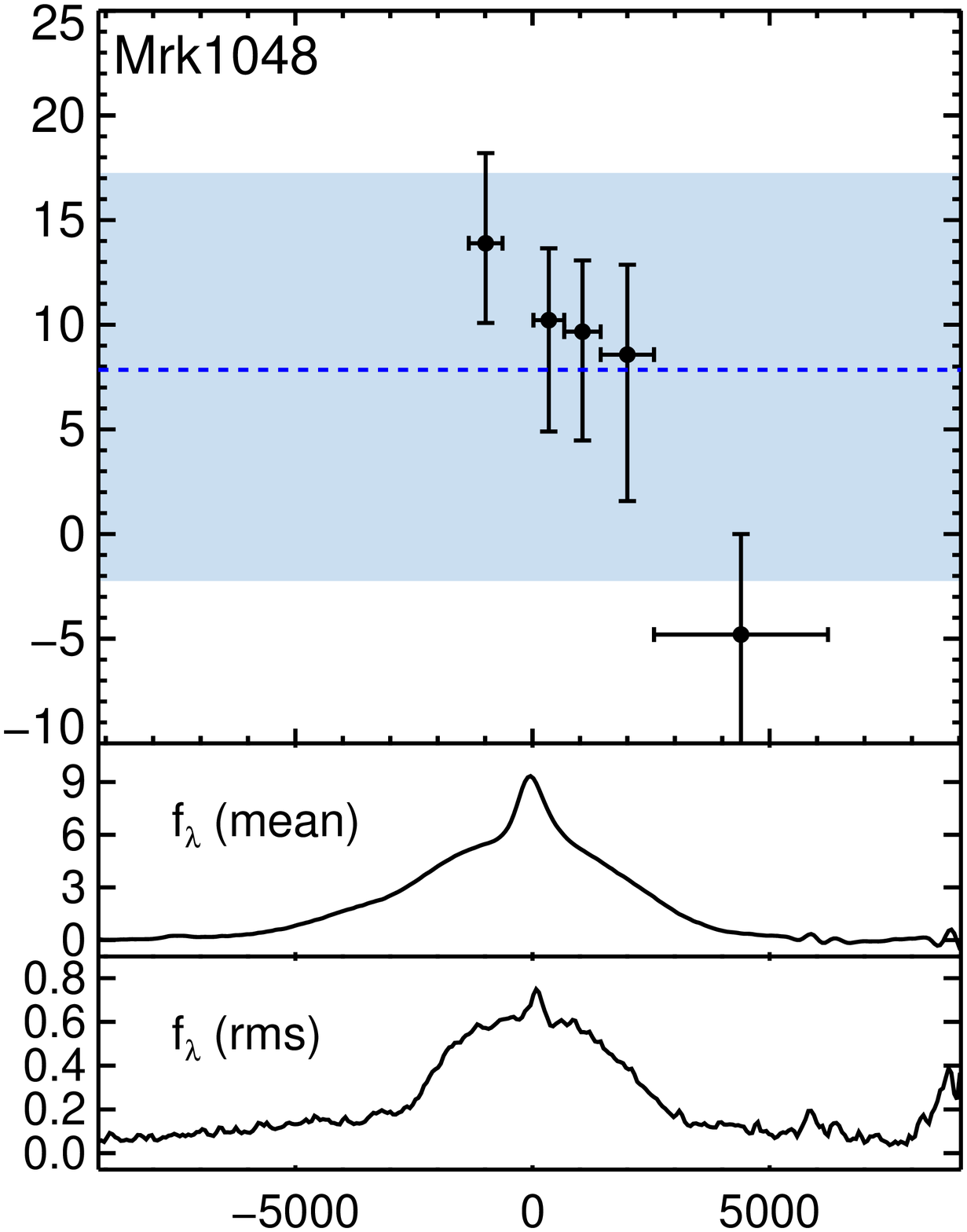}
   \includegraphics[width=.32\textwidth]{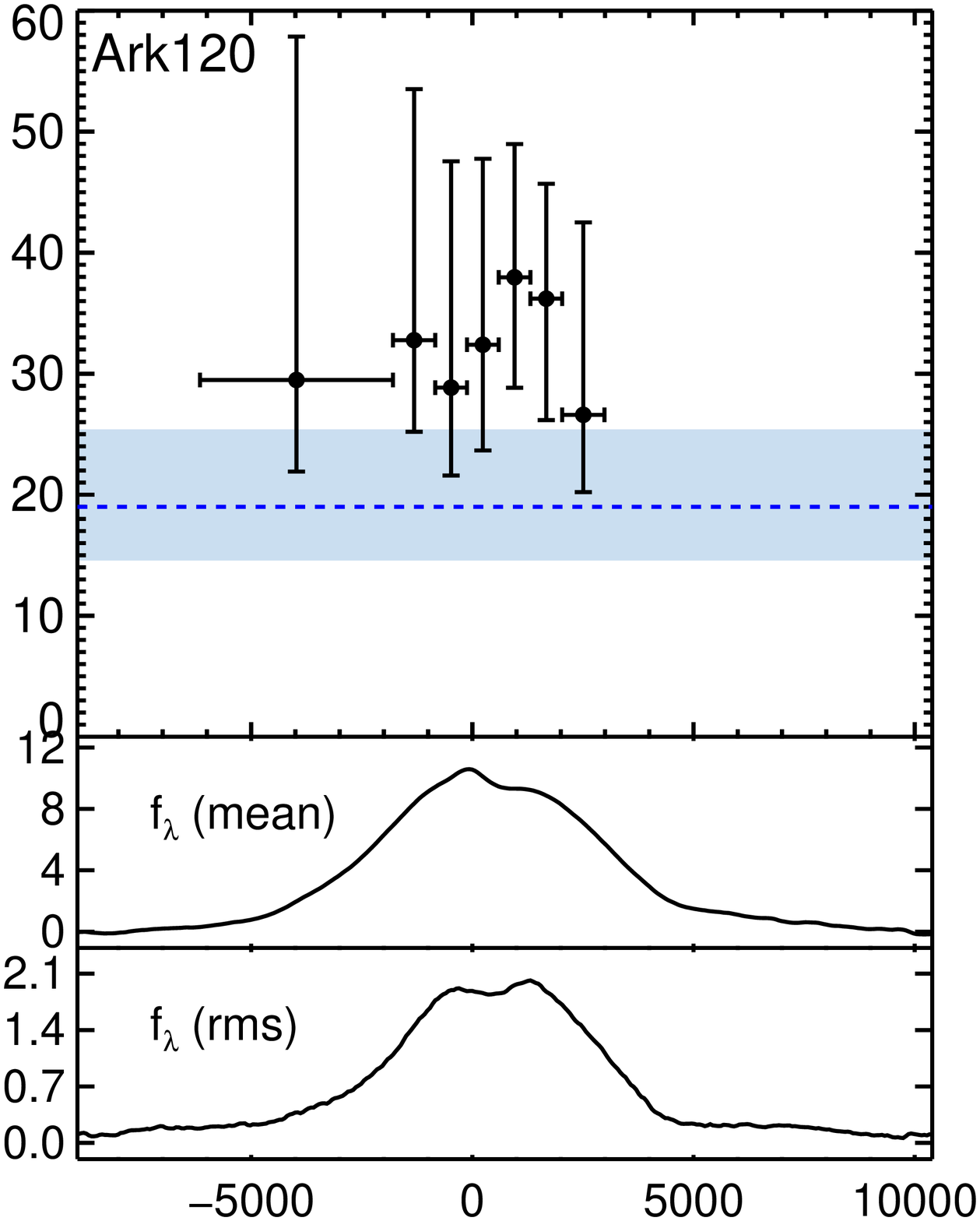}
   \includegraphics[width=.32\textwidth]{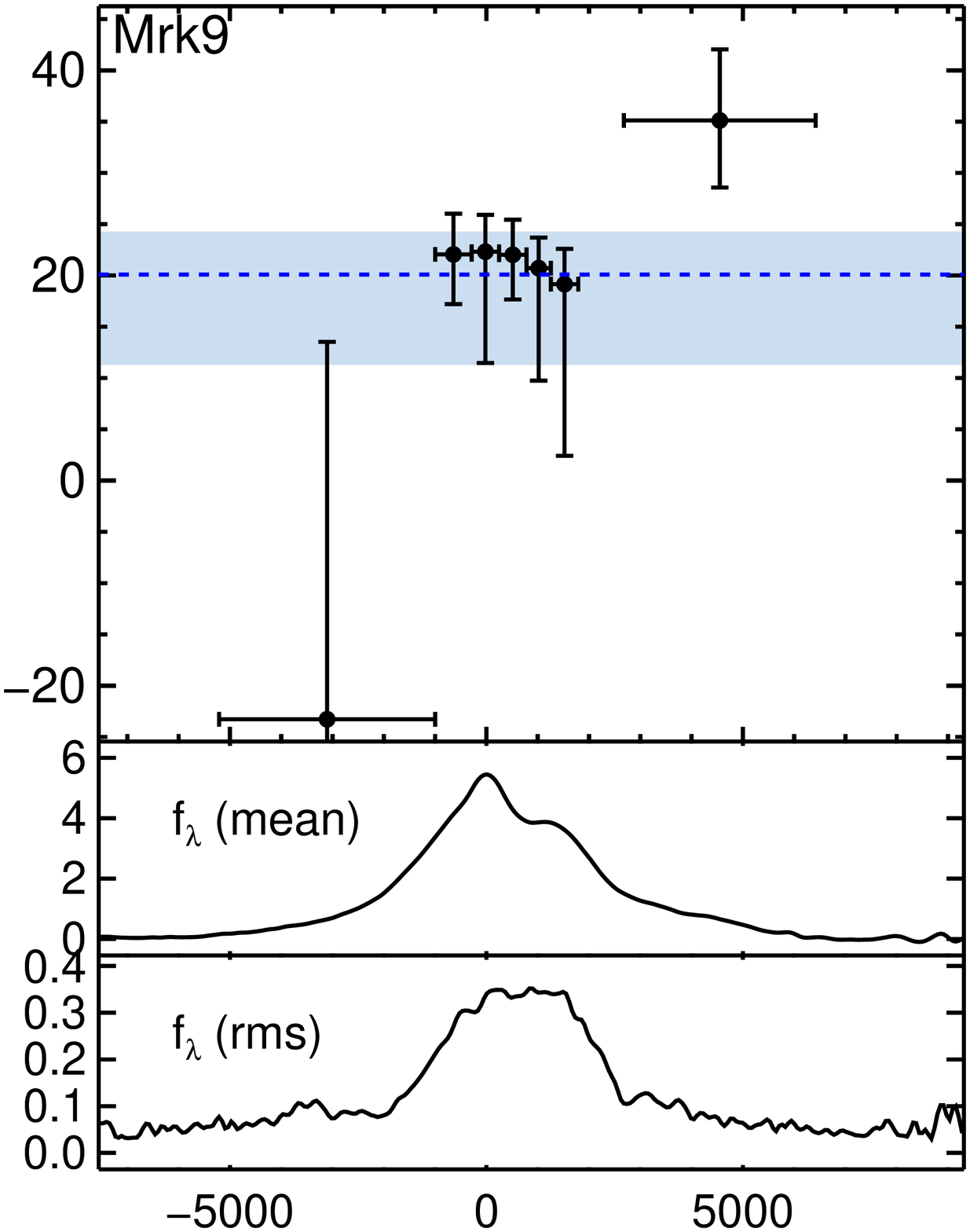}
   \includegraphics[width=.32\textwidth]{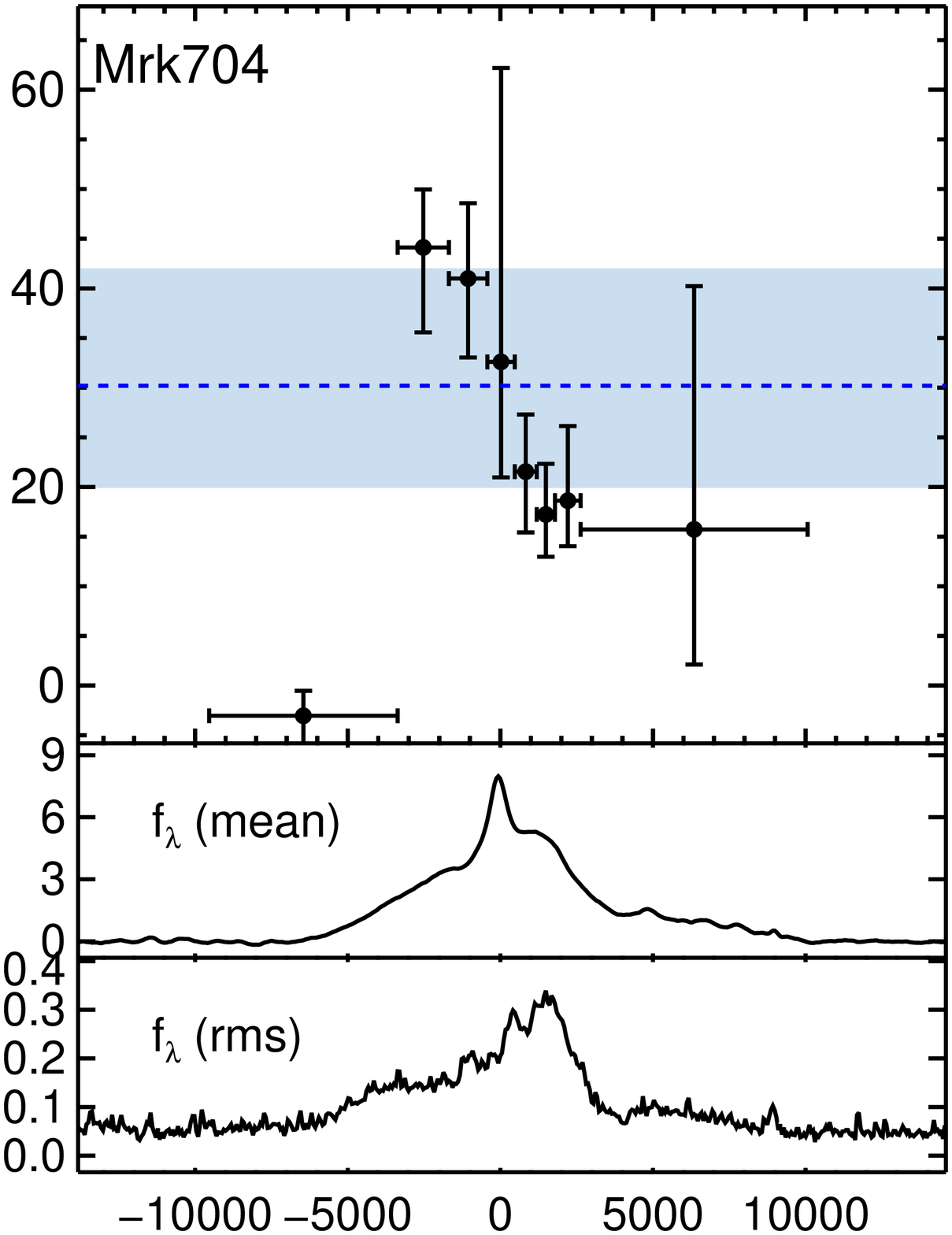}
   \includegraphics[width=.32\textwidth]{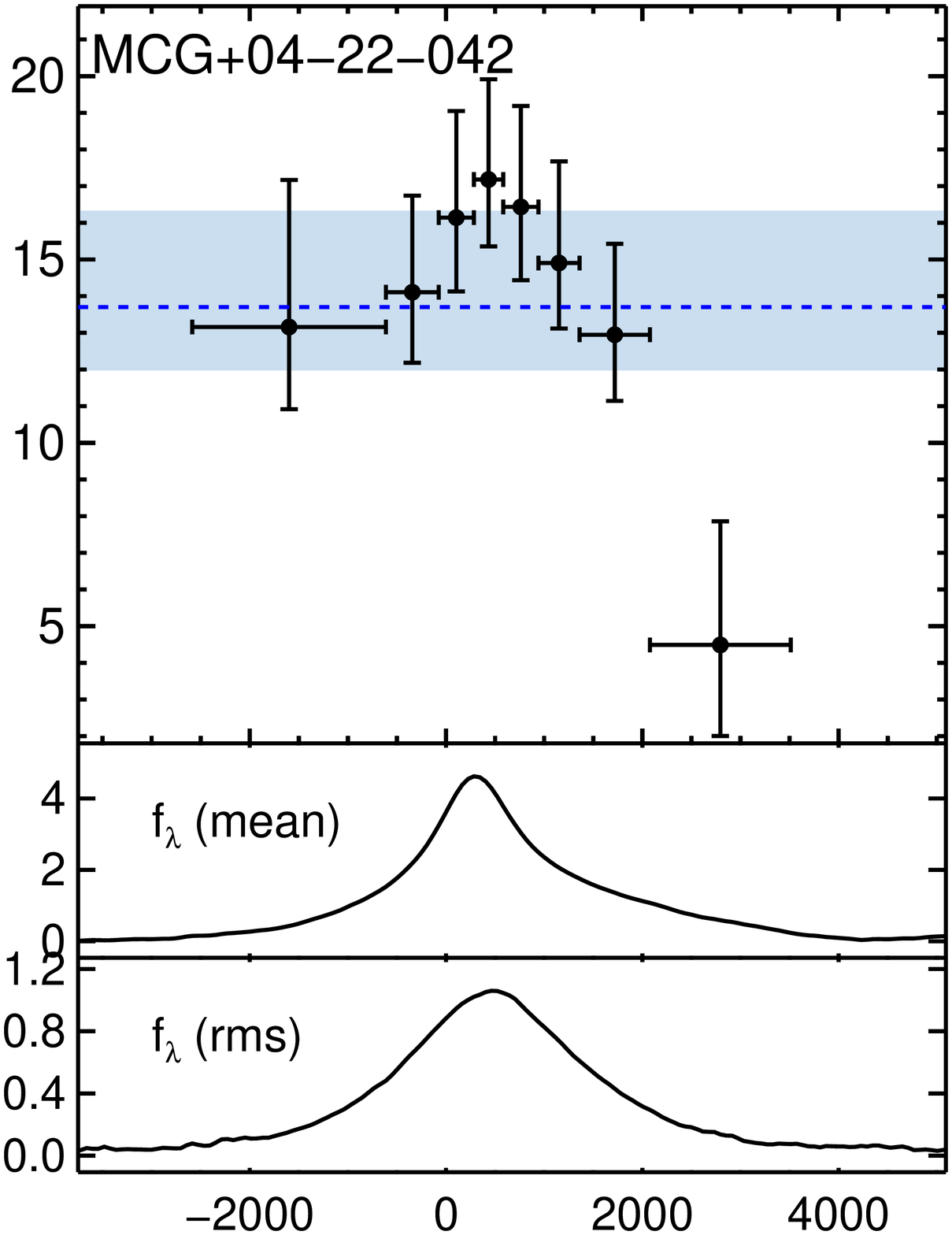}
   \includegraphics[width=.32\textwidth]{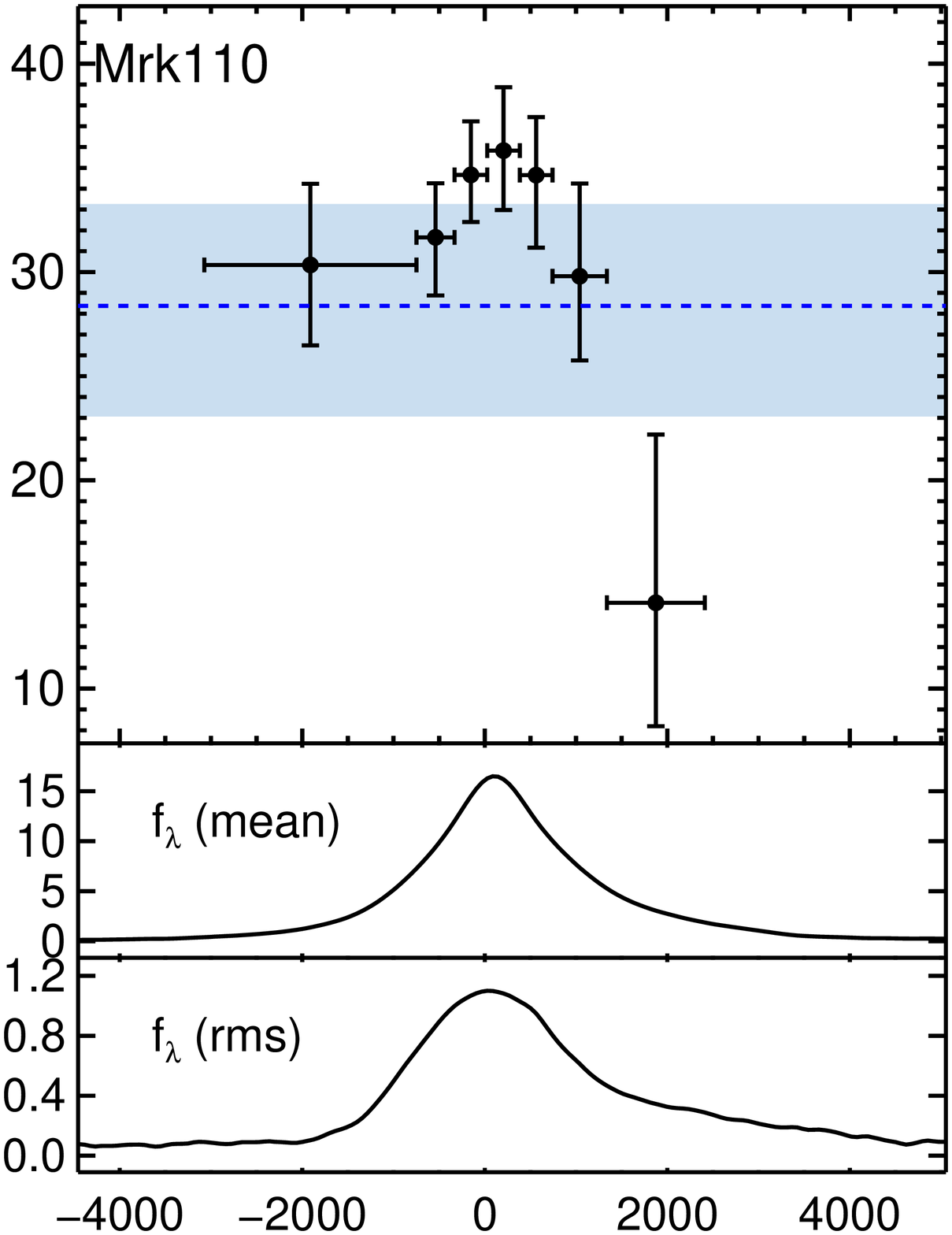}
   \includegraphics[width=.32\textwidth]{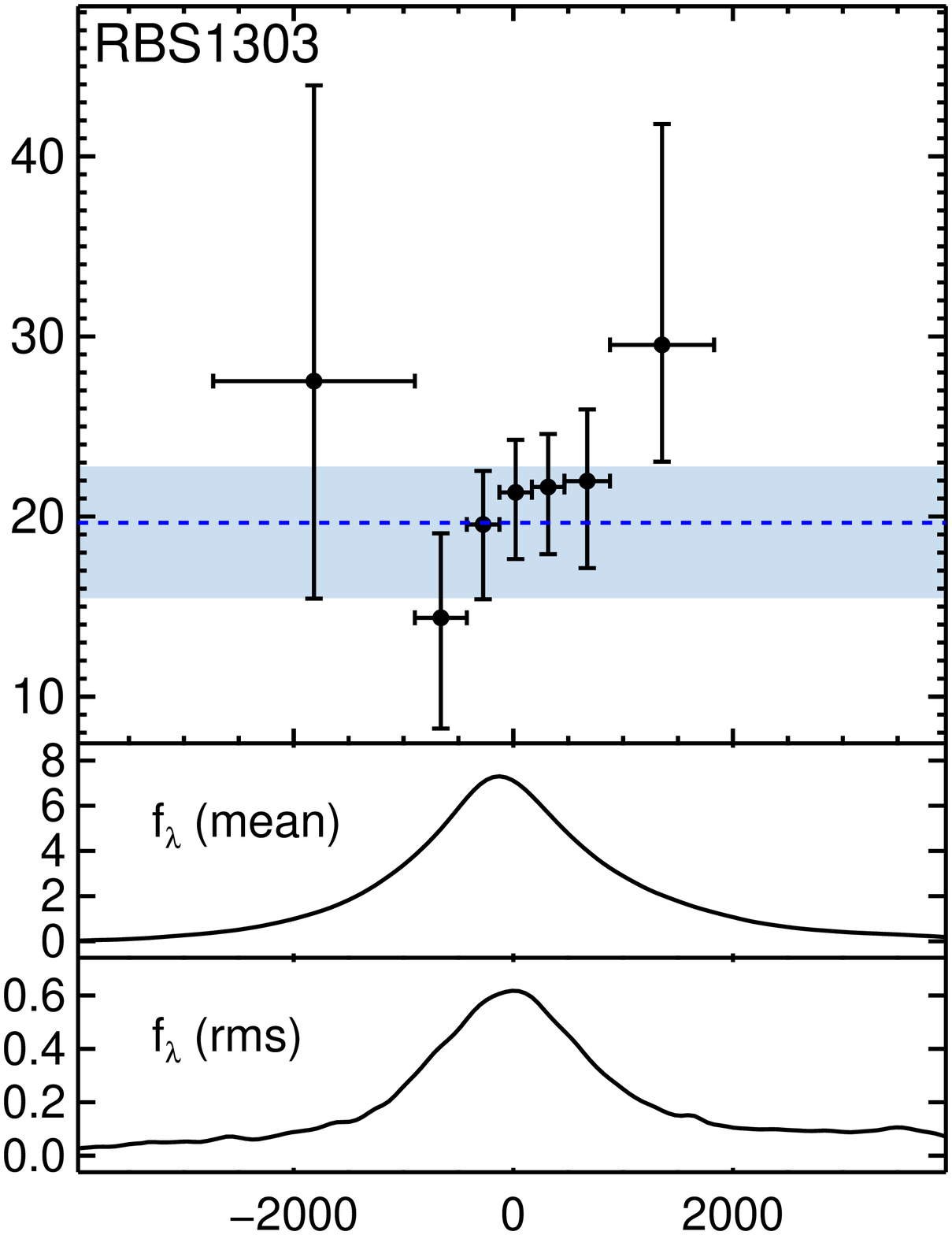}
   \includegraphics[width=.32\textwidth]{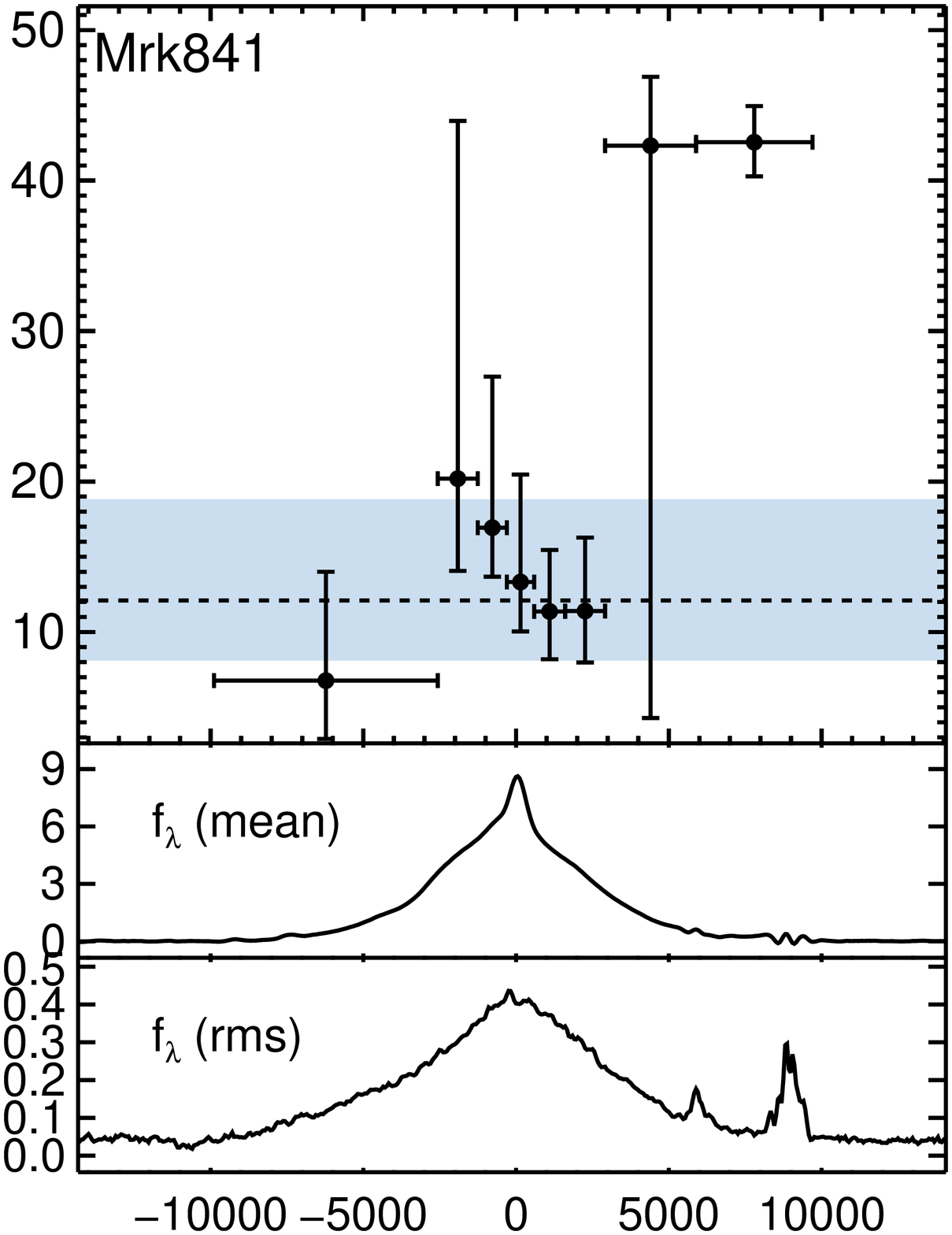}
\caption{Velocity-resolved reverberation lags for our sample. In each
  of the top panels, the \hb~lags in days (y-axis) shown are for bins of equal rms
  flux, plotted against relative velocity in km s$^{-1}$ (x-axis). The blue dotted line indicates the overall integrated \hb~lag
with $\pm1\sigma$ uncertainty spanned by the blue bar. The middle
and bottom panels show the \hb~profile of the mean and rms spectra,
respectively.}  
  \label{fig:velresolve}
\end{figure*}

\begin{figure*}[htbp]
    \ContinuedFloat
    \captionsetup{list=off,format=plain}
   \centering
   \includegraphics[width=.32\textwidth]{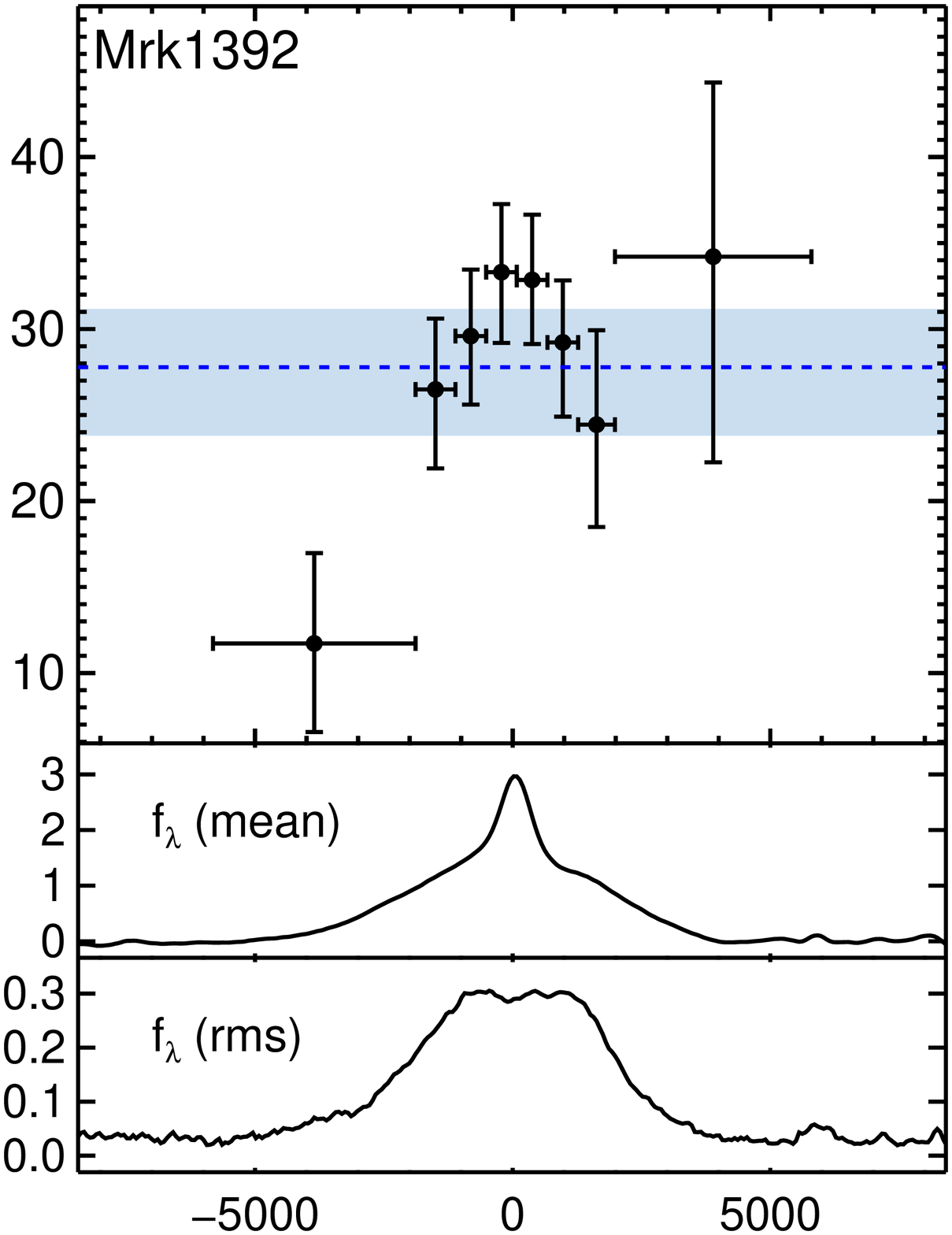}
   \includegraphics[width=.32\textwidth]{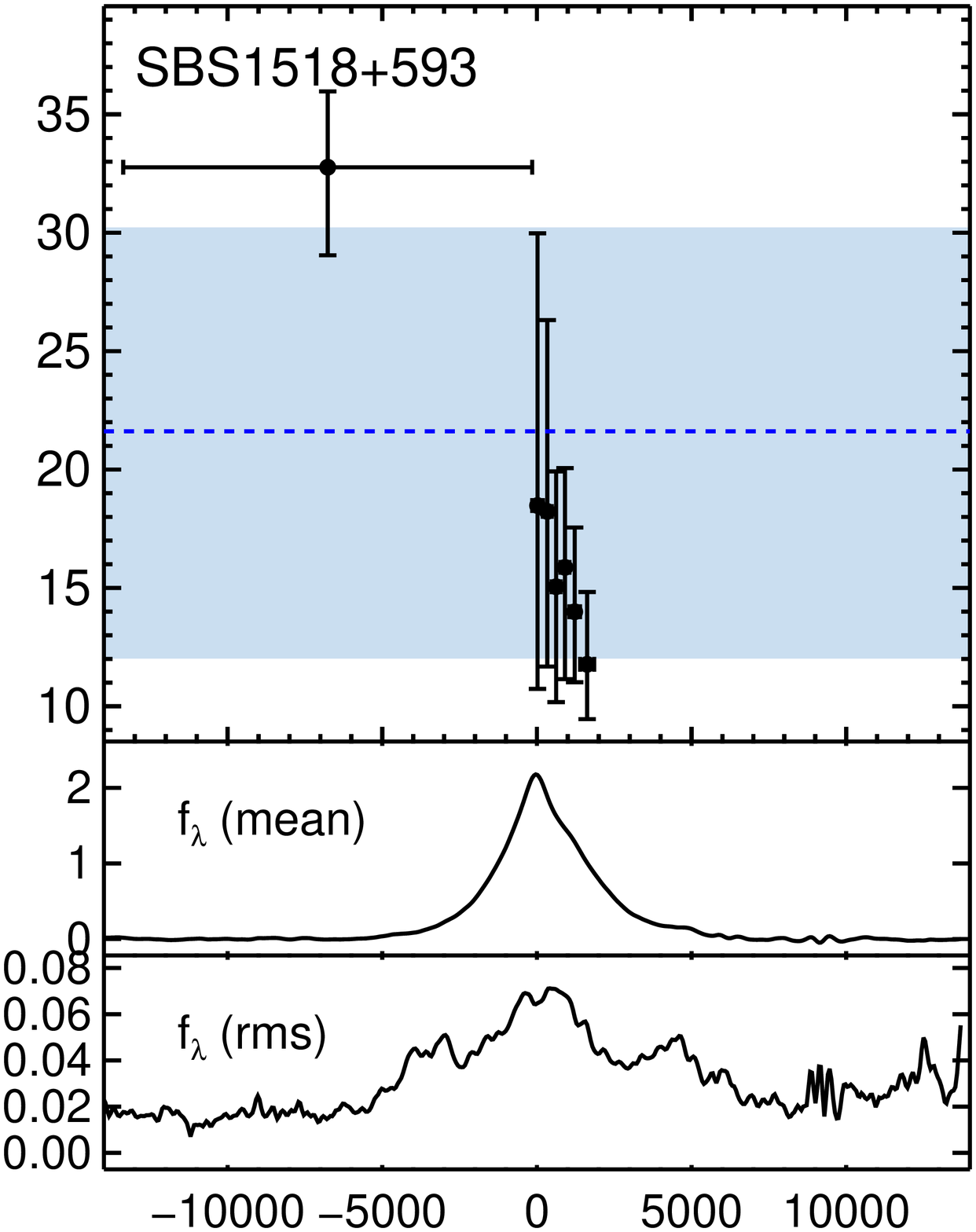}
   \includegraphics[width=.32\textwidth]{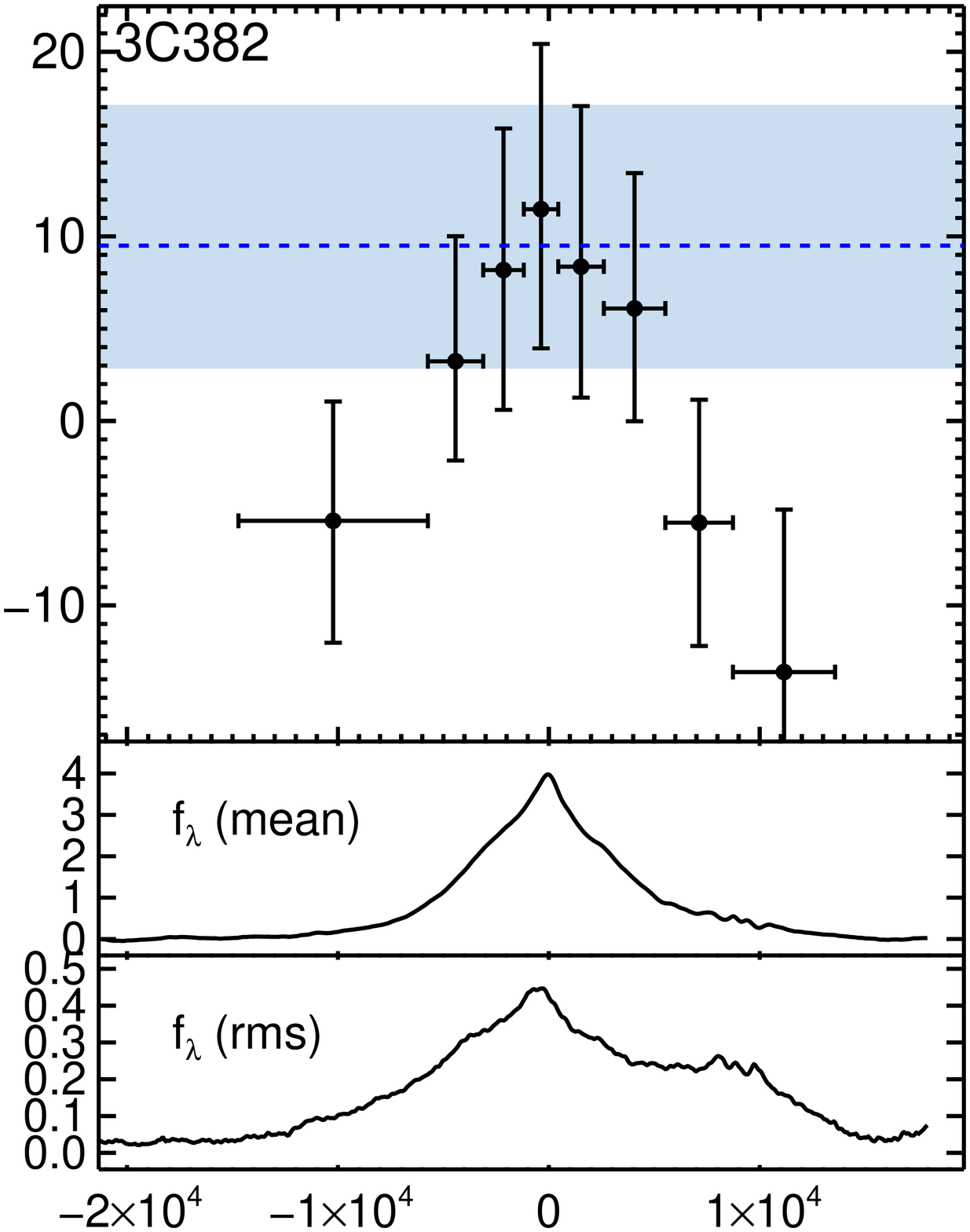}
   \includegraphics[width=.32\textwidth]{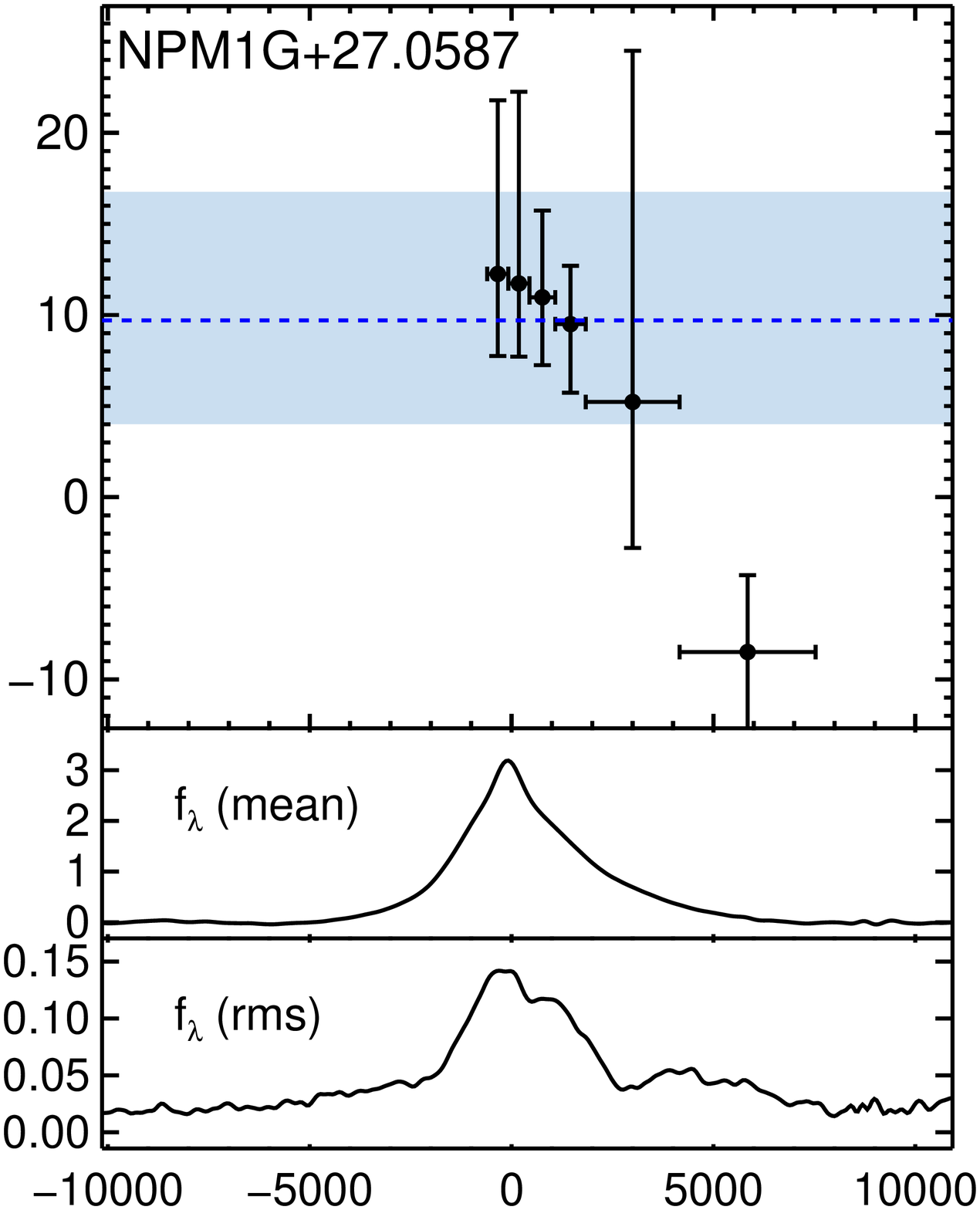}
   \includegraphics[width=.32\textwidth]{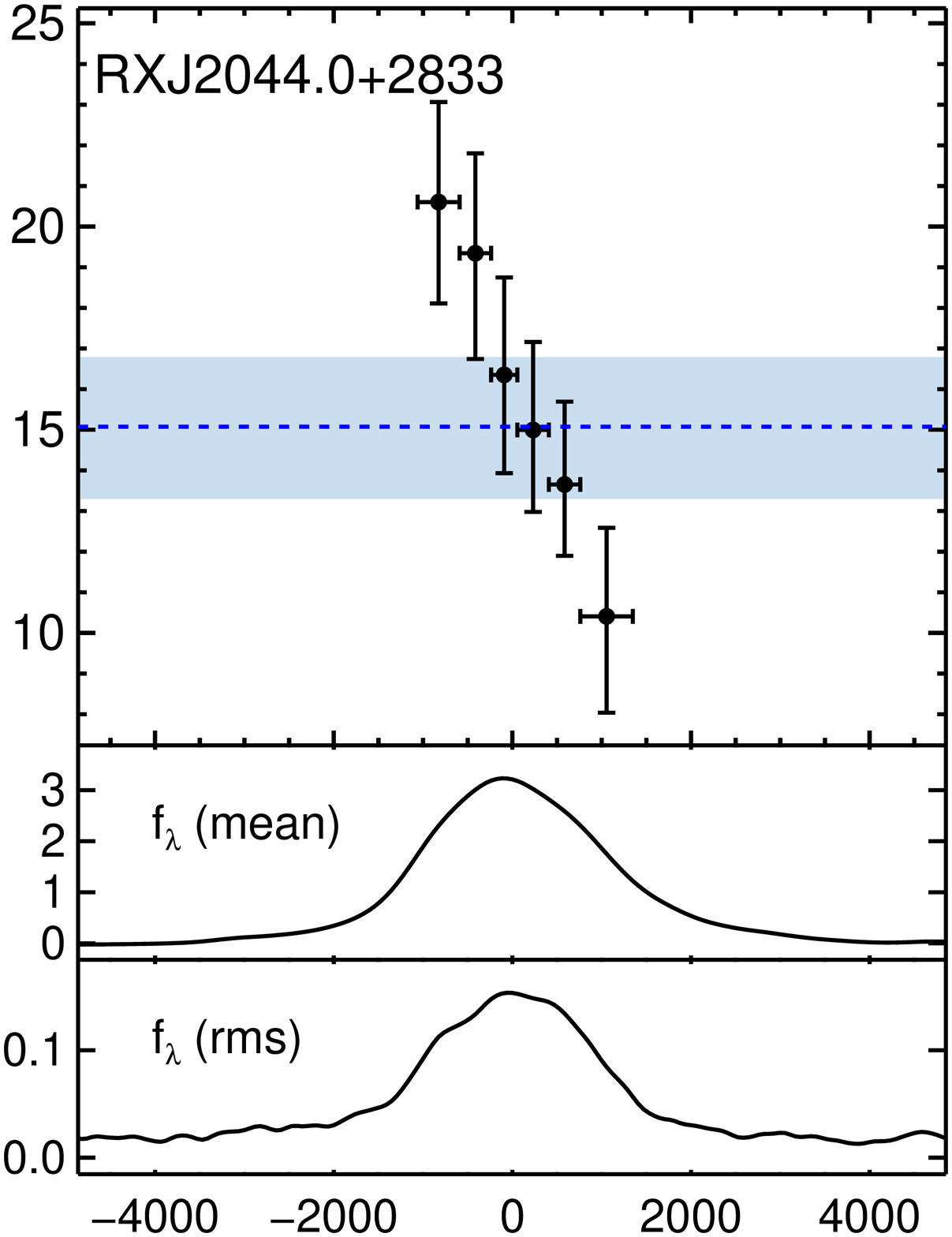}
   \includegraphics[width=.32\textwidth]{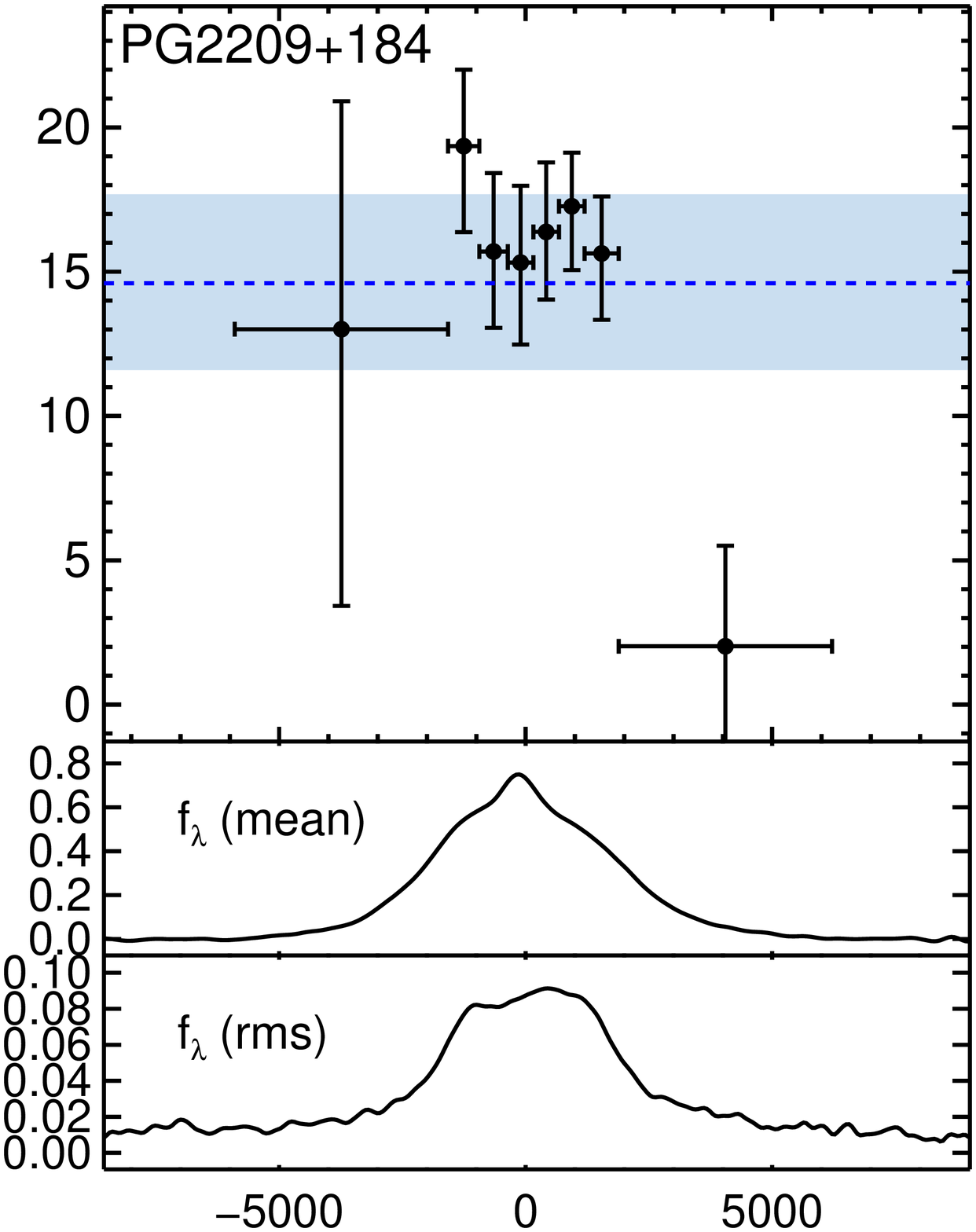}
   \includegraphics[width=.32\textwidth]{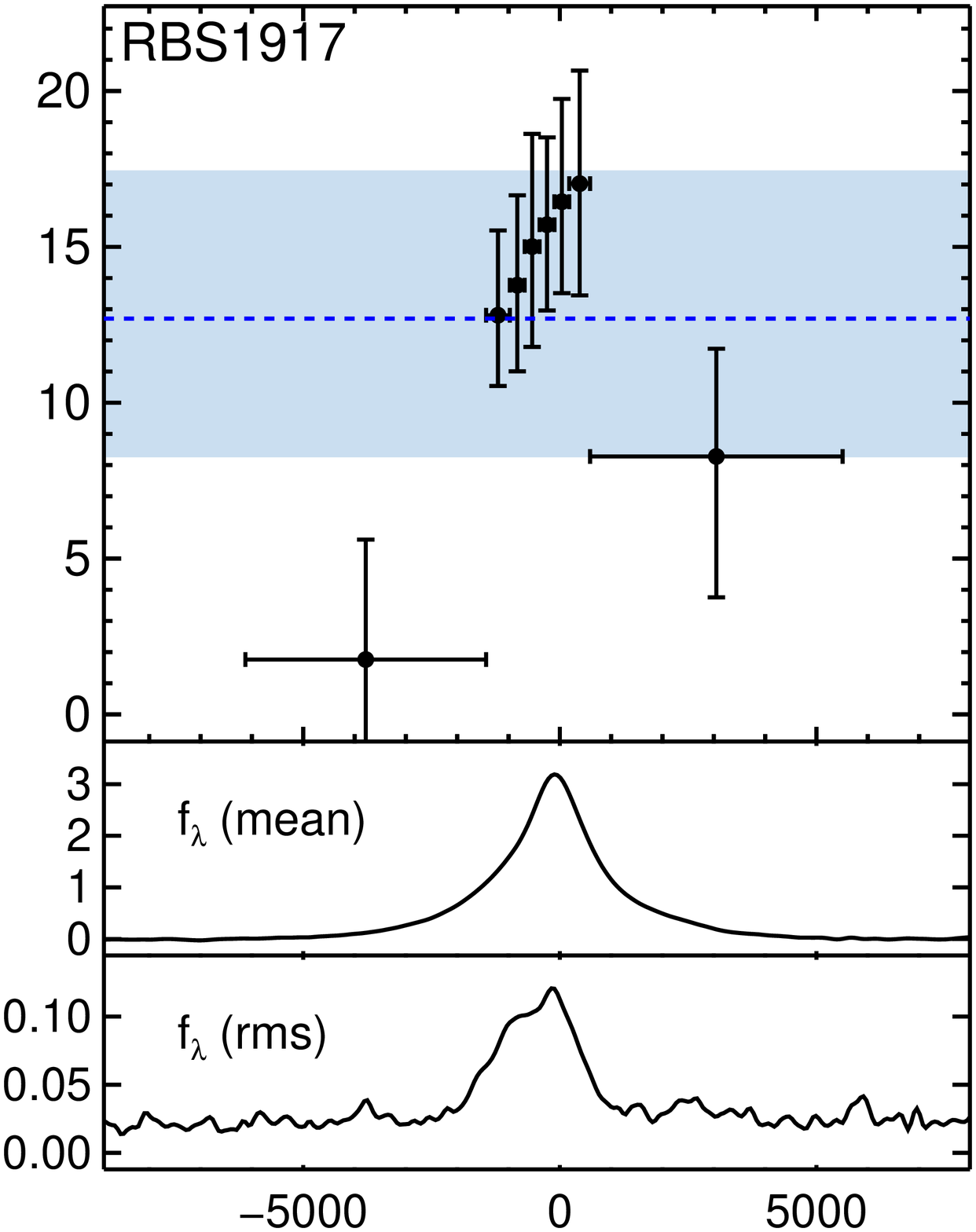}
    \caption{Continued}
\end{figure*}

\begin{deluxetable}{ccccc}[htb]
\tablecaption{Velocity-Resolved Lags\label{tbl:velresolve}}
\tablecolumns{3}
\tablewidth{0pt}
\tablehead{
\colhead{Object} &
\colhead{Velocity Bin} &
\colhead{$\tau_{\rm cen}$} \\
\colhead{} & 
\colhead{(km s$^{-1}$)} &
\colhead{(days)}
}
\startdata
              Mrk110 &      -1910 $\pm$      2324 &  30.34$^{+  3.90}_{-  3.86}$ \\
              Mrk110 &       -539 $\pm$       417 &  31.66$^{+  2.60}_{-  2.79}$ \\
              Mrk110 &       -151 $\pm$       357 &  34.66$^{+  2.57}_{-  2.26}$ \\
              Mrk110 &        205 $\pm$       357 &  35.83$^{+  3.04}_{-  2.85}$ \\
              Mrk110 &        563 $\pm$       357 &  34.65$^{+  2.79}_{-  3.48}$ \\
              Mrk110 &       1040 $\pm$       596 &  29.80$^{+  4.45}_{-  4.04}$ \\
              Mrk110 &       1874 $\pm$      1073 &  14.12$^{+  8.08}_{-  5.93}$ \\
\enddata
\tablecomments{This table is
  published in its entirety in machine-readable format on the online
  version of the journal. Sample entries are shown here for guidance
  regarding the table's form and content.}
\end{deluxetable}

\subsection{Implications for BLR Kinematics}
The velocity-resolved lags allow us to make simple qualitative inferences about the BLR geometry and
kinematics from the transfer-function
models~\cite[e.g.,][]{Welsh91,Horne04,Bentz09c}. Qualitatively, 
symmetric velocity-resolved structure around zero velocity is
consistent with either Keplerian, disk-like rotation or random motion
without net radial inflow or outflow over an extended BLR. In such a case, the
high-velocity wings exhibit the shortest lags because the
highest rotation speeds correspond to material orbiting closest to the
central black hole. Radially infalling gas produces an asymmetric pattern displaying longer lags at the high-velocity blue wing, while the opposite is true in the case of outflow-dominated kinematics. We note that the presence of absorption, noise, or other geometric complexities may
muddle these interpretations, which could be elucidated with
direct modeling using forward-modeling codes such as CARAMEL in follow-up work (Villafa\~na et al., in prep.).

We find a diversity of \edit1{velocity-resolved lag spectra} among our sample \edit1{representing the qualitative signature of} symmetric
(3C 382, MCG +04$-$22$-$042, Mrk 110, Mrk 1392), infalling (Mrk 1048, Mrk 704, Zw 535$-$012, Mrk 841, RXJ 2044.0+2833, NPM1G$+$27.0587, SBS 1518+593), and outflowing (RBS 1303 and RBS 1917) \edit1{scenarios}. 
In the cases of Ark 120, Mrk 9, and PG 2209+184, 
the velocity-resolved structure appears flat across the \hb~profile within the uncertainties, which may be difficult to describe with simple models.
Whether these kinematics exhibit luminosity-dependent
trends and how a subset of these results relate to other RM samples will be further
examined in \S \ref{sec:bh} and \edit1{the} Appendix, 
respectively.  

\section{Black Hole Scaling Relations}
\label{sec:bh}

\subsection{AGN Radius--Luminosity Relationship}
\begin{figure}[htbp]
   \centering
   \includegraphics[width=0.41\textwidth,angle=90]{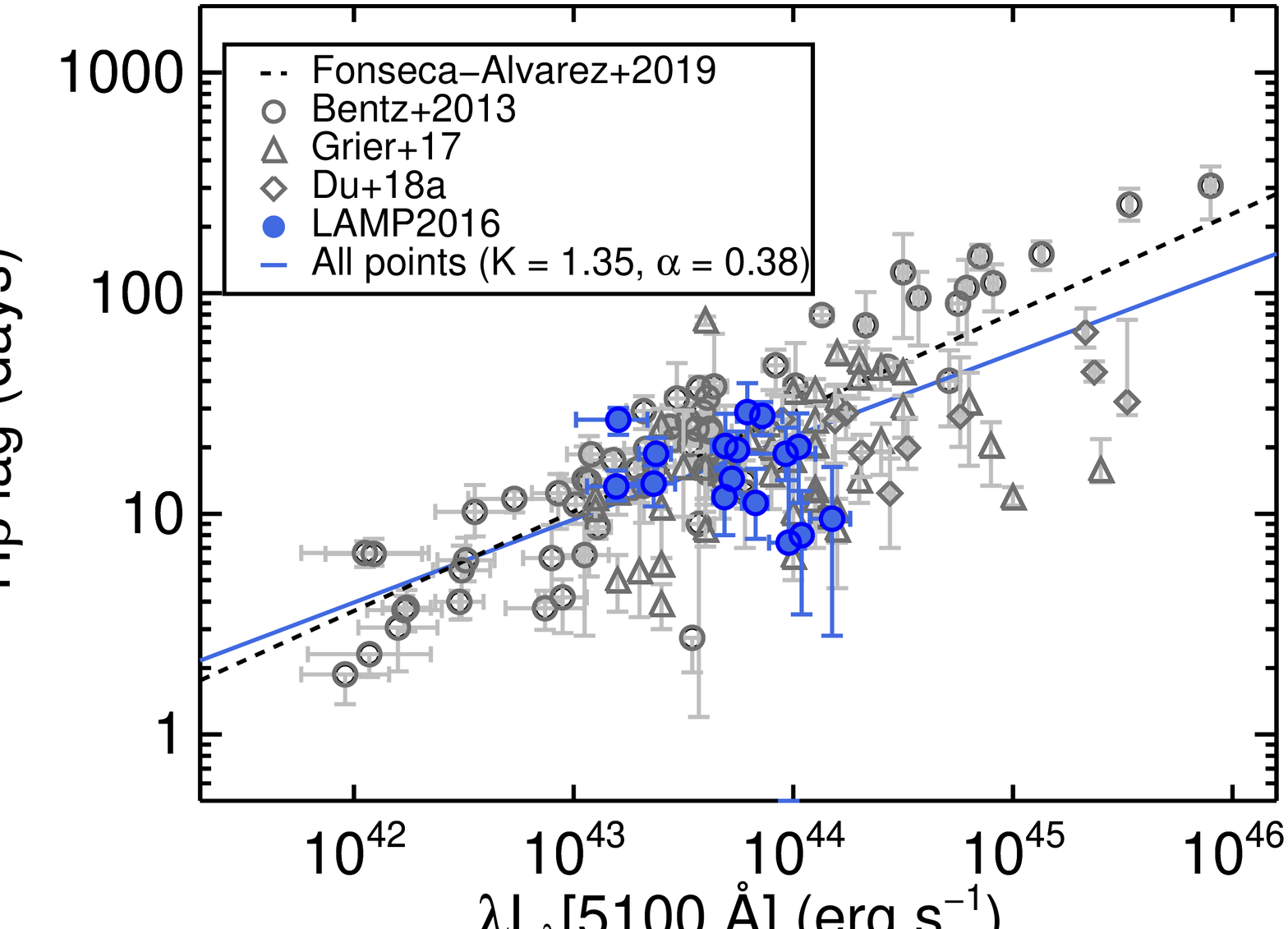}
  \caption{Radius--luminosity relationship of our sample (blue, filled)
  superimposed on top of other work from the literature~\cite[open
  symbols;][]{Bentz13,Grier17,Du18a}. The best-fit relation from SDSS with a median slope $\alpha$ of 0.45
  and a median normalization $K$ of 1.46~\cite[][]{FonsecaAlvarez19} is
  plotted as the dashed line.  The best-fit line in the form of Equation \ref{eqn:radlum} for all the data points combining our current work with those from the literature (blue) has a normalization $K$ = 1.35 and $\alpha$ = 0.38.
} 
  \label{fig:radlum}
\end{figure}

\edit1{The applicability of the radius--luminosity relationship to single-epoch mass determination~\citep{Shen08,Shen12} relies on the small scatter initially quantified~\cite[$\sim$ 0.2 dex;][]{Kaspi05,Bentz09b,Bentz13,KilerciEser15,Martinez-Aldama20}, but recent studies of increasingly diverse AGN samples led to an increase in this scatter~\cite[e.g.,][]{Grier17,Du18a}. The main reason for this increased scatter seems to depend on the Eddington ratio~\citep{Bian12,Du15,Du16a,Martinez-Aldama19,DallaBonta20}, an important third parameter suggesting that there is more diversity in the AGN population than can be explained by the simple two-parameter radius--luminosity relationship. }
Here we compare our sample on the AGN radius--luminosity relationship with
other RM samples from the literature~\cite[][]{Bentz13,Grier17,Du18a} in Figure
\ref{fig:radlum}. The best-fit line from SDSS-RM for this relation is
\begin{equation}
\log(R_{\rm BLR} / {\rm lt day}) = K + \alpha \log(\lambda L_{\lambda}(5100\,\AA)/10^{44}~{\rm erg~s}^{-1}),
\label{eqn:radlum}
\end{equation}
with a median slope $\alpha$ = 0.45 and a median normalization
$K$ = 1.46~\cite[][]{FonsecaAlvarez19}. We computed the AGN luminosity $\lambda L_\lambda$ 
(5100 \AA) based on the AGN continuum flux at 5100\,\AA~as extracted from the decomposition of the mean spectrum. Luminosity
distances were derived using the cosmology calculator ($H_0 = 67.8$\,km\,s$^{-1}$\,Mpc$^{-1}$) provided
by~\cite{Wright06}. All of our AGN fall within 0.5 dex of the
radius--luminosity relation as defined by the literature points within the range of
 $\lambda L_{\lambda}$(5100\,\AA) = $10^{43.5-44.4}$ erg s$^{-1}$. 
 The best-fit line in the form of Equation \ref{eqn:radlum} for all the data points combining our current work with those from the literature has a normalization $K$ = 1.35 and $\alpha$ = 0.38.

\begin{figure}[htbp]
   \centering
   \includegraphics[width=0.41\textwidth,angle=90]{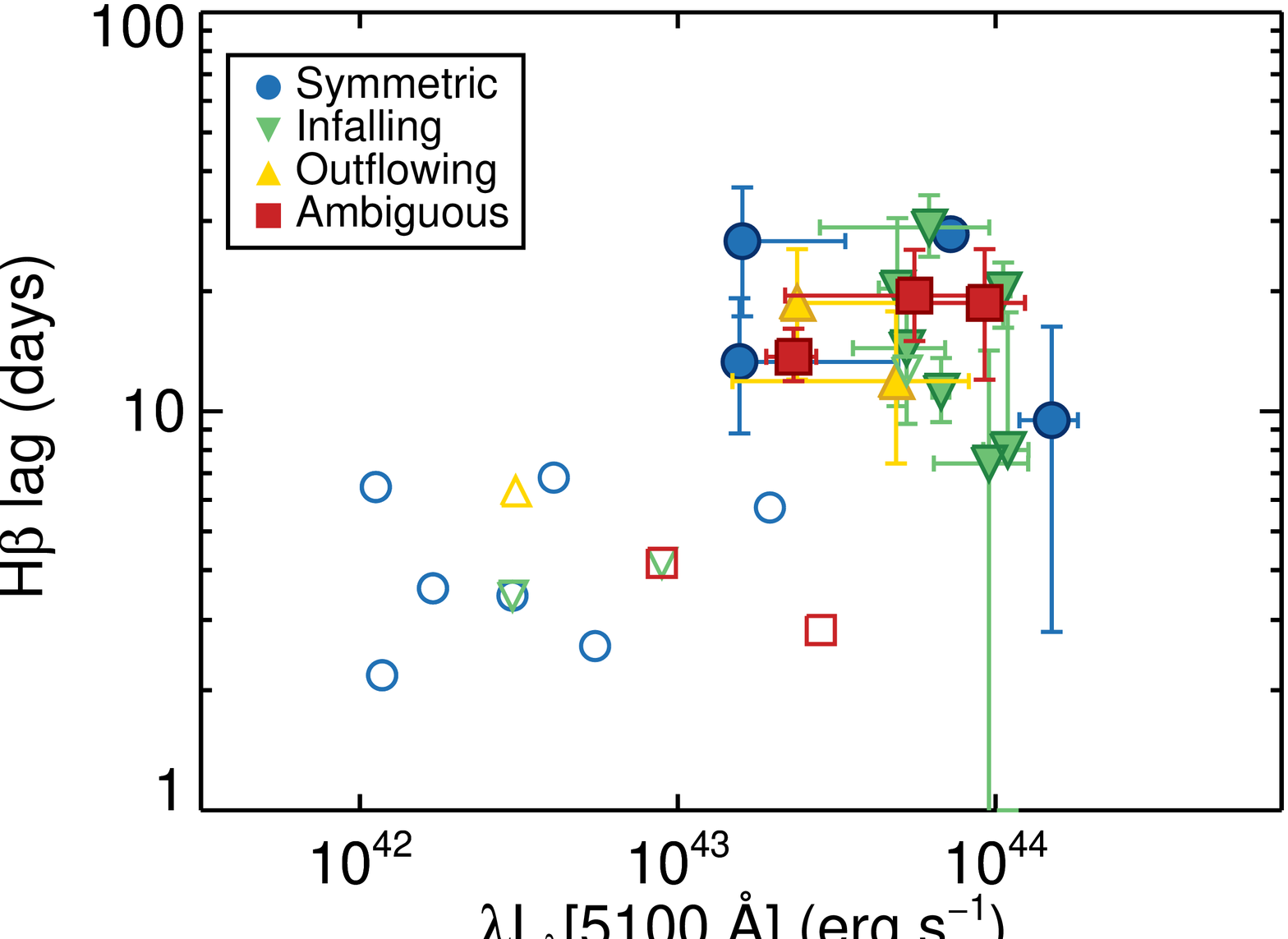}
\caption{Radius--luminosity relationship of our sample (filled and
  color-coded by BLR kinematics) along with other velocity-resolved
  results from \citet{Bentz10a}, \citet{Pancoast14b}, \citet{Grier13a}, \citet{DeRosa18}, and \citet{Du18a} (open symbols with the same
  color coding). No trend is seen for the different BLR kinematics
  within our LAMP2016 sample or when combined with the literature.}
  \label{fig:radlum_kin}
\end{figure}

We further explore whether BLR kinematics might depend on AGN luminosity.
Past work has shown that while inflow or outflow are indicated in a
subset of AGN, the majority of \hb~velocity-delay maps appear roughly
symmetric with rotationally dominated
kinematics~\cite[e.g.,][]{Bentz09c,Bentz10b,Denney10,Barth11b,Horne21}. The AGN
with velocity-resolved reverberation detected to-date are 
mostly within the low-luminosity range [$\lambda L_{\lambda}$ (5100\,\AA) =
$10^{42-43.5}$ erg s$^{-1}$]. Our results add significantly to the
existing number of sources having velocity-resolved
measurements with $\lambda L_{\lambda}$(5100\,\AA) $\approx 10^{44}$ erg
s$^{-1}$. \edit1{While we caution that velocity-resolved lag spectra provide only a qualitative impression of the BLR kinematics,}
we illustrate the BLR kinematics via different colored symbols
shown alongside those from the literature in Figure
\ref{fig:radlum_kin}.
No trend is seen between BLR kinematics and AGN luminosity either within our LAMP2016 sample or when combined with past results from the literature. It is plausible that the BLR shows time-dependent structural changes, from both the kinematic and the geometric points of view, and there may be a lag in its correlation with the \edit1{luminosity} of the AGN.
The diversity in BLR kinematics found in our moderately
high-luminosity sample appears consistent with that seen among AGN with
super-Eddington accretion rates~\cite[e.g.,][]{Du16b}.

\subsection{\hb~Line Widths} 
Following the procedure outlined by~\cite{Barth15}, 
we measured the line width of the broad \hb~component from both the
spectrally-decomposed mean spectrum and the rms spectrum. 
The mean spectrum constructed from the time-series data 
has a continuum level of zero with just residual noise from spectral
fitting, but the continuum of the rms spectrum incorporates
photon-counting errors in addition to residual noise. The noise continuum is
removed after a simple linear fit before making line-width
measurements. 

We computed the line width using two separate parameters according to
\cite{Peterson04}: the full width at half-maximum intensity (FWHM) of the line, and
the line dispersion $\sigma_{\rm line}$.
The FWHM of a single-peaked line is defined as the difference between the wavelengths blueward and redward of the peak $P(\lambda)_{\rm max}$ that
correspond to half of its height. A double-peaked line requires additional attention to defining
$P(\lambda_{\rm blue,red})_{\rm max}$ for the blue and red peaks,
respectively, and thus their corresponding wavelengths at half the
maximum height, but otherwise the procedure in determining the
difference between the resulting wavelengths is similar. As for
$\sigma_{\rm line}$ (square root of the second moment of the \hb~line profile), we
adopted from~\cite{Peterson04} the following: 

\begin{equation}
\sigma^2_{\rm line}(\lambda) = \langle\lambda^2\rangle - \lambda_0^2  \quad,
\end{equation}

where 

\begin{equation}
\langle\lambda^2\rangle = \frac{\int \lambda^2 P(\lambda)
  d\lambda}{\int P(\lambda) d\lambda} 
\end{equation}

and 

\begin{equation}
\lambda_0 = \frac{\int \lambda P(\lambda) d\lambda}{\int P(\lambda)
  d\lambda} \quad.
\end{equation}
In each of these calculations, we also varied the endpoints of the \hb~extraction window randomly within $\pm 5$\,\AA\ ten times in order to account for uncertainties brought about by the precise choice of the \hb~spectral limits. The resulting line-width measurement is the mean from these ten trials.

For error analysis associated with the measured width parameters, we
undertook a Monte Carlo bootstrapping procedure and simulated 100 realizations of
the mean and rms spectra, respectively, based on random subset
sampling of the existing nightly spectra. For a source with $n$ epochs of observations, we simulate new mean and rms spectra of the broad \hb~component from combining $n$ randomly-drawn time-series spectra with repetition allowed. We repeated our measurement
of the FWHM and $\sigma_{\rm 
  line}$ on these new spectra, where the means and
the standard deviations of the resulting distributions were taken as
the line widths and their corresponding uncertainties, respectively.  

The final line-width measurements are corrected for instrumental
broadening subtracted in quadrature from the observed line width
$\Delta \lambda_{\rm obs}$: 
\begin{equation}
\Delta\lambda^2_{\rm intrinsic} = \Delta \lambda^2_{\rm obs} - \Delta
\lambda^2_{\rm inst} \quad.
\end{equation}
Given that we used the same blue-side spectral setup as the LAMP2011 campaign,
we adopt from~\cite{Barth15} the instrumental FWHM of 380 km s$^{-1}$
and dispersion $\sigma_{\rm inst} \approx 162$ km s$^{-1}$ for our small
corrections. The resulting FWHM and $\sigma_{\rm line}$ measured from
both the mean and the e-rms spectra are presented in Table
\ref{tbl:linewidths}. 

\begin{deluxetable*}{lr@{ $\pm$ }lr@{ $\pm$ }lr@{ $\pm$ }lr@{ $\pm$ }l}[htb]
\tablecaption{Rest-Frame Line Widths of the Broad \hb~Component\label{tbl:linewidths}}
\tablecolumns{9}
\tablewidth{0pt}
\tablehead{
\colhead{} &
\multicolumn{4}{c}{Mean} &
\multicolumn{4}{c}{rms} \\
\colhead{Object} &
\multicolumn{2}{c}{FWHM} &
\multicolumn{2}{c}{$\sigma_{\rm line}$} &
\multicolumn{2}{c}{FWHM} &
\multicolumn{2}{c}{$\sigma_{\rm line}$}
}
\startdata
        Zw 535$-$012 &  2705 &    33 &  1474 &    28 &  2005 &    80 &  1259 &   112 \\ 
              I Zw 1 &  2187 &   920 &  1644 &   102 &  2665 &   700 &   916 &   165 \\ 
            Mrk 1048 &  4830 &    80 &  1840 &    58 &  4042 &   406 &  1726 &    76 \\ 
             Ark 120 &  5656 &    44 &  2125 &    41 &  4765 &    84 &  1882 &    42 \\ 
             Mrk 376 & 10055 &   341 &  2662 &    38 & 14438 &  1439 &  3633 &   210 \\ 
               Mrk 9 &  3751 &    71 &  1555 &    68 &  3159 &   144 &  1326 &    69 \\ 
             Mrk 704 &  8597 &   448 &  2002 &    63 &  8597 &   448 &  2009 &   102 \\ 
MCG $+$04$-$22$-$042 &  2658 &    57 &  1141 &    39 &  2120 &    39 &   977 &    29 \\ 
             Mrk 110 &  2048 &    13 &  1203 &    32 &  2052 &   100 &  1314 &    69 \\ 
            RBS 1303 &  2286 &    21 &  1243 &    26 &  1738 &   113 &  1292 &   156 \\ 
             Mrk 684 &  2174 &    33 &  1216 &    74 &  2391 &   632 &  2010 &   773 \\ 
             Mrk 841 &  7073 &   311 &  2139 &    55 &  7452 &   660 &  2278 &    96 \\ 
            Mrk 1392 &  4267 &    25 &  1635 &    13 &  3690 &   138 &  1501 &    38 \\ 
      SBS 1518$+$593 &  3312 &    24 &  1782 &    21 &  4656 &  1692 &  2249 &   255 \\ 
              3C 382 &  7772 &    54 &  4218 &    67 & 13926 &  2337 &  5616 &    97 \\ 
    NPM1G $+$27.0587 &  3501 &    28 &  1683 &    42 &  2893 &   177 &  1735 &   136 \\ 
   RXJ 2044.0$+$2833 &  2196 &    31 &   989 &    32 &  2047 &    72 &   870 &    50 \\ 
       PG 2209$+$184 &  4045 &    34 &  1573 &    40 &  3247 &    88 &  1353 &    64 \\ 
       PG 2214$+$139 &  5133 &    17 &  1947 &    55 &  4867 &   473 &  1984 &    90 \\ 
            RBS 1917 &  2399 &    11 &  1180 &    50 &  1653 &   287 &   851 &   154 \\ 
             Mrk 315 &  3097 &   252 &  1824 &    67 &  3411 &  1438 &  2031 &   260 
\enddata
\tablecomments{All line-width measurements are in km s$^{-1}$.}
\end{deluxetable*}

\subsection{Virial Black Hole Masses}
The mass of a black hole may be determined most robustly using 
dynamical modeling codes such as CARAMEL~\cite[e.g.,][]{Pancoast11}. Nonetheless, virial
estimates are useful within uncertainties that depend on the BLR
geometry. The black hole mass and its associated uncertainty are computed 
and propagated from the virial equation
\begin{equation}
M_{\rm virial} = f \frac{c \tau v^2}{G} \quad , 
\end{equation}
\begin{eqnarray}
\sigma_M = \sqrt{\left(\frac{\delta M}{\delta \tau} \sigma_\tau\right)^2+\left(\frac{\delta M}{\delta v} \sigma_v\right)^2} \nonumber \\
 = M \sqrt{\left(\frac{\sigma_\tau}{\tau}\right)^2 + 4\left(\frac{\sigma_v}{v}\right)^2} \quad ,
\end{eqnarray}
where $f$ is the scaling factor that depends on the geometry and
kinematics of the BLR, $c$ is the speed of light, $\tau \pm \sigma_\tau$ is the time delay with uncertainties, $v \pm \sigma_v$ is the velocity
of the BLR gas with uncertainties as measured by the line width, and $G$ is the
gravitational constant. The empirically fitted virial
factor from~\cite{Woo15} is consistent with 
that from CARAMEL dynamical modeling~\cite[][]{Pancoast14b}, so we have adopted for our black hole mass calculations $\langle f \rangle = 10^{0.65} = 4.47$ from~\cite{Woo15}, $\tau =
\tau_\mathrm{cen}$, and $v = \sigma_\mathrm{line}$(rms) \edit1{from~\cite{DallaBonta20}} to facilitate comparisons with other studies.  We
present both the virial products $c \tau_{\rm cen} \sigma^2_{\rm line}
/ G$ and the derived $f$-dependent black hole masses in Table \ref{tbl:bhmass}. Comparisons of our results to existing prior black hole mass measurements for several sources can be found in \edit1{the} Appendix. 
Overall, we find agreements between our measurements and those from the literature within uncertainties.

\begin{deluxetable}{lcc}[htb]
\tablecaption{Virial Products and Derived BH Masses \label{tbl:bhmass}}
\tablecolumns{3}
\tablewidth{0pt}
\tablehead{
\colhead{Object} &
\colhead{$c \tau_{\rm cen} \sigma^2_{\rm line} / G$} &
\colhead{$M_{\rm BH}^\dag$} \\
\colhead{} &
\colhead{(10$^7~M_\odot$)} &
\colhead{(10$^8~M_\odot$)}
}
\startdata
        Zw 535$-$012 &   0.85$^{+  0.36}_{-  0.18}$ &   0.38$^{+  0.16}_{-  0.08}$ \\
            Mrk 1048 &   0.50$^{+  0.60}_{-  0.64}$ &   0.22$^{+  0.27}_{-  0.29}$ \\
             Ark 120 &   1.62$^{+  0.55}_{-  0.38}$ &   0.73$^{+  0.25}_{-  0.17}$ \\
               Mrk 9 &   0.91$^{+  0.21}_{-  0.41}$ &   0.41$^{+  0.09}_{-  0.18}$ \\
             Mrk 704 &   2.30$^{+  0.91}_{-  0.79}$ &   1.03$^{+  0.41}_{-  0.36}$ \\
MCG $+$04$-$22$-$042 &   0.34$^{+  0.07}_{-  0.05}$ &   0.15$^{+  0.03}_{-  0.02}$ \\
             Mrk 110 &   0.77$^{+  0.14}_{-  0.15}$ &   0.35$^{+  0.06}_{-  0.07}$ \\
            RBS 1303 &   0.57$^{+  0.09}_{-  0.12}$ &   0.25$^{+  0.04}_{-  0.06}$ \\
             Mrk 841 &   1.04$^{+  0.58}_{-  0.35}$ &   0.47$^{+  0.26}_{-  0.16}$ \\
            Mrk 1392 &   1.40$^{+  0.17}_{-  0.20}$ &   0.63$^{+  0.08}_{-  0.09}$ \\
      SBS 1518$+$593 &   1.23$^{+  0.49}_{-  0.56}$ &   0.55$^{+  0.22}_{-  0.25}$ \\
              3C 382 &   3.12$^{+  2.51}_{-  2.19}$ &   1.39$^{+  1.12}_{-  0.98}$ \\
    NPM1G $+$27.0587 &   0.51$^{+  0.37}_{-  0.30}$ &   0.23$^{+  0.16}_{-  0.13}$ \\
   RXJ 2044.0$+$2833 &   0.27$^{+  0.04}_{-  0.04}$ &   0.12$^{+  0.02}_{-  0.02}$ \\
       PG 2209$+$184 &   0.66$^{+  0.14}_{-  0.14}$ &   0.29$^{+  0.06}_{-  0.06}$ \\
            RBS 1917 &   0.32$^{+  0.12}_{-  0.12}$ &   0.14$^{+  0.06}_{-  0.05}$ \\
\enddata
\tablecomments{$^\dag$Assuming $f = 4.47$~\cite[][]{Woo15}.}
\end{deluxetable}

\section{Conclusions}
\label{sec:conclusions}

We present the first results from our 100-night LAMP2016 campaign
where we monitored a sample of 21 luminous Seyfert 1 nuclei with AGN luminosity
$\lambda L_{\lambda}$(5100\,\AA) $\approx 10^{44}$ erg s$^{-1}$ during April
2016 -- May 2017 at Lick Observatory. Our analysis here provided 
the \hb~emission-line light curves and integrated \hb~lag
detections measured against $V$-band continuum photometric light
curves for the full sample. We further assessed the significance of
the lag determinations, and computed, for a subset of sources with
good-quality lags, their velocity-resolved reverberations, inferred
BLR kinematics, broad \hb~line widths, and virial black hole mass estimates.

The unusually rainy weather at Mount Hamilton during the winter months of 2016--2017 hampered our ability to monitor the variability of our AGN as closely as would have been ideal, leaving large gaps in several of our emission-line light curves. Consequently, we were not able to measure reliable lags for six of our AGN owing to inadequate light curves. Our overall results nearly double the number of existing velocity-resolved lag measurements, particularly at the moderately high-luminosity regime among previous reverberation-mapped samples, and revealed a
diversity of \edit1{signatures potentially indicative of different} BLR gas kinematics. 
\edit1{Given the complexity of the multiple factors involved, the lack of any clear correlation between AGN luminosity and velocity-resolved lag structure suggests that luminosity itself is not the sole factor responsible for setting the kinematic state of the BLR. It may also}
depend on other properties of the accreting black holes.
Follow-up direct dynamical modeling work will shed light on the
detailed kinematics and provide robust dynamical black hole masses to
help further calibrate widely-adopted black hole scaling relations. 

\begin{acknowledgments}

We thank the anonymous referee for thoughtful comments and great suggestions that significantly improved this manuscript.  We are grateful to M. Fausnaugh for offering help with their \texttt{mapspec} code
used to test different spectral
scaling of our Shane/Kast data, and M. Bentz for providing her table
of values for the radius--luminosity relationship.
We acknowledge the following individuals for their time and effort
contributed to the Lick observing campaign: (Shane) Zachary Parsons, Estefania
Padilla Gonzalez, Noah Rivera, Cristilyn Gardner, Jake Haslemann, Sean
Lewis, and Ellen Glad; (Nickel) Nick Choksi, Sameen Yunus, Jeff Molloy, Andrew
Rikhter, and Haynes Stephens. 
Photometric data collection at MLO was supported by NSF grant AST-1210311; we thank Robert Quimby, Emma Lee, Joseph Tinglof, Eric McLaughlin, Amy Igarashi, and Tariq Johnson for assistance with these observations.

We are deeply grateful to the UCO/Lick staff for help with scheduling and supporting the
observations. Research at Lick Observatory is partially supported by a
generous gift from Google. The Kast red CCD detector upgrade, led by
B. Holden, was made
possible by the Heising-Simons Foundation, William and Marina Kast, and the
University of California Observatories. 
KAIT and its ongoing operation were made possible by donations from Sun Microsystems, Inc., the Hewlett-Packard Company, AutoScope Corporation, Lick Observatory, the NSF, the University of California, the Sylvia \& Jim Katzman Foundation, and the TABASGO Foundation.
Data presented herein were obtained using the UCI Remote Observing Facility, made possible by a generous gift from John and Ruth Ann Evans.

Research at UC Irvine has been supported by NSF grants AST-1412693 and AST-1907208. 
V.U acknowledges funding support from the University of California Riverside's Chancellor's
Postdoctoral Fellowship and NASA Astrophysics Data Analysis Program Grant \#80NSSC20K0450. Her work was conducted in part at the Aspen Center for Physics, which is supported by NSF grant PHY-1607611; she thanks the Center for its hospitality during the ``Astrophysics of Massive Black Holes Merger" workshop in June and July 2018. 
T.T. acknowledges support by the Packard Foundation through a Packard research fellowship.
V.N.B. and I.S. gratefully acknowledge assistance from NSF Research at Undergraduate Institutions (RUI) grants AST-1312296 and AST-1909297. Note that findings and conclusions do not necessarily represent views of the NSF. 
G.C. acknowledges NSF support under grant AST-1817233. 
J.H.W. acknowledges the funding from the Basic Science Research Program through the National Research Foundation of Korean Government (NRF-2021R1A2C3008486). 
A.V.F.'s group at U.C. Berkeley is grateful for support from the TABASGO Foundation, the Christopher R. Redlich Fund, the Miller Institute for Basic Research in Science (in which he is a Miller Senior Fellow), and many individual donors.
K.H. acknowledges support from STFC grant ST/R000824/1.
We acknowledge the generous support of Marc J. Staley, whose fellowship partly funded B.E.S. whilst contributing to the work presented herein as a graduate student.
I.S. acknowledges support from the Deutsche Forschungsgemeinschaft (DFG, German Research Foundation) under Germany’s Excellence Strategy --- EXC 2121 "Quantum Universe" --- 390833306.
Research by S.V. is supported by NSF grants AST–1813176 and AST-2008108.

This work makes use of observations from the LCOGT network.
The Liverpool Telescope is operated on the island of La Palma by
Liverpool John Moores University in the Spanish Observatorio del Roque
de los Muchachos of the Instituto de Astrofisica de Canarias with
financial support from the UK Science and Technology Facilities
Council. 
Based on observations acquired at the Observatorio Astron\'{o}mico
Nacional in the Sierra San Pedro M\'{a}rtir (OAN-SPM), Baja California,
M\'{e}xico, we thank the daytime and night support staff at the OAN-SPM
for facilitating and helping obtain our observations.
Some of the data used in this paper were acquired with the RATIR instrument, funded by the University of California and NASA Goddard Space Flight Center, and the 1.5-meter Harold L.\ Johnson telescope at the Observatorio Astron\'omico Nacional on the Sierra de San Pedro M\'artir, operated and maintained by the Observatorio Astron\'omico Nacional and the Instituto de Astronom{\'\i}a of the Universidad Nacional Aut\'onoma de M\'exico. Operations are partially funded by the Universidad Nacional Aut\'onoma de M\'exico (DGAPA/PAPIIT IG100414, IT102715, AG100317, IN109418, IG100820, and IN105921). We acknowledge the contribution of Leonid Georgiev and Neil Gehrels to the development of RATIR.
This research was made possible through the use of the AAVSO
Photometric All-Sky Survey (APASS), funded by the Robert Martin Ayers
Sciences Fund and NSF grant AST-1412587 and contributed by observers worldwide.
We acknowledge the use of The AGN Black Hole Mass Database as a
compilation of some of the reverberation mapped black hole masses
prior to 2015~\cite[][]{Bentz15}.
This research has made use of the
NASA/IPAC Extragalactic Database (NED), which is operated by the Jet
Propulsion Laboratory, California Institute of Technology, under
contract with the National Aeronautics and Space Administration. 

\end{acknowledgments}

\facility{Shane (Kast double spectrograph), KAIT, Nickel, LCOGT,
  MLO:1m, OANSPM:1.5m, FLWO:1.2m, BYU:0.9m, Liverpool:2m, AAVSO}

\software{mpfit~\citep{Markwardt2009}, CARMA~\citep{Kelly09,Kelly14}, IDL Astronomy User's Library~\citep{Landsman1993}, IRAF~\citep{Tody86}, CARAMEL~\cite[][Villafa\~na et al., in prep.]{Pancoast11,Pancoast14a}, mapspec~\citep{Fausnaugh17_mapspec}}

\appendix
\restartappendixnumbering

\section{Comparison With Prior $M_{\rm BH}$ Measurements}
\label{sec:compBH}
\edit1{RM studies over the years have shown the intrinsic scatter among BH mass measurements from the RM technique to be inherently small, demonstrating its robustness for measuring BH masses. In the well-studied case of NGC 5548 with many epochs of RM-based BH masses~\cite[see][for a recent compilation]{DallaBonta20}, the distribution of these masses exhibits an intrinsic scatter of about 0.2 dex, suggesting that reverberation BH mass measurements for individual sources may typically fall within this range.}
\edit1{Here we examine the level of agreement between our results and those from the literature for the AGN where BH mass measurements from previous RM campaigns are available.
Given the expected $\sim0.2$ dex scatter in mass estimates obtained from light curves at different epochs, our measurements for the majority of these sources are generally consistent with past $M_\mathrm{BH}$ estimates, with a factor of 2.7 difference (0.44 dex) in the most discrepant case.}

\subsection{Ark 120}
Ark 120 was monitored by~\cite{Peterson98}
and~\cite{Peterson04} who had measured a rest-frame \hb~lag of
47.1$^{+8.3}_{-12.4}$ and 37.1$^{+4.8}_{-5.4}$ days and
$\sigma_{\rm line}$ of 1959$\pm$109 and 1884$\pm$48 km s$^{-1}$ from two different epochs,
respectively.
Adopting the same $\langle f \rangle$ of 4.47
from~\cite{Woo15} as in this study, the black hole masses calculated 
are 1.57$^{+0.45}_{-0.59} \times$ 10$^8$ M$_\odot$ 
and 1.15$^{+0.21}_{-0.23} \times$ 10$^8$ M$_\odot$, respectively.

In comparison, we measured a shorter lag of 18.7$^{+5.9}_{-4.5}$
days in \hb, resulting in a smaller black hole mass of
\edit1{0.73$^{+0.25}_{-0.17} \times 10^8$ M$_\odot$}.  In a concurrent
campaign, the MAHA group determined an \hb~lag of 16.2$^{+3.2}_{-3.1}$ days
and consequently a black hole mass of 0.68$^{+0.14}_{-0.13} \times
10^8$ (using the same virial factor) in Ark 120~\cite[][]{Du18b}. 
\edit1{Our measurement ($\log (M_\mathrm{BH}/M_\odot) = 7.86$) is very similar to the MAHA result ($\log (M_\mathrm{BH}/M_\odot) = 7.83$), and roughly agree with the previous masses ($\log (M_\mathrm{BH}/M_\odot) = 8.20$) from over a decade ago within $\pm$ 0.2 dex,} indicating that the state of the AGN and the intrinsic BLR properties \edit1{stay generally consistent} in the past two decades.

 \subsection{Mrk 704}
 Mrk 704 was among five bright Seyfert 1 galaxies that were targeted
 by~\cite{DeRosa18} in an RM campaign in early 2012.~\cite{DeRosa18}
 measured an \hb~lag of 12.65$^{+1.49}_{-2.14}$ days and $\sigma_{\rm
   line}$ of 1860$^{+108}_{-130}$ km s$^{-1}$, resulting in a black
 hole mass of 0.43$^{+0.16}_{-0.12} \times 10^8$ M$_\odot$, \edit1{or $\log (M_\mathrm{BH}/M_\odot) = 7.63$}. Four years
 later, we determined a longer lag of 28.9$^{+10.2}_{-10.0}$ days and
 a larger $\sigma_{\rm line}$ of 2009$\pm$102 km s$^{-1}$. Our black
 hole mass measurement of \edit1{1.03$^{+0.41}_{-0.36} \times 10^8$ M$_\odot$}, \edit1{or $\log (M_\mathrm{BH}/M_\odot) = 8.01$,} 
 \edit1{is more than twice as large as the result from~\cite{DeRosa18}, but our measurement's uncertainty is substantial such that there is statistical agreement within 1.2$\sigma$.}

 \subsection{Mrk 110}
Like Ark 120, Mrk 110 was another RM target monitored by~\cite{Peterson98}
and~\cite{Peterson04} within the time period and had a rest-frame \hb~lag of
20.4$^{+10.5}_{-6.3}$, 24.3$^{+5.5}_{-8.3}$, and 33.3$^{+14.9}_{-10.0}$ days with $\sigma_{\rm line}$ of 1115$\pm$103, 1196$\pm$141, 755$\pm$29 km s$^{-1}$ from three different epochs, respectively.
Adopting the same $\langle f \rangle$ of 4.47
from~\cite{Woo15} as in this study, the black hole mass
calculated correspond to
2.2$^{+1.5}_{-1.1} \times$ 10$^7$ M$_\odot$,
3.0$^{+1.4}_{-1.8} \times$ 10$^7$ M$_\odot$, and  
1.7$^{+0.9}_{-0.6} \times$ 10$^7$ M$_\odot$, or a combined value of 
2.3$^{+0.8}_{-0.7} \times$ 10$^7$ M$_\odot$ \edit1{($\log (M_\mathrm{BH}/M_\odot) = 7.36$).}

With a measured \hb~lag of 27.8$^{+4.3}_{-5.1}$ days and a $\sigma_{\rm
  line}$ of 1314$\pm$69 km s$^{-1}$, we obtained a larger black hole mass of
\edit1{3.5$^{+0.6}_{-0.7} \times 10^7$ M$_\odot$ ($\log (M_\mathrm{BH}/M_\odot) = 7.54$), consistent with the previous estimate within statistical uncertainties (1$\sigma$) and $\pm$0.2 dex.}

 \subsection{SBS 1518+593}
 SBS 1518+593 was one target from the MAHA
 campaign carried out by~\cite{Du18b} during a similar time frame as
 our LAMP study.~\cite{Du18b} established an \hb~lag of
 19.7$^{+9.9}_{-6.0}$ days and $\sigma_{\rm line}$ of 1038$\pm$20 km
 s$^{-1}$. The inferred black hole mass adjusting for $\langle f
 \rangle$ is 1.98$^{+1.00}_{-0.61}  \times 10^7$ M$_\odot$, \edit1{or $\log (M_\mathrm{BH}/M_\odot) = 7.30$}.
 Measuring a similar lag of 20.1$^{+8.4}_{-8.9}$ days and larger
 $\sigma_{\rm line}$ of 2249$\pm$255 km s$^{-1}$, our black hole mass
 for this AGN was \edit1{5.5$^{+2.2}_{-2.5} \times 10^7$  M$_\odot$}, \edit1{or $\log (M_\mathrm{BH}/M_\odot) = 7.74$. Our measurements agree statistically within 1.3$\sigma$ given the large uncertainties, with a difference roughly within $\pm$0.2 dex.} However, our \edit1{velocity-resolved lag spectra} disagree qualitatively --- ~\cite{Du18b} found the BLR kinematics to be in virialized motion while we see infalling behavior. This discrepancy may be due to differences in our respective campaigns' monitoring durations, where they observed SBS 1518+593 for approximately 180 days while our coverage spanned 360 days. This discrepancy may also be due to the fact that our \hb~light curve exhibits much noise and scatter, which may in turn affect our velocity-resolved measurement.

\subsection{Zw 535-012}
 The black hole mass in Zw 535-012 was previously determined to be 2.6 $\times$
 10$^7$\,M$_\odot$~\citep{Wang09}, \edit1{or $\log (M_\mathrm{BH}/M_\odot) = 7.41$}. It was computed using the scaling law from~\cite{Greene05}, assuming $\log
 L$(\hb) = 42.1 erg s$^{-1}$ and FWHM(\hb) = 2555 km s$^{-1}$ from a
 single-epoch spectroscopic observation in 2006. In our campaign, we obtained a black hole mass of
\edit1{3.8$^{+1.6}_{-0.8} \times 10^7$\,M$_\odot$, or $\log (M_\mathrm{BH}/M_\odot) = 7.58$}, which is 1.4 times larger than the previous estimate but consistent within \edit1{statistical uncertainties (0.6$\sigma$) and $\pm$0.2 dex}.

\end{document}